\documentclass[11pt,preprint]{aastex}

\usepackage{array}
\usepackage{geometry}
\usepackage{amsmath}
\usepackage{amssymb}
\usepackage{graphicx}
\usepackage{float}
\usepackage{color}
\usepackage{enumitem}

\usepackage{hyperref}

\usepackage[]{natbib}

\usepackage{tikz}
\usepgflibrary{arrows}

\newcommand{\rulesep}{\unskip\ \vrule\ }

\shorttitle{Stokes-V DKIST observations for the 3D Coronal Magnetic Fields}

\begin{document}

\title{Justification of the use of Stokes-V Cryo-NIRSP/DKIST observations for the 3D Reconstruction of the Coronal Magnetic Field}

\author{Maxim Kramar\footnote{e-mail: kramar@hawaii.edu}, Haosheng Lin}
\affil{Institute for Astronomy, University of Hawaii at Manoa, 34 Ohia Ku St, Pukalani, HI 96768, USA}

\begin{abstract}
This study presents a justification of the use of Stokes-V Cryo-NIRSP/DKIST observations for the 3D Reconstruction of the Coronal Magnetic Field. A magnetohydrodynamic (MHD) model of the solar corona during solar minimum generated by Predictive Science Inc. was used to synthesize spectropolarimetric measurements of the Fe XIII 1075 nm coronal emission line as observed from the Earth. 

Stokes-Q,U data is ``taken'' for half a solar rotation period (about two weeks) by Upgraded Coronal Multichannel Polarimeter (UCoMP) with field of view (FOV) up to $2\ R_\odot$. 
Stokes-V data is ``taken'' once a day for two days by Cryo-NIRSP/DKIST with a total FOV equal to the two Cryo-NIRSP FOVs. 

We demonstrated that even this amount Stokes-V observations with Cryo-NIRSP FOV coverage can remarkably improve the 3D coronal magnetic field reconstruction over an active region compared with tomography reconstruction based only on UCoMP linear polarization data. 
\end{abstract}


\section{Introduction}

Energetic solar eruptions are among the most dramatic and enigmatic stellar processes. The energetic particles and the disturbances in the heliosphere's magnetic fields caused by these eruptions have direct impact on near-earth space weather,  affecting power and communication infrastructures in space and on the ground that modern technology-based societies rely on to function. Detailed observational information of the physical properties of the solar atmosphere, including the magnetic field and thermodynamic structures, before, during, and after the eruptions are needed to help theoretical investigation to decipher the physical processes involved. They are also crucially needed for the development of the predictive capability to forecast the timing of the onset and the intensities and effects of the eruptions in near-earth space. 

Due to the low optical density of the coronal atmosphere, direct observational inference of the physical properties of the corona are inevitably constrained by the entanglement of coronal signals due to line of sight (LOS) integration of signals, except in situations where localized emission sources can be reasonably assumed, such as bright loops in active regions. To dis-entangle the line-of-sight (LOS) integrated signals to unveil the underlying 3D structure of the corona, tomographic inversion is needed. 

Tomography is the determination of the structure of a 3D object using measurements of line-of-sight integrated signals obtained from many different viewing directions. The reliability of scalar field tomographic inversion when sufficient measurements are available has been well-established, as exemplified in Computed Tomography (CT) scan used daily in hospitals around the world. Vector tomographic inversion of coronal magnetic fields was pioneered by Kramar and Inhester starting from 2006 \citep{Kramar_2006,Kramar_2007, Kramar_2013}, and culminated in the recent work that utilized the synoptic linear polarization data of the Fe XIII 1075 nm line obtained by {\it Coronal Multichannel Polarimeter} \citep[CoMP;][]{Tomczyk_2008SoPh} to demonstrate the first ``direct observation'' of the 3D vector magnetic field structure of the solar corona \citep{Kramar_2016ApJL}. While other methods, namely, magnetic field extrapolations and MHD simulation 
\citep{Wiegelmann_2017SSRv,Yeates_2018SSRv} have been extensively used to model the coronal magnetic fields, it is important to note that tomographic inversion is constrained by signals originating from the corona that respond directly to coronal magnetic fields. In comparison, current MHD simulations and in particular magnetic field extrapolation are only based on photospheric magnetic boundary conditions and are not constrained by coronal observations. Therefore, tomographic inversion provide a new tool for the study of the coronal magnetic fields and will provide new insight into the physics of the solar corona. 

The work of \cite{Kramar_2016ApJL} is a major step forward in our quest for direct observational inference of the coronal magnetic fields. However, the fidelity of the tomographically derived coronal magnetic field model is subject to several limiting factors related to the existing observing capabilities. Primarily, tomographic inversion with only CEL linear polarization data is not sensitive to certain magnetic field geometry \citep{Kramar_2013}. 
Also, since CEL linear polarization signals are not sensitive to the strength of the coronal magnetic field, constraints on the coronal magnetic field strengths in the inversion process have to be provided by other observations - primarily the photospheric magnetic field serving as the boundary condition of the reconstruction volume (Note that CEL circular polarization signals will provide {\it in-situ} constraints of the magnetic field strength, however, capabilities for synoptic observations of CEL circular polarization are still not available to date).
Nevertheless, the validity of the inverted results can be assessed by independent observations such as EUV CEL images that indirectly inform on the magnetic field configurations of the corona. Another limitation is that tomography relies on observations of the object under study from multiple observing directions. With earth-based observations from only one line of sight, we rely on the rotation of the sun to simulate observations from different line of sights. Therefore, current earth-bound tomographic inversion can only retrieve the psuedo-static component of the coronal magnetic fields.  Despite of these limitations, tomography is a new and powerful method to derive the 3D structure of the corona for the study of coronal physics. And with careful screening of the observational data and acute understanding of the strengths and limitations of the method, significant progresses can be made.

\subsection{Challenges of Coronal Measurements of the Magnetically Dependent Radiation.}

Although tools for measurement and intepretation of vector magnetic field measurements of the photospheric magnetic fields are well established, observational determination of the three-dimensional (3D) magnetic field structure of the corona remains one of the most difficult observational problems of solar physics. Information about the coronal magnetic fields $\vec{B}(\vec{r}; t)$, temperature $T(\vec{r}; t)$, and density $n(\vec{r}; t)$ are encoded in the continuous and polarized spectra of coronal emission lines (CELs) in ultraviolet (UV), visible, and IR wavelengths through normal and saturated Hanle effects and the Zeeman effect \citep{Judge_1998, Casini_1999, Lin_Casini_2000}. Here $\vec{r}$ is the 3D position vector of the coronal volume elements, and $t$ is time. The degree of linear polarization (LP) of the CELs due to the saturated Hanle effect are at the order $10^{-2}$ to $10^{-1}$, and CEL linear polarization of the full corona (up to $1.4 \ R_\odot$) are now been obtained routinely with small aperture coronagraphs \citep[e.g.][]{Tomczyk_2008SoPh}. However, the CEL LP signals are insensitive to the strength of the magnetic fields. On the other hand, the circular polarization (CP) signals of forbidden CELs are proportional to the longitudinal components of the coronal magnetic fields. However, due to the high temperature, low density, and weak magnetic field conditions of the corona, the amplitude of the CEL CP signals are at the order of 0.01 \% per Gauss for the Fe xiii 1075 nm line, making direct detection of these signals a very challenging observational problem. Although \cite{Lin_Penn_Tomczyk_2000,Lin_Kuhn_Coulter_2004} have demonstrated that direct detection of these extremely weak signals can be achieved with modern high performance c.     oronal spectropolarimeters, routine observations have not been achieved to date. However, with a 4-m aperture, the Daniel K. Inouye Solar Telescope (DKIST, \cite{Keil_2004IAUS}) is expected to allow for direct measurements of CEL circular polarization routinely by Cryo-NIRSP and DL-NIRSP instruments \citep{Woeger_2021AAS_DKIST,Rimmele_2020SoPh_DKIST}.

\subsection{Challenges in Interpretation of the Coronal Measurements.}

Due to the low optical density of the coronal atmosphere, direct observational inference of the physical properties of the corona are inevitably constrained by the entanglement of coronal signals due to line of sight (LOS) integration of signals, except in situations where localized emission sources can be reasonably assumed, such as bright loops in active regions (references)...

To demonstrate the complexity of the problem, 
we used a sophisticated MHD model from Predictive Science, Inc\footnote{\url{http://www.predsci.com}} and 
calculated Stokes emissivities and plasma parameters along a number of LOSes. 
The Stokes emissivities were calculated according to expressions in \cite{Casini_1999} and using 
the {\it Coronal Line Emission} (CLE 2.0) code \citep{Judge_Casini_2001ASPC}. 
Although recently \cite{Schad_Dima_2021ascl_pycelp} developed a more precise code for the calculation of the Coronal Emission Line Polarization (pyCELP), we use the CLE 2.0 code here which is not critical for the purpose of the paper. 

\begin{figure}[h!]
\includegraphics*[bb=65 567 635 1148,width=0.49\linewidth]{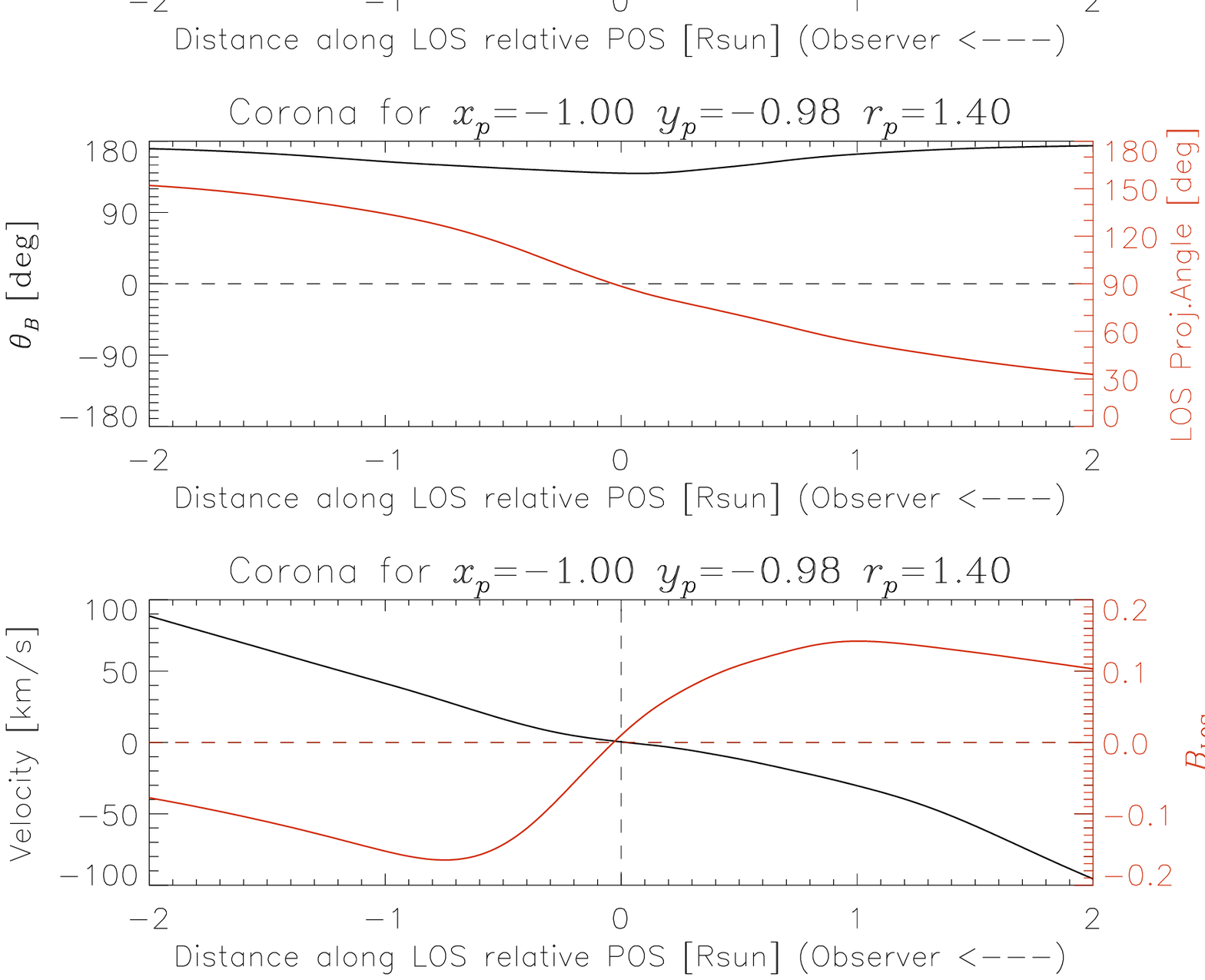}
\rulesep
\includegraphics*[bb=65 567 635 1148,width=0.49\linewidth]{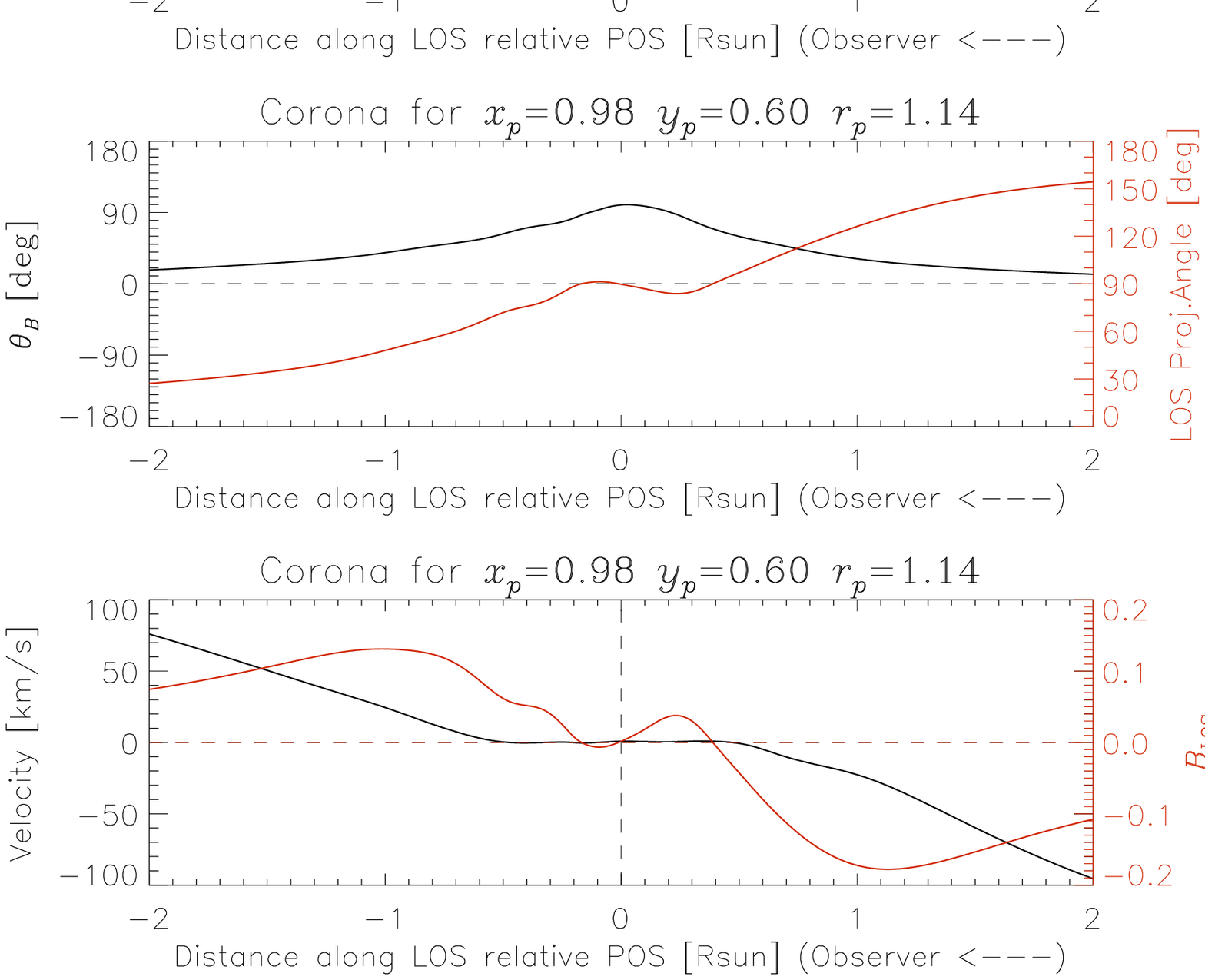}
\caption{Distributions of the electron density and temperature (top row), Stokes-I,Q,U emissivities (second row), and the magnetic field vector orientation angles (third row) over LOS for two different LOSes with image projection coordinates of $(-1.00;-0.98)$ and $(0.98;0.60)$ $R_\odot$. 
The absolute values of the Stokes parameters units here may not accurately represent the real situation which is not a point of concern for the purpose of this paper.}
\label{Fig_CLE_LOS}
\end{figure}

Figure \ref{Fig_CLE_LOS} shows distributions of the electron density, temperature, Stokes-I,Q,U emissivities over two particular LOSes. 
On the left panel, 
the density and temperature concentrated near POS resulting Stokes-I emissivity also to be concentrated near the  POS. 
On the left panels, the linear polarization (Stokes-Q,U) signal is also concentrated near the  POS. 
This may allow to apply a simplified approach, such as POS approximated solution, for inferring the magnetic field near POS. 
However, Stokes-Q,U signals depend also on the magnetic field vector orientation resulting in that their maximal contribution over LOS are not obligatory coincide with the density distribution over the LOS. This makes interpretation of the LP signal very difficult even when the density distribution over the LOS are concentrated near POS. This case is shown on the right panel. 


To our knowledge, only tomography methods can resolve this LOS problem. 
In next section, we describe the tomography methods to infer the 3D distribution of coronal plasma quantities: 
electron density, temperature, and vector magnetic field. 


\begin{figure}[h!]
\includegraphics*[bb=37 19 1390 978,width=\linewidth]{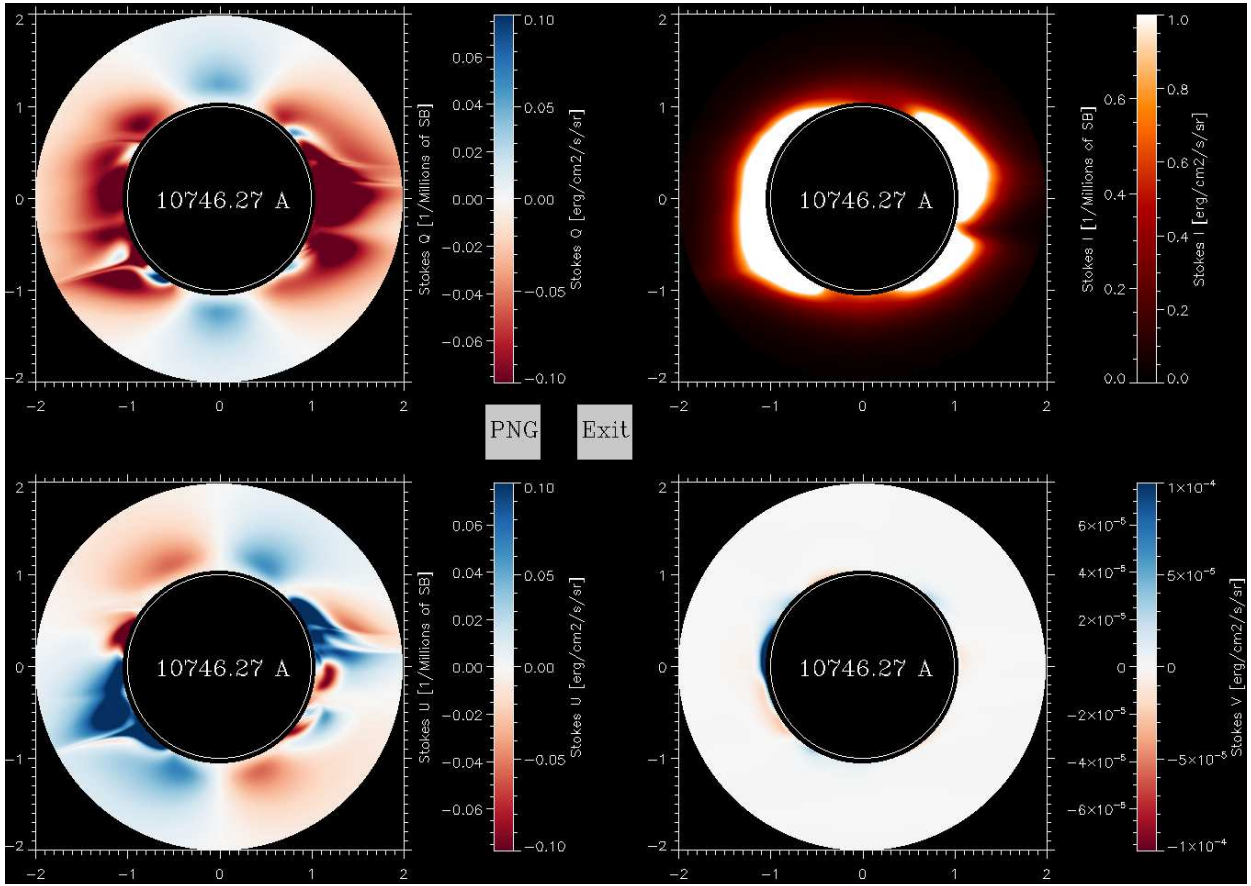}
\caption{Full Stokes synthetic observation maps of the Fe XIII 10747 \AA \ line (integrated over the line profile) for CR 2207.}
\label{Fig_CLE_LOS_map}
\end{figure}

%

\clearpage

\section{Vector Tomography of Coronal Magnetic Fields}

\subsection{General Formulation}
\label{Sec_Vec_Tomo}

Given CEL polarization observations from sufficient number of LOSs, the 3D magnetic field structure of the corona, $\vec B(\vec r)$ can be reconstructed by tomographic inversion method. The tomographic inversion function $F$ for the reconstruction of the coronal magnetic field $\vec B(\vec r)$ in its most extensive form, with observations of multiple spectral line, is given by 
\begin{eqnarray}
F(\vec B (\vec r)) = {\sum\limits_{i,j,l,s}} \left[ S^{\rm \ obs}_{l,s} (\hat e^i_{\rm LOS}, \vec \rho_j)
                                                                     -S^{\rm \ sim}_{l,s} (\hat e^i_{\rm LOS}, \vec \rho_j, \vec B(\vec r)) \right]^2 
                                   + \mu \sum \limits_q \left [ \nabla \cdot \vec{B}(\vec{r}) \right ]^2  
\label{MinFun_TomoVec}
\end{eqnarray}
where $S^{\rm \ obs}_{l,s}$ is the observed Stokes vector of spectral line $l$ or at the wavelength $\lambda_l$, with $s = 0, 1, 2, 3$ corresponds to the Stokes $I$, $Q$, $U$, and $V$ parameters, respectively; $S^{~\rm sim}_{l,s}$ is the synthesized Stokes data given $\vec B(\vec r)$. The summation integrate over all the LOSs, all plane-of-the-sky positions, all spectral lines, and all Stokes parameters.  Gauss's law, $ \nabla \cdot \vec{B}(\vec{r}) = 0$, with its squares summed over every voxel element $q$, is used as the regularization function to provide further constraint on $\vec B(\vec r)$ to help stabilize the inversion process. The simulated Stokes parameters are calculated using the emissivities of the Stokes parameters of the CEL,
\begin{equation}
S^{\rm \ sim}_{l,s} ({\hat e}_{\textrm{LOS}},\vec{\rho}) = k  \int \limits_\textrm{LOS} \varepsilon_{l,s}(\vec{r}) \mathrm{d} \ell ,
\label{Stokes_Para}
\end{equation}
where the coefficient $k$ accounts for instrument properties (resolution, efficiency, etc.). 

The theories of the polarization of the forbidden coronal emission lines have been investigated by several authors, and expressions of $\varepsilon_s$ for all the polarization states can be found in more recent publications in \cite{Judge_1998, Casini_1999} and \cite{Lin_Casini_2000}. 
The emissivities of all four components of the Stokes vector are influenced by both electron density and temperature $N_e(\vec r)$ and $T(\vec r)$. Therefore, the knowledge of 3D distributions of $N_e(\vec r)$ and $T(\vec r)$ are a prerequisite for the correct reconstruction of the magnetic fields. 
Our reconstruction method can use $N_e(\vec r)$ and $T(\vec r)$ obtained independently from multi-channel UV or multi-line coronal intensity observations to reduce the complexity of the vector tomographic inversion problem for $\vec B(\vec r)$. 
This approach was successfully implemented in \cite{Kramar_2016ApJL}.

\subsection{Vector Tomography based on Linear Polarization Data}
\label{Sec_Vec_Tomo_Hanle}

While the formulation presented above is valid for vector tomographic inversion of coronal $\vec B(\vec r)$ with full-Stokes polarization data input. However, no synoptic observation of CEL Stokes $V$ exist to date primarily due to their extremely low amplitudes \citep{Lin_Penn_Tomczyk_2000, Lin_2004}. Nevertheless, CoMP has been installed at the Mauna Loa Solar Observatory (MLSO) and conducting synoptic full-corona observation of the Fe XIII 1075 nm intensity and linear polarization out to $1.4 R_{\odot}$ since 2011 \citep{Tomczyk_2008SoPh}, and has provided valuable data for coronal research using forward modeling tool to synthesize the LOS-integrated linear polarization signals of coronal models. 
We have developed and tested a vector tomographic inversion code based only on CEL linear polarization observations and photospheric magnetic field boundary condition \citep{Kramar_2013} to take advantage of the new capability and the valuable dataset that have been accumulated, and applied this Linear Polarization Tomographic Inversion (LPTI) code to observations of CR 2112 to demonstrate the first direct observation of the 3D coronal magnetic fields \citep{Kramar_2016ApJL}. 

CEL linear polarization signals are not sensitive to $|\vec{B}|$. Lacking direct observational input of CEL Stokes $V$ data which respond directly to the line-of-sight component of the coronal magnetic field vector, the LPTI algorithm constraints $|\vec{B}|$ with the inclusion of the photospheric magnetic field as the inner boundary condition. The 2013 and 2016 studies provided qualitative description and validation of the inversion results by comparison of field lines morphology between the test and reconstructed coronal magnetic fields. 

\subsection{Vector Tomography based on Circular Polarization Data}
\label{Sec_Vec_Tomo_Zeeman}

Vector Tomography inversion based on Circular Polarization Data (Stokes-V component) can be performed similarly to the tomographic inversion of LP measurements using the minimization function (\ref{MinFun_TomoVec}). 
In this case quantities $S_{l,3}^\textrm{obs}$ and $S_{l,3}^\textrm{sim}$ are Stokes-V observed and LOS integrated simulated data: 
\begin{equation}
S_{l,3}^\textrm{sim}=V_{l}^\textrm{sim}=
k  \int \limits_\textrm{LOS} \varepsilon_V(\vec{r}) \mathrm{d} \ell ,
\label{StokesV_Para}
\end{equation}
where $\varepsilon_V(\vec{r})$ is local Stokes-V emissivity at the coronal position $\vec{r}$ and 
according to \cite{Casini_1999} it can be expressed as follows: 
\begin{equation}
\varepsilon_V(\vec{r})=
-\omega_\textrm{L} C_{JJ0} \phi\prime (\omega_0-\omega) 
\left[ \bar{g}+E_{JJ0} \sigma_0^2(\alpha_0 J) \right] \cos{\Theta}
\label{StokesV_Emiss}
\end{equation}
Here, $\phi(\omega_0-\omega)$ is the Voigt function describing the line profile around center frequency $\omega_0$, 
coefficient $C_{JJ0}$ depends on the population density of the excited level and Einstein coefficient for spontaneous recombination. 
Coefficients $E_{JJ0}$ is expressed through 6-$j$ and 9-$j$ symbols and Lande-factor. 
Term $\sigma_0^2(\alpha_0 J)$ represents a ratio of population density for two multipole orders. 
We see that coefficients $C_{JJ0}$ and $\sigma_0^2(\alpha_0 J)$ depend on the local density and temperature. 
Therefore, the determination of the 3D coronal electron density and temperature is required before the inversion of (\ref{MinFun_TomoVec}). 

In practice, as a first step in this inversion effort, we suggest to employ another approach. 
The second term in (\ref{StokesV_Emiss}) $E_{JJ0} \sigma_0^2(\alpha_0 J)$ is usually has low significance 
(although for some ions and magnetic field configurations it can be high enough). 
Then the term $C_{JJ0} \phi (\omega_0-\omega)$ is essentially Stokes-I emissivity \citep{Casini_1999} and therefore we get the "classic" magnetograph formula: 
\begin{equation}
\varepsilon_V(\vec{r},\omega_0-\omega)=\omega_\textrm{L} \bar{g} 
\frac{d \varepsilon_I(\vec{r},\omega_0-\omega)}{d \omega} \cos{\Theta}
\label{Magnetograph_Formula}
\end{equation}
Integrating the magnetograph formula (\ref{Magnetograph_Formula}) over LOS and neglecting the plasma velocity we get: 
\begin{equation}
V_{l,s}^\textrm{sim}(\omega)=k 
\int \bar{g} \frac{d \varepsilon_I(\vec{r},\omega_0(\vec{v})-\omega)}{d \omega} \omega_\textrm{L} \cos{\Theta} \mathrm{d} \ell , 
\label{Magnetograph_Formula_LOS}
\end{equation}
which means that a measured Stokes-V signal is proportional to a weighted LOS integral of the LOS projection of the local magnetic field vector with the weighting coefficient being the derivatives over the line profile of local Stokes-I emissivities. Or, simplifying: 
\begin{equation}
V_{l,s}^\textrm{sim}(\omega)=
\int K(\vec{r}) B(\vec{r}) \cos{\Theta} \mathrm{d} \ell = 
\int K(\vec{r}) B_\textrm{LOS}(\vec{r}) \mathrm{d} \ell ,
\label{StokesV_Tomo_Simplified}
\end{equation}
where $B$ and $B_\textrm{LOS}$ is the absolute value of the magnetic field vector and its projection on LOS, respectively. 

Employing this simplified description of the Stokes-V measurements allowed \cite{Kramar_2006} to investigate basic fundamental properties and limitations of the tomographic inversion for the coronal magnetic field based on this type of data. 

\subsection{Full-Stokes Vector Tomography}
\label{Sect_Vec_Tomo_full_Stokes}

\begin{figure}[H]
\includegraphics*[bb=23 549 603 742,width=0.5\linewidth]{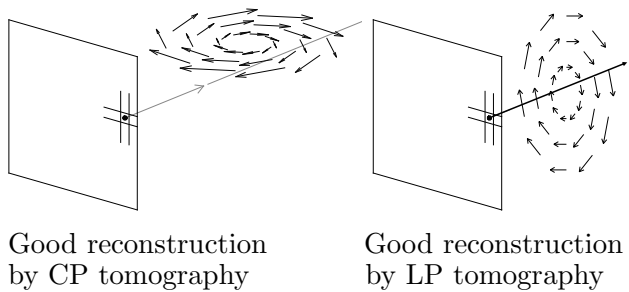}\\
\put(0,  0){Good reconstruction \ \ \ \ \ \ \ \ \ Good reconstruction}
\put(0,-12){by CP tomography \ \ \ \ \ \ \ \ \ \ \ by LP tomography}
\caption{The limitation of vector tomography based on partial polarization data observed from the ecliptic plane. LP tomography observations cannot resolve the field configuration on the left \citep{Kramar_2013}, while CP observation cannot be used to tomographically reconstruct the field configuration on the right \citep{Kramar_2006}.}
\label{Fig_LP_CP}
\end{figure}

While the Hanle effect vector tomographic inversion results have demonstrated the potential of tomography as a powerful tool for the study of coronal magnetic fields, vector tomography based on partial Stokes vector inputs is still subject to several limitations. 
Summarizing results from \cite{Kramar_2006,Kramar_2013}, 
Figure \ref{Fig_LP_CP} shows qualitative schematics for the two basic field configurations that are most sensitive to different types of the vector tomography depending on which type of partial polarization data the tomography is based on -- LP data (Hanle effect) or CP data (Zeeman effect). 
With current instrumentation efforts such as DKIST Cryo-NIRSP, DL-NIRSP, and Massively-multipleXed Coronal Spectropolarimetric Magnetometer (mxCSM; \cite{Lin_2016FrASS}) in progress to provide full-Stokes CEL observations on a routine basis, the development of a full-Stokes vector tomographic inversion algorithm is critical for the DKIST and mxCSM instruments to achieve their science objectives, and for the future of a space-based coronal magnetic fields mission. 

In this study we investigate how DKIST Cryo-NIRSP Stokes-V measurements can improve the 3D coronal magnetic field reconstruction using the Full-Stokes Vector Tomography method. 


\section{Quantitative Validation of Vector Tomography}
\label{Sec_VecTomo_validation}

To quantify the accuracy of the tomographic inversion based on LP (Hanle effect) and CP (Zeeman effect) data, 
we produced a synthetic Stokes data by integrating the local Stokes emissivities, 
$\varepsilon_{Q,U}$, over LOS-es using 3D coronal electron density, temperature, and vector magnetic field from the MHD model produced by Predictive Science Inc\footnote{\url{http://www.predsci.com}} (PSI). 

The field of view (FOV) for the synthetic LP observations is a disk with inner and outer heliocentric distances of $1.04$ and $2.0\ R_\odot$, respectively. 
This range can be routinely covered by modern coronal polarimetric observations with UCoMP or mxCSM. 

As the goal of this paper is to investigate the accuracy of the tomographic inversion, 
we do not need precise calculations of the $C_{JJ_0}$, $D_{JJ_0}$, $\sigma_0^2(\alpha_0 J)$ parameters. 
Therefore, similarly to \cite{Kramar_2013}, for calculations of Stokes-Q,U data we use approximated formulas from \cite{Querfeld_1982} 
which are based on the solution of the reduced version of \cite{House_1977} calculations. 

For simulation of Stokes-V data, we use the magnetograph formula approximation expressed by (\ref{StokesV_Tomo_Simplified}) in a similar manner as it was done in \cite{Kramar_2006}. 
Because the goal of this assessment is to explore the accuracy and possibility of the vector tomography method, 
this approach allows us to avoid the population levels calculations. 

The reconstruction domain consists of uniform rectangular cubic cells within 
a sphere with inner and outer heliocentric distances of $1$ and $2.5\ R_\odot$, respectively. 
As the inner and outer spherical boundaries cut through some cubical cells, 
the shape of these cut cell and magnetic flux through their boundaries are correspondingly corrected.

\subsection{The Ground Truth Model}
\label{Sec_VecTomo_Model}

The 3D coronal electron density, temperature, and vector magnetic field are result from the thermodynamic MHD model routinely produced by Predictive Science Inc. 
The PSI thermodynamic MHD model uses an improved equation for energy transport in the corona that includes parallel thermal conduction along the magnetic field lines, radiative losses, and parameterized coronal heating \citep{Lionello_2009}. 
The model is based on photospheric magnetic field data from SDO/HMI (ref). 
This thermodynamic MHD model produces more accurate estimates of plasma density and temperature in the corona than previous polytropic model. 
A detailed description is given by \cite{Mikic_2007,Lionello_2009}. 
Its application to the total solar eclipse of 1 August 2008 was described by \cite{Rusin_2010}. 
Recent development of the PredSci MHD model was tested by 21 August 2017 eclipse observations \citep{Mikic_2018NatAs}. 

\begin{figure}[h!]
\includegraphics*[bb=84 304 557 620,width=0.32\linewidth]{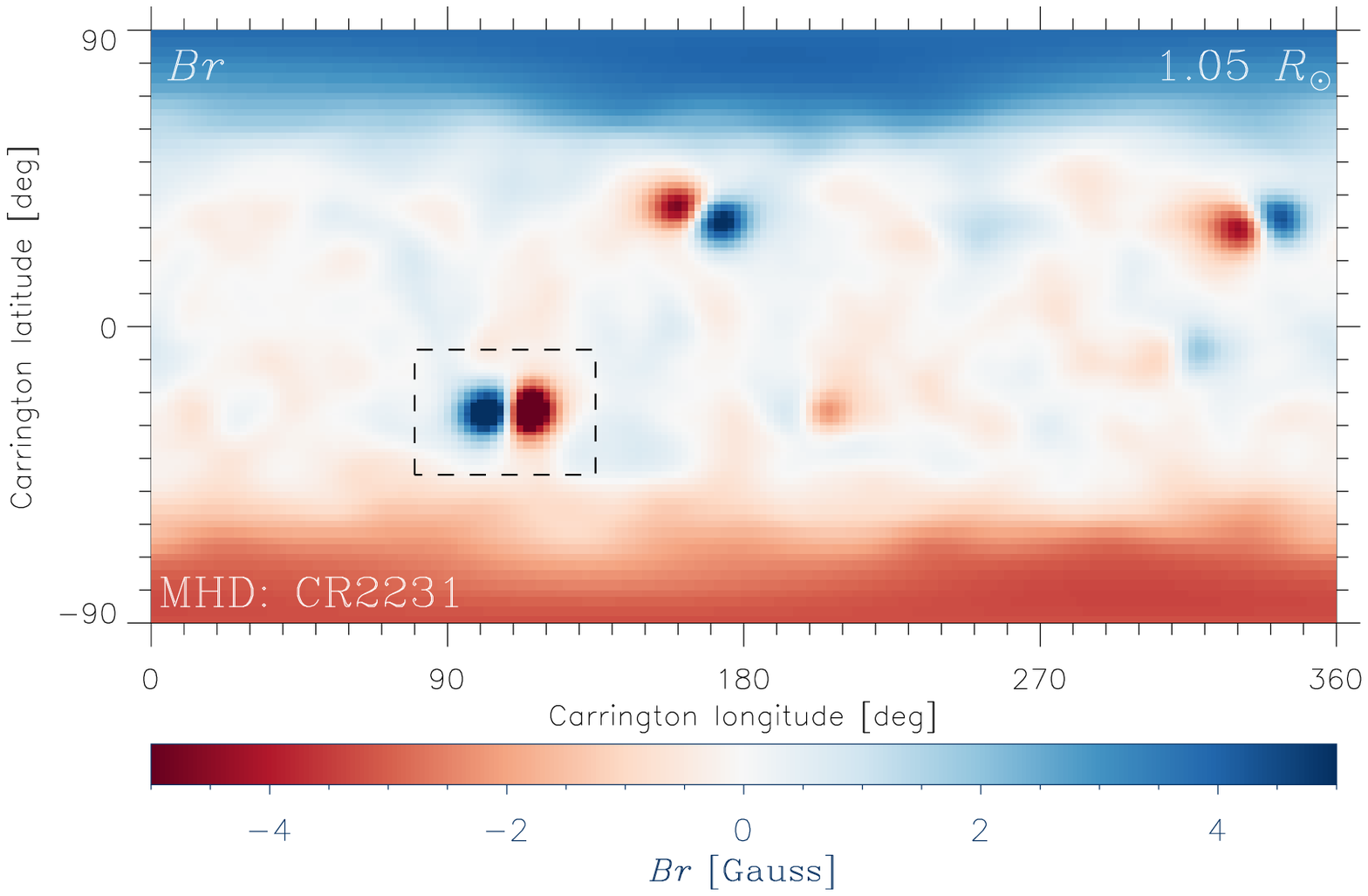}
\includegraphics*[bb=84 304 557 620,width=0.32\linewidth]{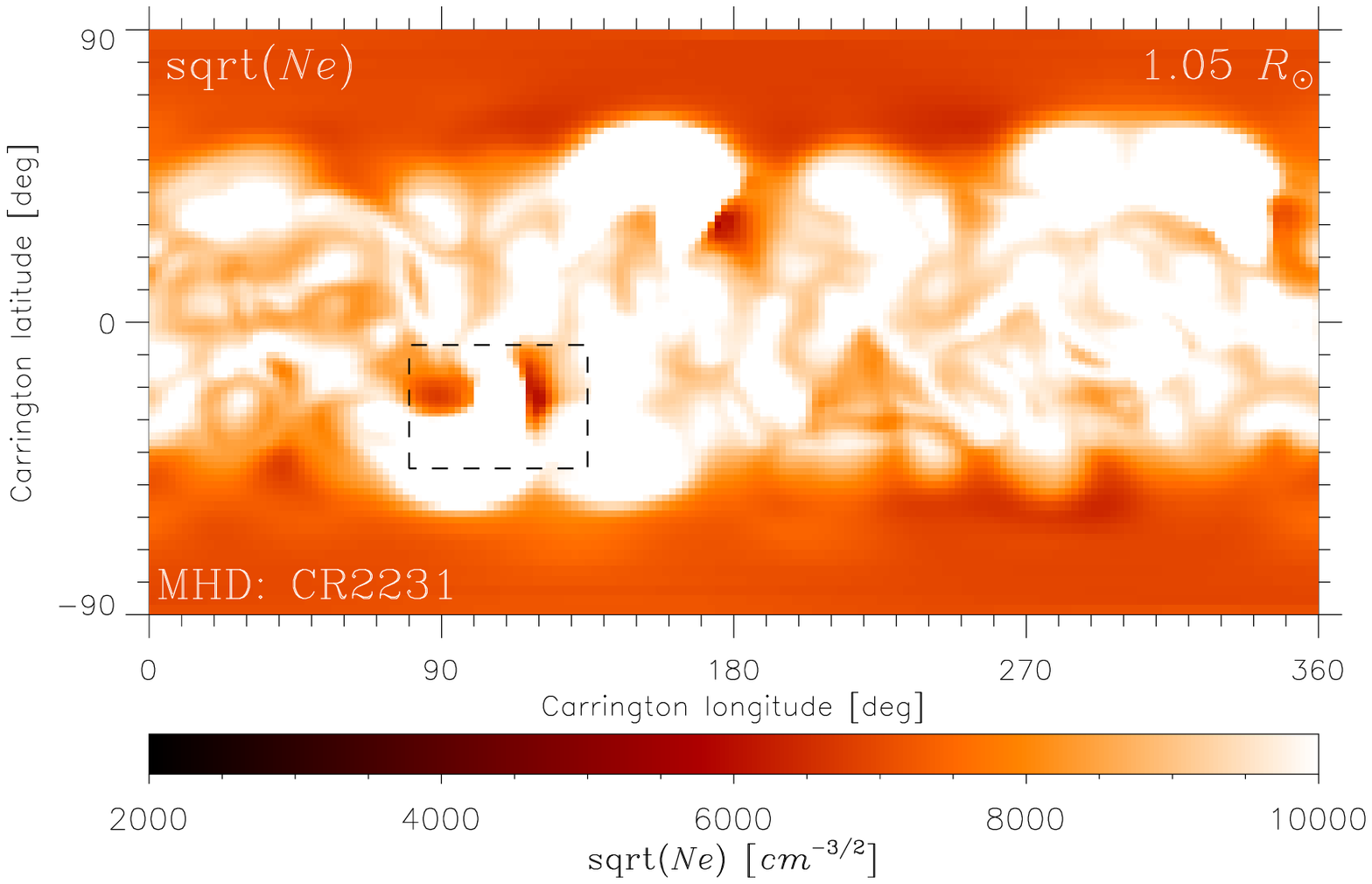}
\includegraphics*[bb=84 304 557 620,width=0.32\linewidth]{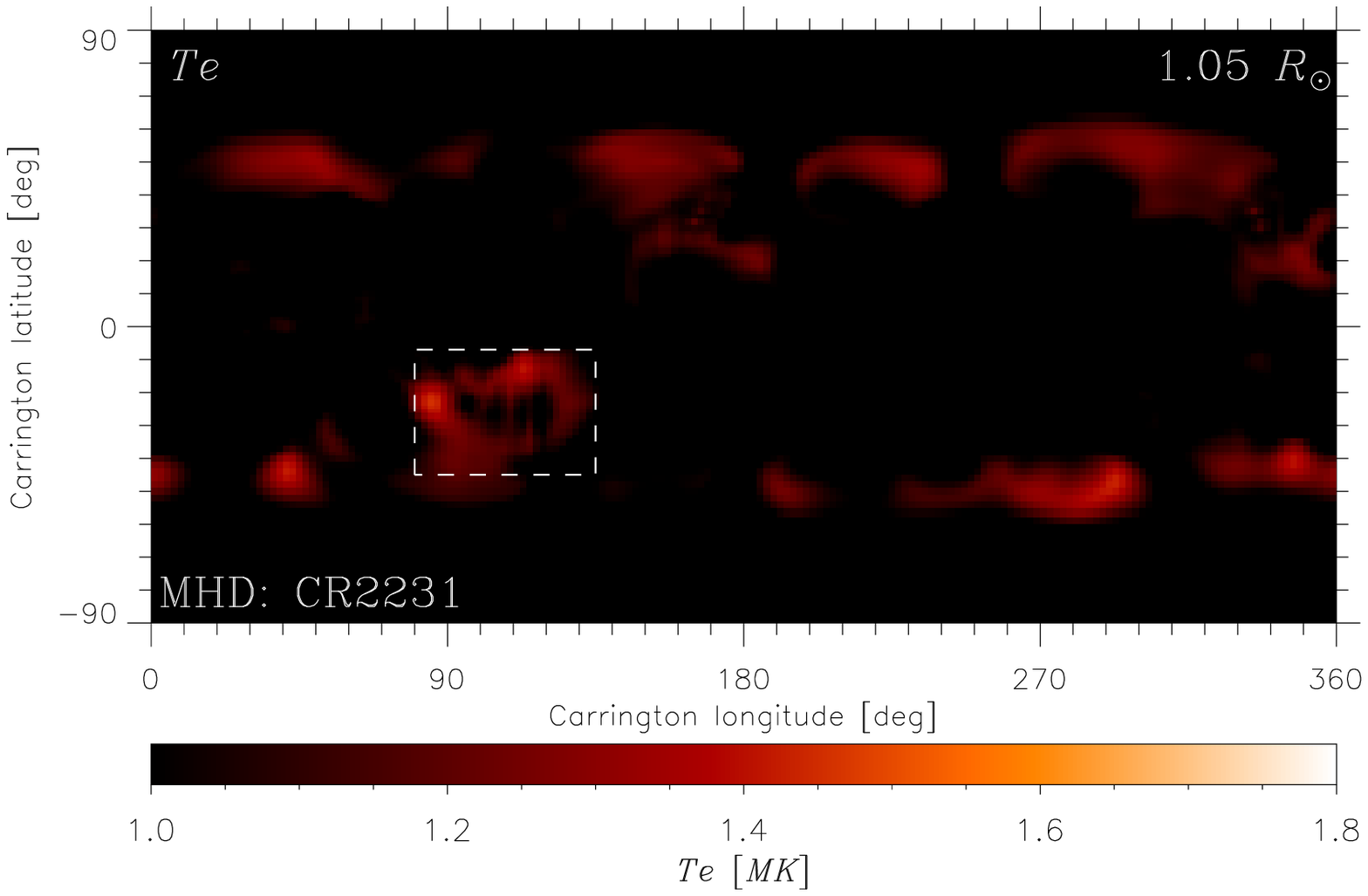}\\
\includegraphics*[bb=84 304 557 620,width=0.32\linewidth]{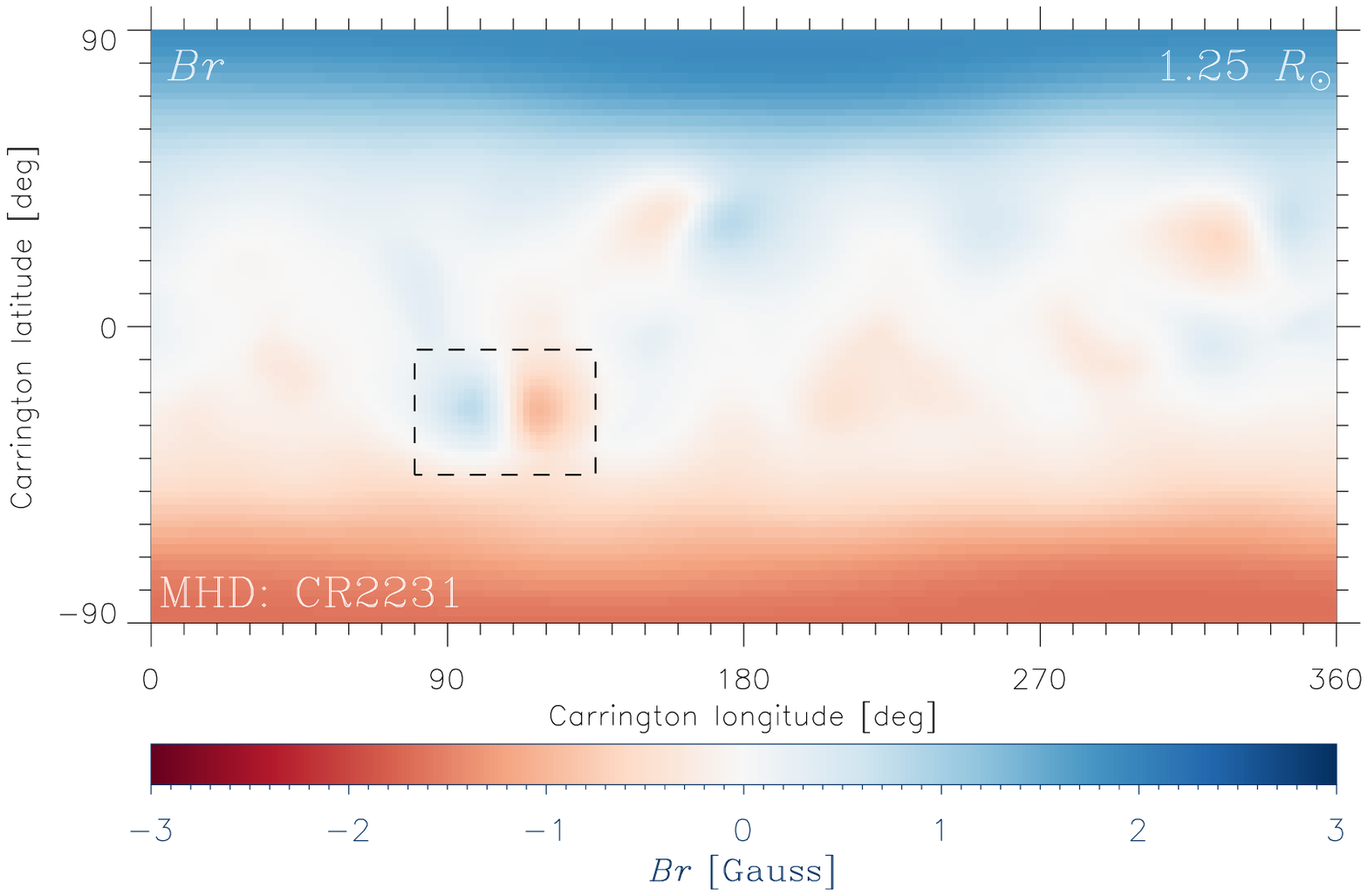}
\includegraphics*[bb=84 304 557 620,width=0.32\linewidth]{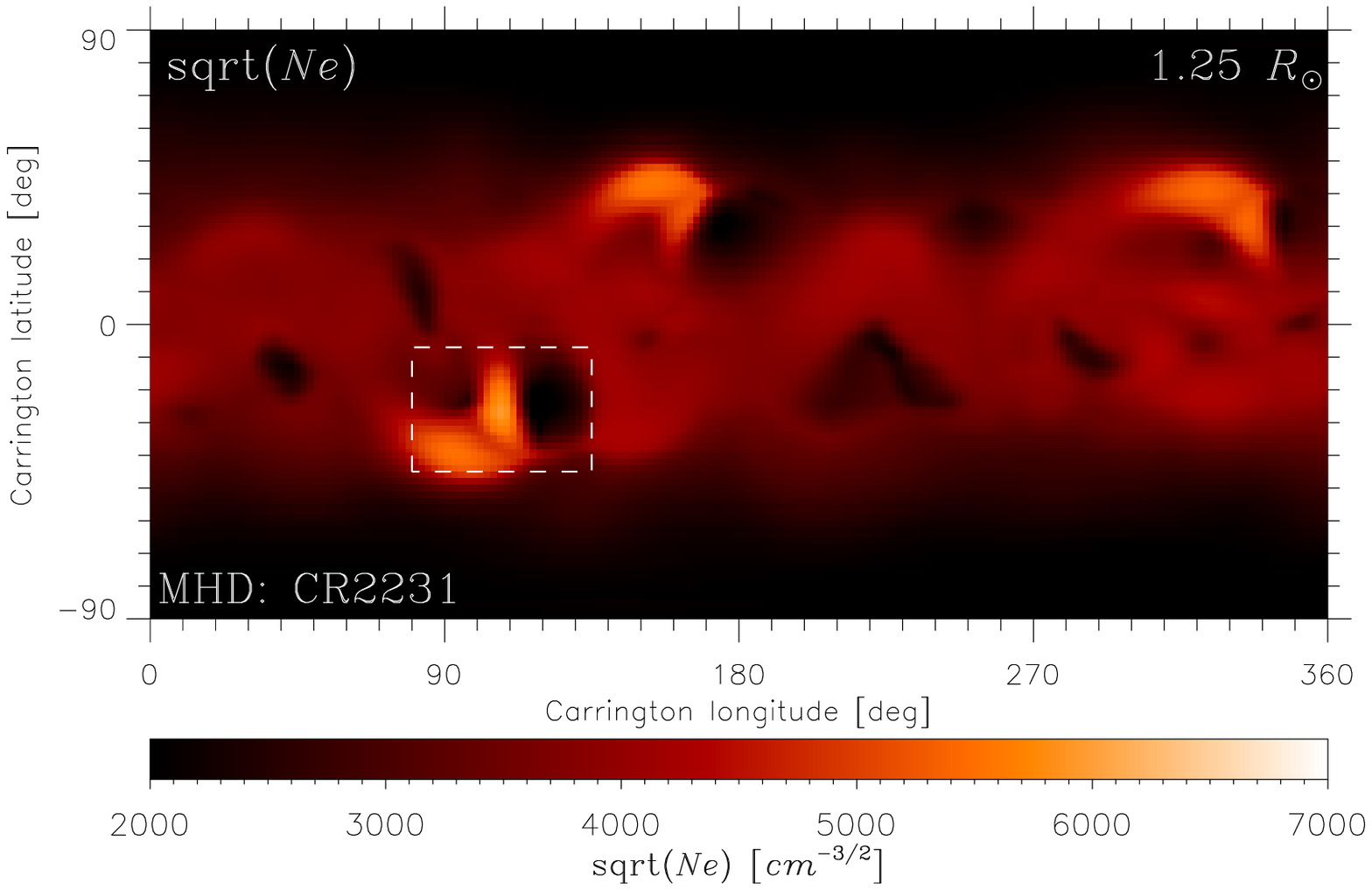}
\includegraphics*[bb=84 304 557 620,width=0.32\linewidth]{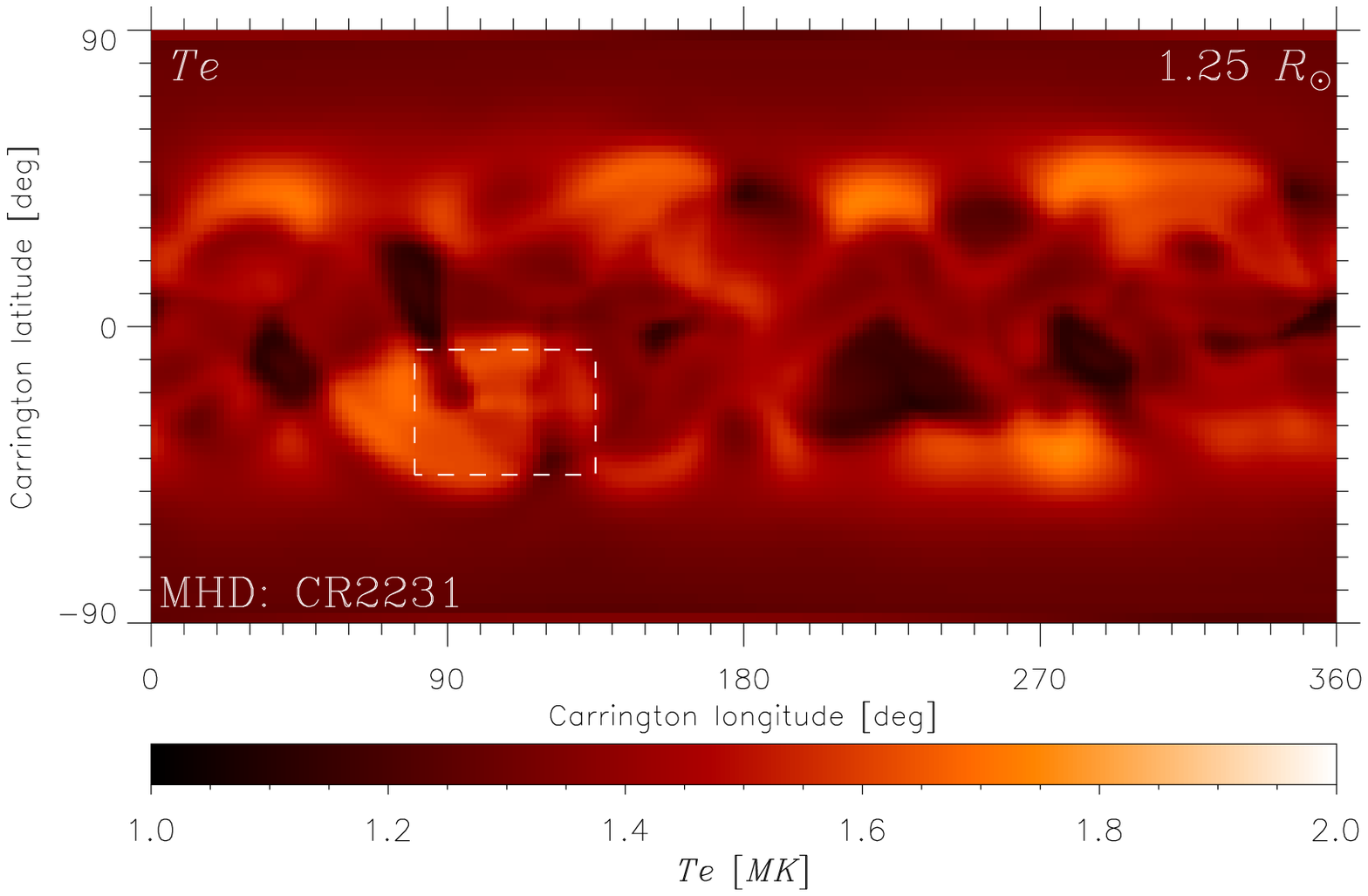}
\caption{Spherical cross-sections of the radial component of the magnetic field vector, electron density and temperature for the MHD model at $1.05$ and $1.25\ R_\odot$. The rectangular box in dash line mark active region 1 (AR1) discussed in text.}
\label{Fig_MHD_sphsect}
\end{figure}

Figure \ref{Fig_MHD_sphsect} shows spherical cross-sections of the radial component of the magnetic field vector, electron density and temperature of the PSI MHD coronal model for Carrington Rotation (CR) 2231 at two heights to provide context information of the corona that was used to validate the tomography algorithm. CR 2231 occurs during solar minimum with relatively simple coronal magnetic field structure. It has a strong global dipole moment and only a few active regions (ARs). The maximum intrinsic field strength $|\vec{B}|$ in the largest active region AR1 located between $90$ to $120^\circ$ longitude and $-15$ to $-35^\circ$ latitude in the southern hemisphere is about $25\ \textrm{Gauss}$. 
The model temperature and density 3D distribution were treated as known quantities. 

\subsection{The Simulated Stokes Data}
\label{Sec_Sim_Data}

To simulate the local Stokes emissivities, $\varepsilon_s(\vec{r})$, at any point in the corona $\vec{r}$, 
we employ approximated expressions for it derived by \cite{Querfeld_1982}. 
The detailed description of the local Stokes-Q,U emissivities simulation is described in \cite{Kramar_2013} and therefore we do not describe it again here. 

For simulation of Stokes-V data, we use the magnetograph formula approximation expressed by (\ref{StokesV_Tomo_Simplified}). 
Because the goal of this assessment is to explore the accuracy limits (theoretical accuracy) of the vector tomography method, 
this approach allows us to avoid the population levels calculations. 
When dealing with real data, we will need to perform the population levels calculations. 
The detailed description of the CP simulation is described in \cite{Kramar_2006} and therefore we do not describe it again here. 

The simulated Stokes observational data were obtained by integrating the local Stokes emissivities, $\varepsilon_s(\vec{r})$, over each LOS according to equation (\ref{Stokes_Para}). 
As the purpose of this paper is to quantify accuracy limits (theoretical accuracy), 
we do not introduce any noise and the LOS integration is confined within a sphere with radius of $2.5\ R_\odot$ (see Figure). 

The field of view (FOV) of the simulated LP data is within the heliocentric distances from $0.05$ to $2.0\ R_\odot$. 
This FOV size is about the same as in the modern real instruments such as UCoMP and mxCSM. 

For the CP data, the FOV area is set to be about double of maximal Cryo-NIRSP FOV which is about 0.25x0.19 $R_\odot$. 
Only two vantage directions were used for CP data separated by $15^\circ$ (or by about one day). 
The FOV scans for the CP data were placed at the active region AR1 (see Figures \ref{Fig_MHD_sphsect} and \ref{Fig_Tomo_Err_Map_All}).

\begin{figure}[h!]
\includegraphics*[bb=0 3 1062 972,width=\linewidth]{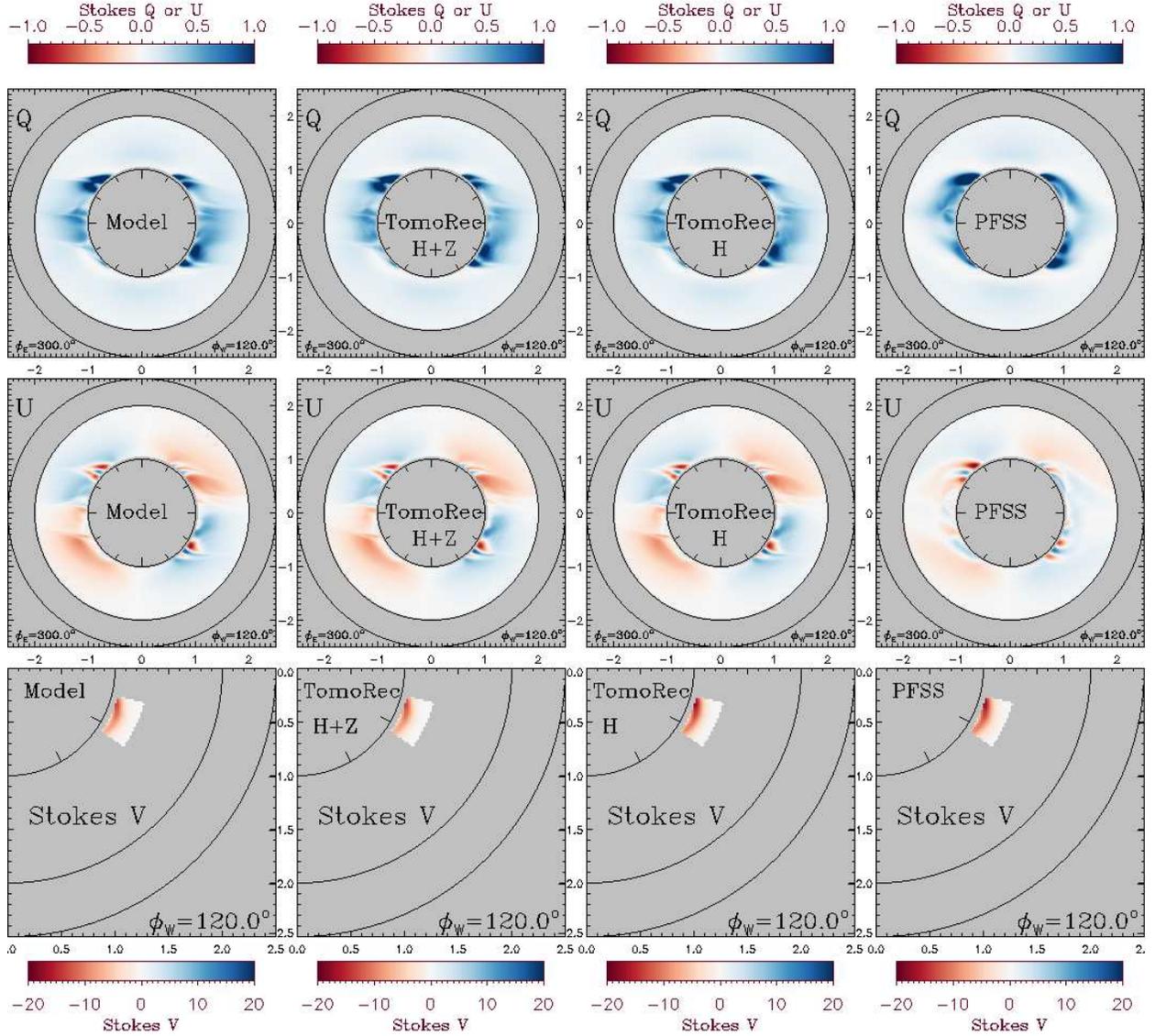}
\caption{Maps of the simulated Stokes Q (upper row), Stokes U (middle row), and Stokes V (lower row) data produced from the MHD model (first column),  tomographically reconstructed field based on full Stokes data (second column) and only LP data (third column), and PFSS model (fourth column). The LP reference frames are oriented radially from the Sun disk center for each image pixel. The Stokes maps are displayed in relative code units.}
\label{Fig_Tomo_Err_Data_Map}
\end{figure}

Example of maps of the simulated Stokes Q and U data produced from the MHD model are shown on the first column of Figure \ref{Fig_Tomo_Err_Data_Map}.

\clearpage
\subsection{The Inversion procedure}
\label{Sec_LP_Tomo_Proc}

Equation (\ref{MinFun_TomoVec}) describes basic general approach to the vector tomographic inversion with one physically based regularization term. 
In order to improve the solution, we introduced here two more regularization terms: 
second order smoothness factor, $F_\textrm{sm}$, and requirements that the magnetic field becomes radial at $2.5\ R_\odot$ (outer boundary condition), $F_\textrm{rad}$. 
The first order smoothness factor is just requirements that difference between two neighbour cells should be minimal. 
The second order smoothness factor implies that differences of differences between a cell and its neighbour cells should be minimal. 

The requirements that the magnetic field becomes radial at some distance (Source Surface) comes from observations (ref) and has been widely used in routine potential field models (PFSS). 
Although there are evidences that the Source Surface lies at about $3\ R_\odot$ \citep{Kramar_2014}, 
we use the value of $2.5\ R_\odot$ here which is not critical for the goal of this paper. 

In result, the final minimization function for the LP tomography used in the tests is: 
\begin{eqnarray}
F(\vec B (\vec r)) = \mu_\textrm{H} 
{\sum\limits_{i,j,s}} \left[ S^{\rm \ obs}_{s} (\hat \vec{e}^i_{\rm LOS}, \vec \rho_j)
-S^{\rm \ sim}_{s} (\hat \vec{e}^i_{\rm LOS},\vec{\rho_j}, \vec B(\vec{r})) \right]^2 + \nonumber \\
\mu_\textrm{Z} 
{\sum\limits_{i,j}} \left[ V^{\rm \ obs} (\hat \vec{e}^i_{\rm LOS}, \vec \rho_j)
-V^{\rm \ sim} (\hat \vec{e}^i_{\rm LOS},\vec{\rho_j}, \vec B(\vec{r})) \right]^2 + \nonumber \\
+\ \mu_\textrm{divB} \sum \limits_q \left [ \nabla \cdot \vec{B}(\vec{r}) \right ]^2
+ \mu_\textrm{sm} F_\textrm{sm}(\vec{B}(\vec{r}))
+ \mu_\textrm{rad} F_\textrm{rad}(\vec{B}(\vec{r}))
\label{MinFun_LP_Tomo}
\end{eqnarray}
where $S^{\rm \ obs}_{l,s}$ is the observed Stokes vector, with $s = 1, 2$ corresponds to the Stokes $Q$ and $U$ parameters, respectively; $S^{~\rm sim}_{s}$ is the synthesized Stokes data given $\vec B(\vec r)$. The summation integrate over all the LOSs, all plane-of-the-sky positions, all spectral lines, and all Stokes parameters.  Gauss's law, $ \nabla \cdot \vec{B}(\vec{r}) = 0$, with its squares summed over every voxel element $q$, is used as the regularization function to provide further constraint on $\vec B(\vec r)$ to help stabilize the inversion process. 

The inversion was performed using the conjugate gradient method.

\subsubsection{Starting Field Model}

To obtain best inversion result, it is important to have starting field as close as possible to the real field (Ground Truth field in our experiment). 
Currently, a most widely and robustly used coronal magnetic model is the Potential Field with Source Surface (PFSS) model. 
For calculations of the PFSS model we employ the Finite Difference Iterative Potential-field Solver 
(FDIPS; \cite{Toth_2011ApJ_FDIPS,FDIPS_2016ascl_soft06011T}). 
The PFSS model is based on the same photospheric boundary data as was used for the Ground Truth Model and also in $F_\textrm{divB}$ term in the minimizing function (\ref{MinFun_LP_Tomo}). 
The Source Surface is placed at the heliocentric distance of $2.5\ R_\odot$. 
We use the PFSS model as the starting field for the inversion procedure.

\subsection{Summary of the data simulation and inversion parameters}

Summary of the data simulation and inversion parameters: 

\begin{table}[h!]
\caption{Summary of the data simulation and inversion parameters.}\vspace{0.5ex}
\label{Table_sim_param}
\centering 
\resizebox{\linewidth}{!}{\begin{tabular}{l l l}
\hline\hline
Parameter & Value & Comments \\
\hline
Stokes-Q,U FOV & From $1.05$ to $2.0\ R_\odot$ & LP Data simulation\\
Stokes-Q,U duration & two weeks  & LP Data simulation\\
\hline
Stokes-V FOV & two Cryo-NIRSP FOV (0.25x0.19 $R_\odot$) & CP Data simulation\\
Stokes-V duration & once a day for two days & CP Data simulation\\
\hline
Image resolution & 50 pixels per $R_\odot$ & Data simulation\\
Rec. domain size & From $1$ to $2.5\ R_\odot$; $7827960$ cells & Rec. domain is spherical\\
Rec. cell size & $0.02\ R_\odot$ & Grid cells are rectangular\\ [0.5ex]
\hline
\end{tabular}}
\end{table}

\subsection{Reconstruction Result.}
\label{Sec_VecTomo_Result}

\begin{figure}[h!]
\begin{center}
\includegraphics*[bb=66 365 541 716,width=0.49\linewidth]{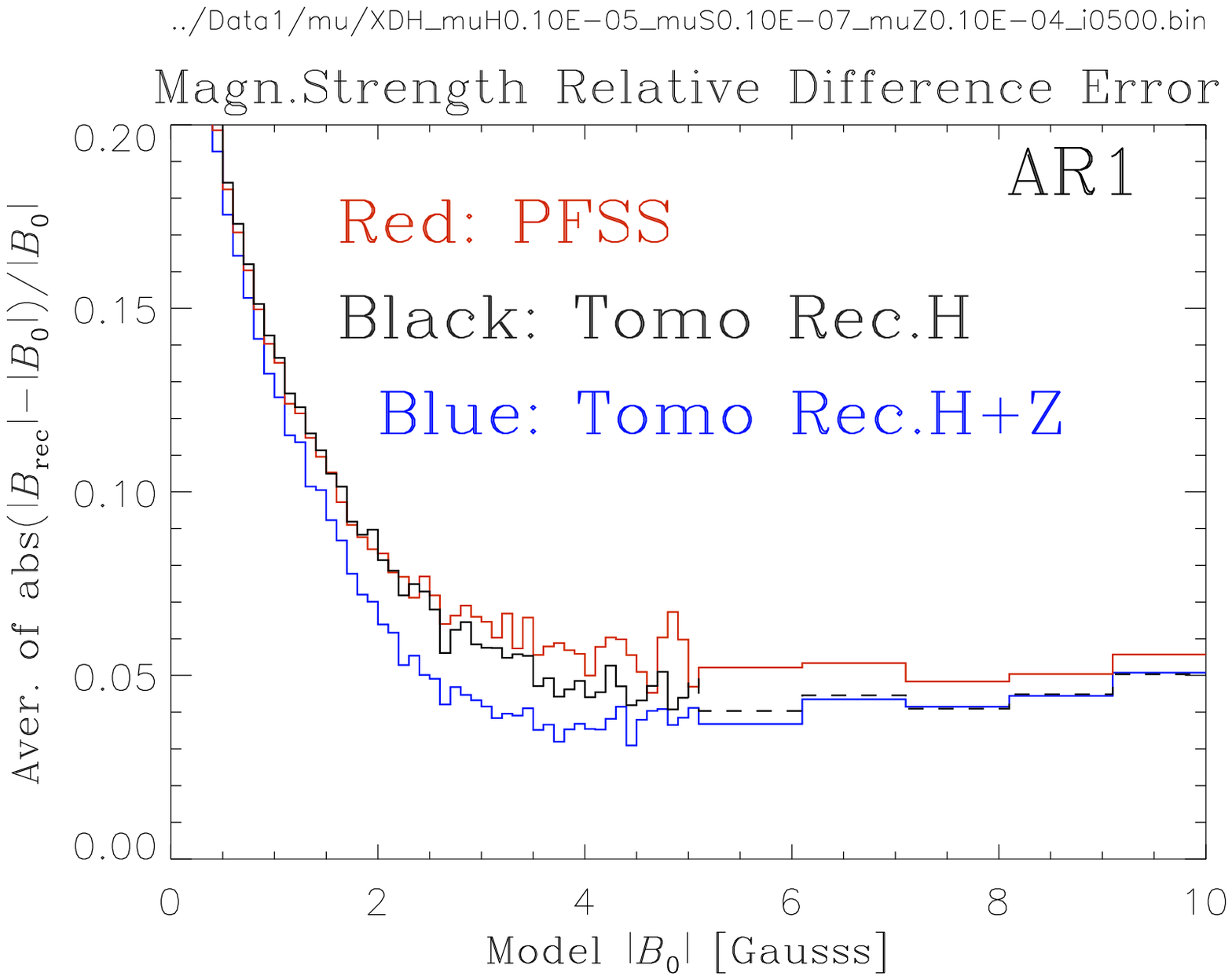}
\includegraphics*[bb=66 365 541 716,width=0.49\linewidth]{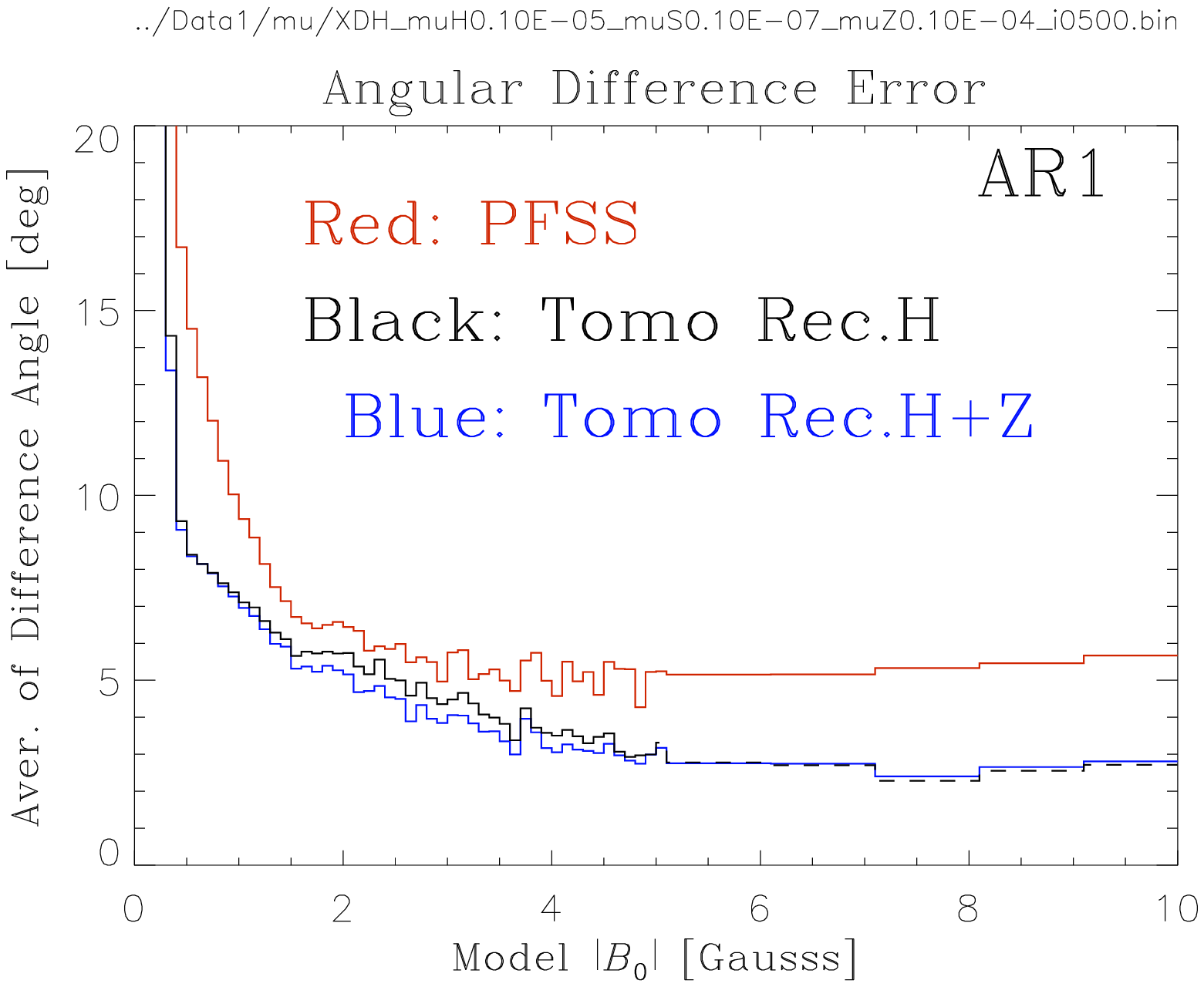}
\end{center}
\caption{Histograms of the accuracy distribution of the relative error of the magnetic field absolute value (left) and the directional difference (right) from the GT model for the tomography reconstruction based on the full Stokes data (blue), tomography reconstruction based on only LP data (black), and starting PFSS (red) model. Number of voxels in AR1 is 79162.}
\label{Fig_Tomo_Err_hist}
\end{figure}

Figure \ref{Fig_Tomo_Err_Data_Map} is a clear demonstration of the robustness of the inversion algorithm. However, it does not guarantee that the inversion algorithm has successfully recovered $\vec B_0$ since the existence of a unique solution has not been (and maybe {\it cannot} be) established. Nevertheless, the accuracy of the tomography inversion algorithm can be assessed by comparing voxel by voxel the strength and orientation of $\vec B$ and $\vec B_0$ over the entire tomographic reconstruction volume. 
Figure \ref{Fig_Tomo_Err_hist} shows the scattered plot of the relative magnetic field strength error $\varepsilon_B = (\vert B_0 \vert  - \vert B \vert ) / \vert B_0 \vert$ (left panel) and pointing errors $\triangle\gamma  =  \cos^{-1} ((\vec B_0 \cdot \vec B)/(\vert B_0 \vert \cdot \vert B \vert))$ (right panel) of the tomographically inverted field based on the full Stokes data (Blue; ``H+Z'' label),  tomographically inverted field based on only LP data (Black; ``H'' label), and starting PFSS field (Red). 
The curves show $\varepsilon_B(\vert B_0\vert)$ and $\triangle\gamma (\vert B_0\vert)$, the averaged values of the relative magnetic field strength and orientation errors as functions of the model field strength, respectively. 
The errors were calculated only over the active region AR1 where the CP data were taken. 

Figure \ref{Fig_Tomo_Err_hist} shows that the tomography with only LP data can significantly improve the reconstruction of the field vector direction but the improvement in the field strength is relatively small. 
However, the full Stokes tomography shows significant improvement in the reconstruction of both - the field strength and its direction. 
It should be noted that this improvement is achieved by inclusion of the Stokes-V observation over the relatively limited FOV and were conducted only twice over two days period. 
{\it This will require for only about two hours of observation time a day.}

\begin{figure}[h!]
\includegraphics*[bb=87 308 426 703,width=0.24\linewidth]{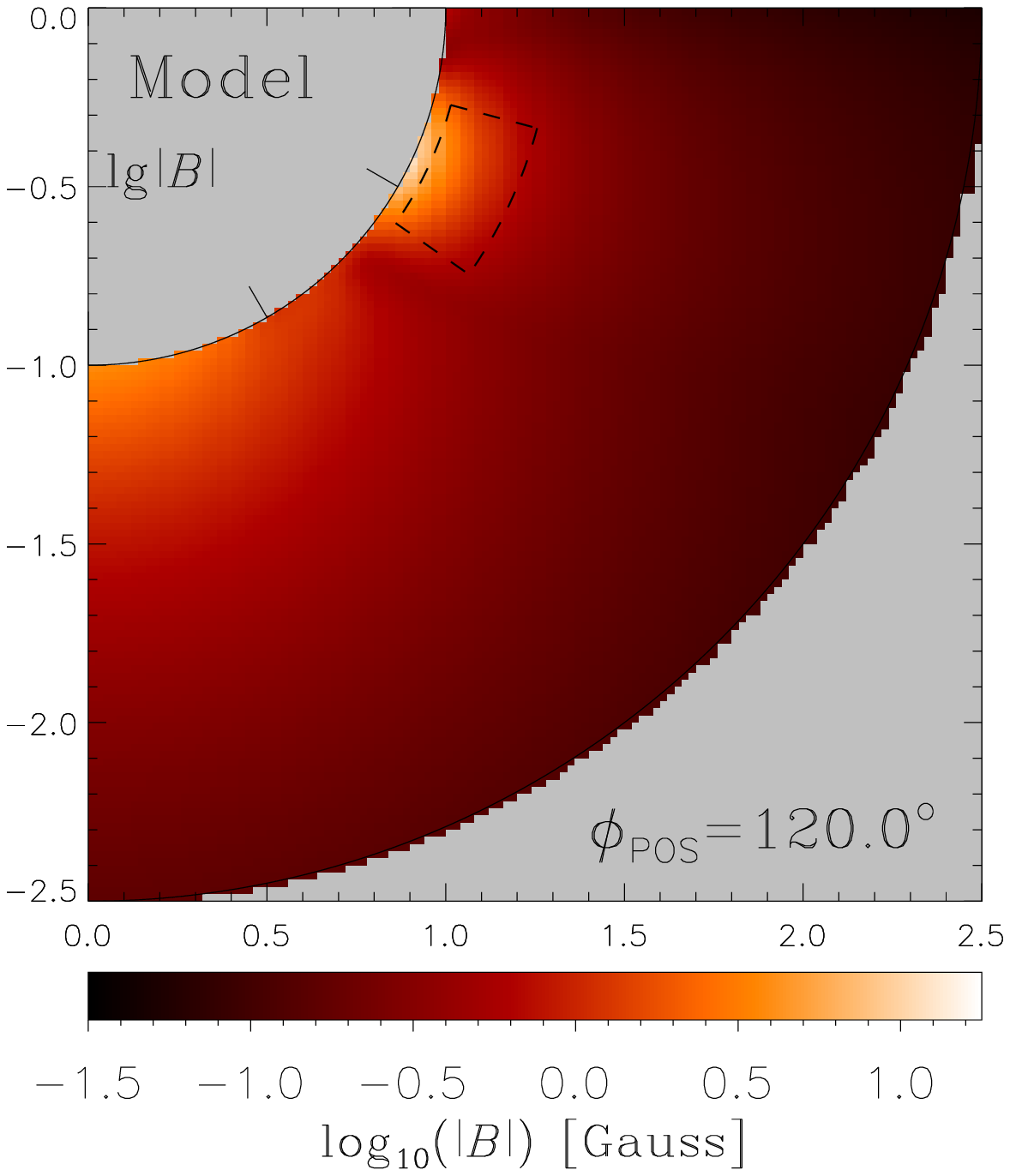}
\includegraphics*[bb=87 308 426 703,width=0.24\linewidth]{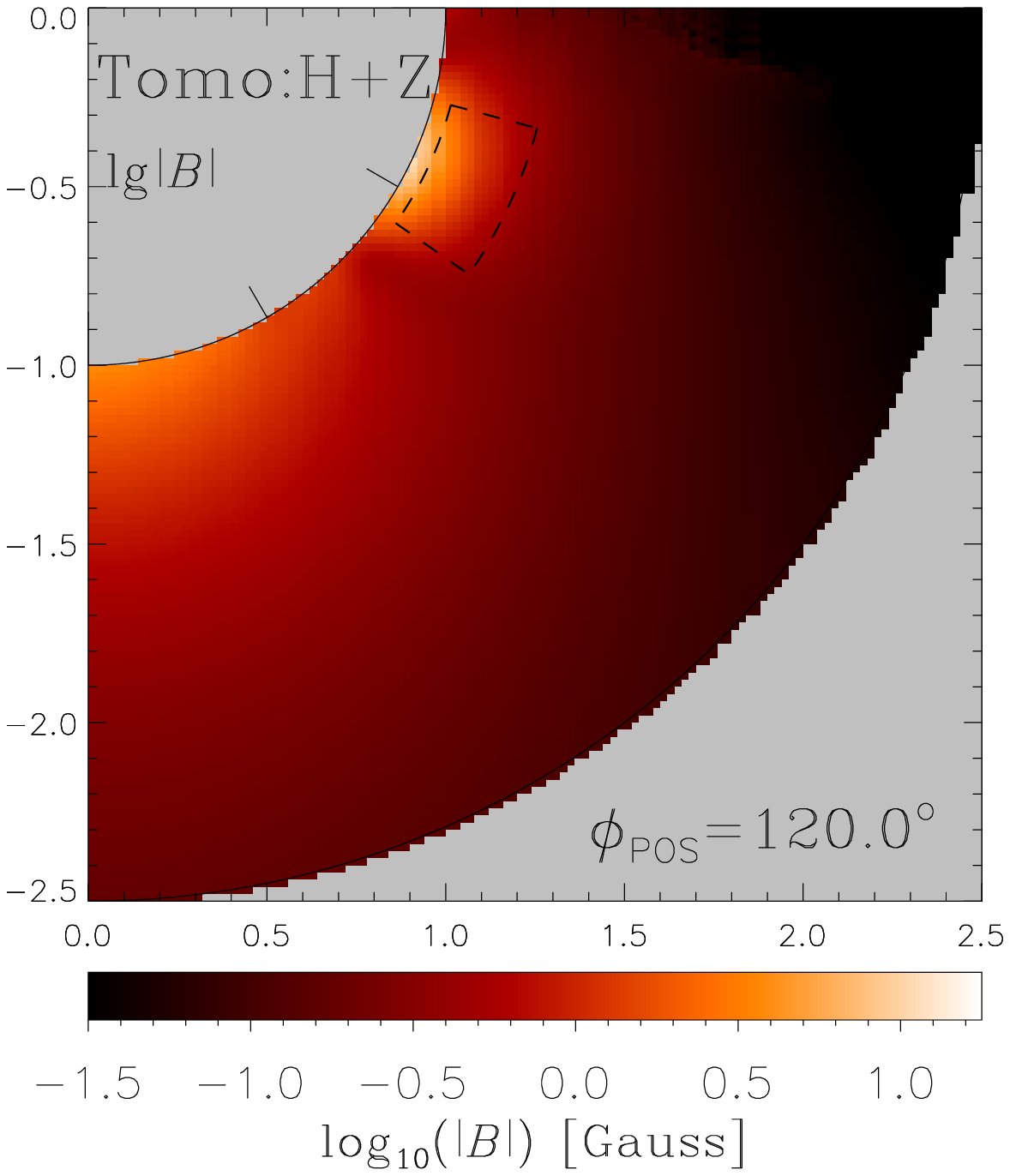}
\includegraphics*[bb=87 308 426 703,width=0.24\linewidth]{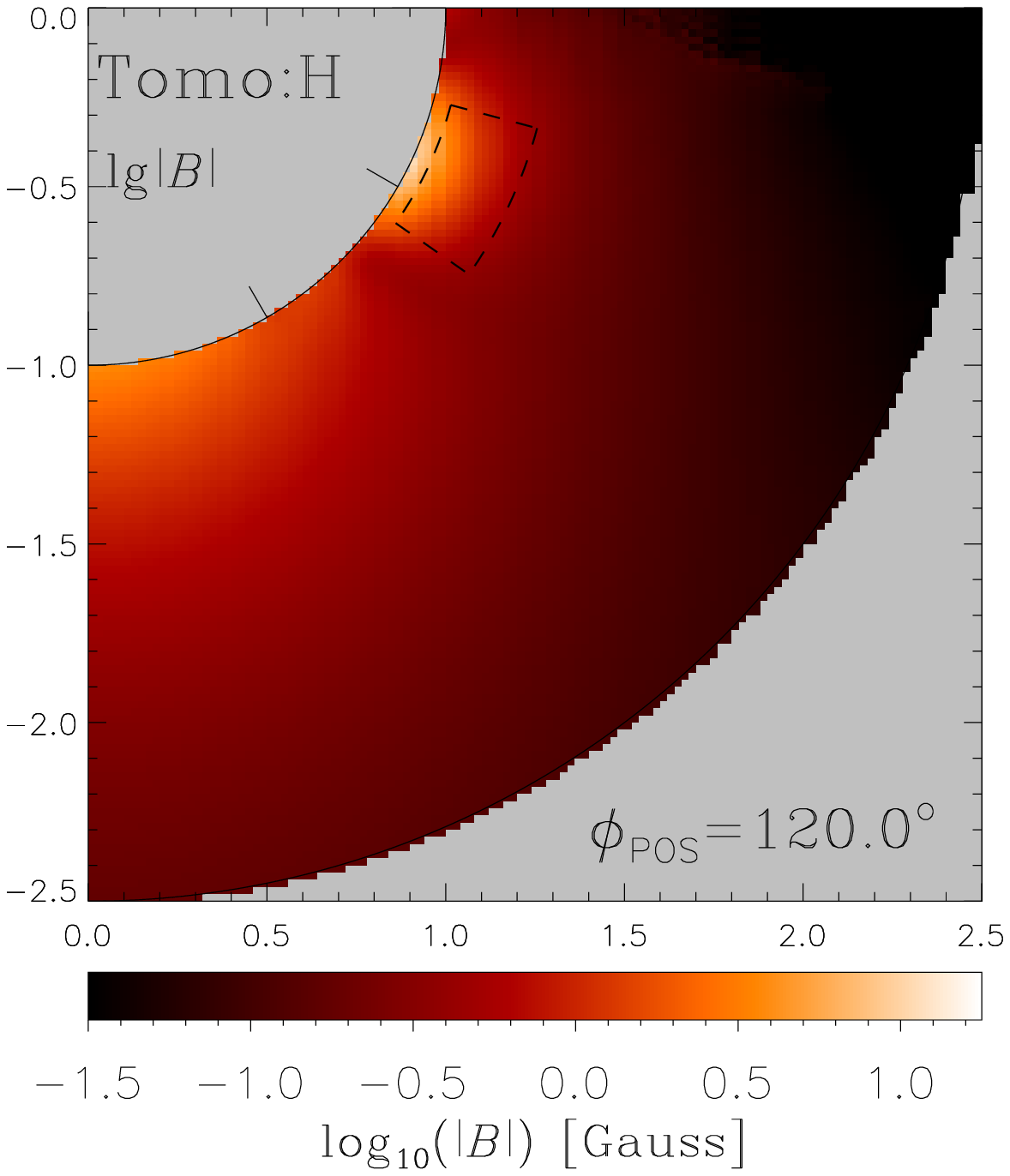}
\includegraphics*[bb=87 308 426 703,width=0.24\linewidth]{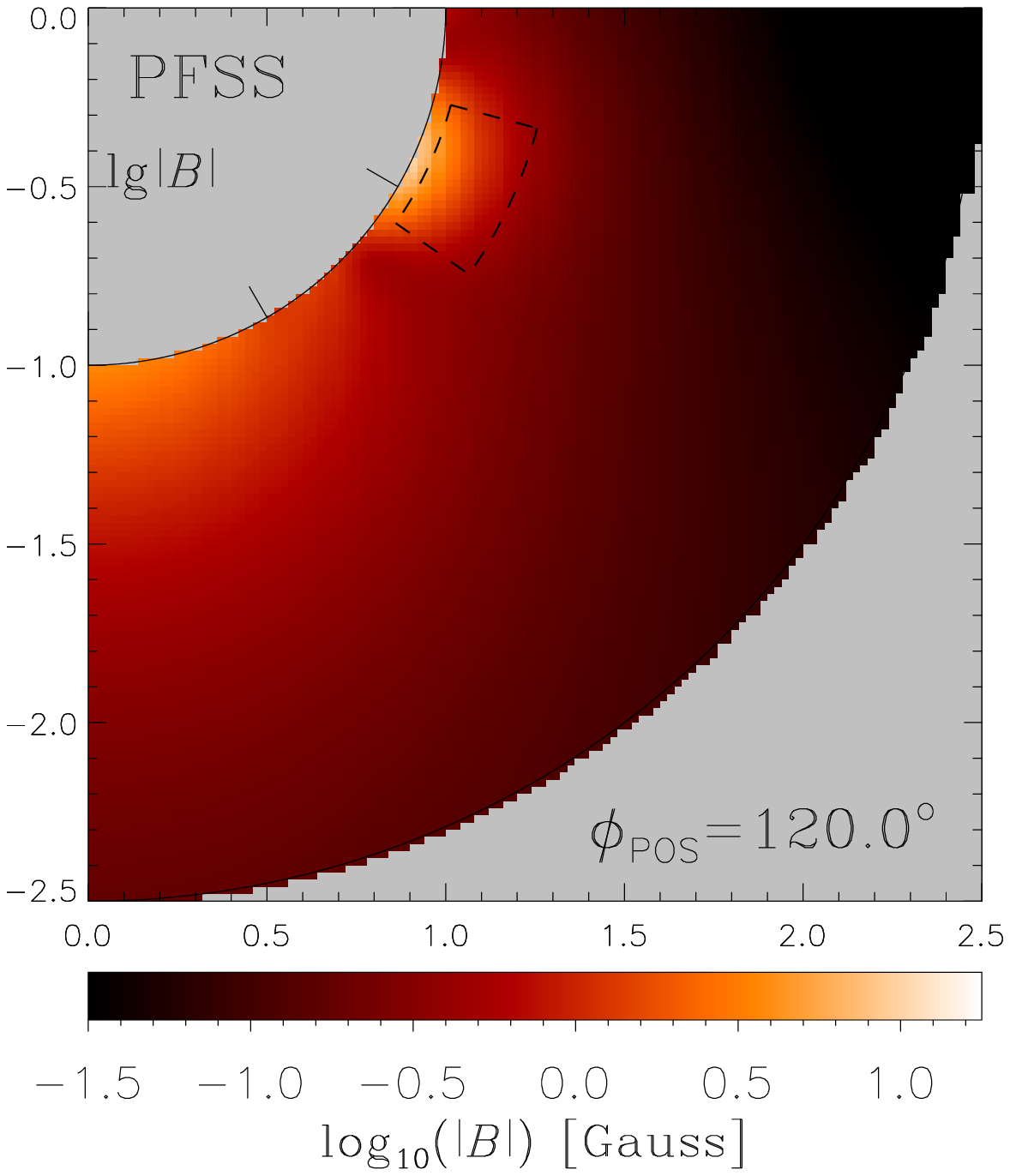}\\
\includegraphics*[bb=87 308 426 703,width=0.24\linewidth]{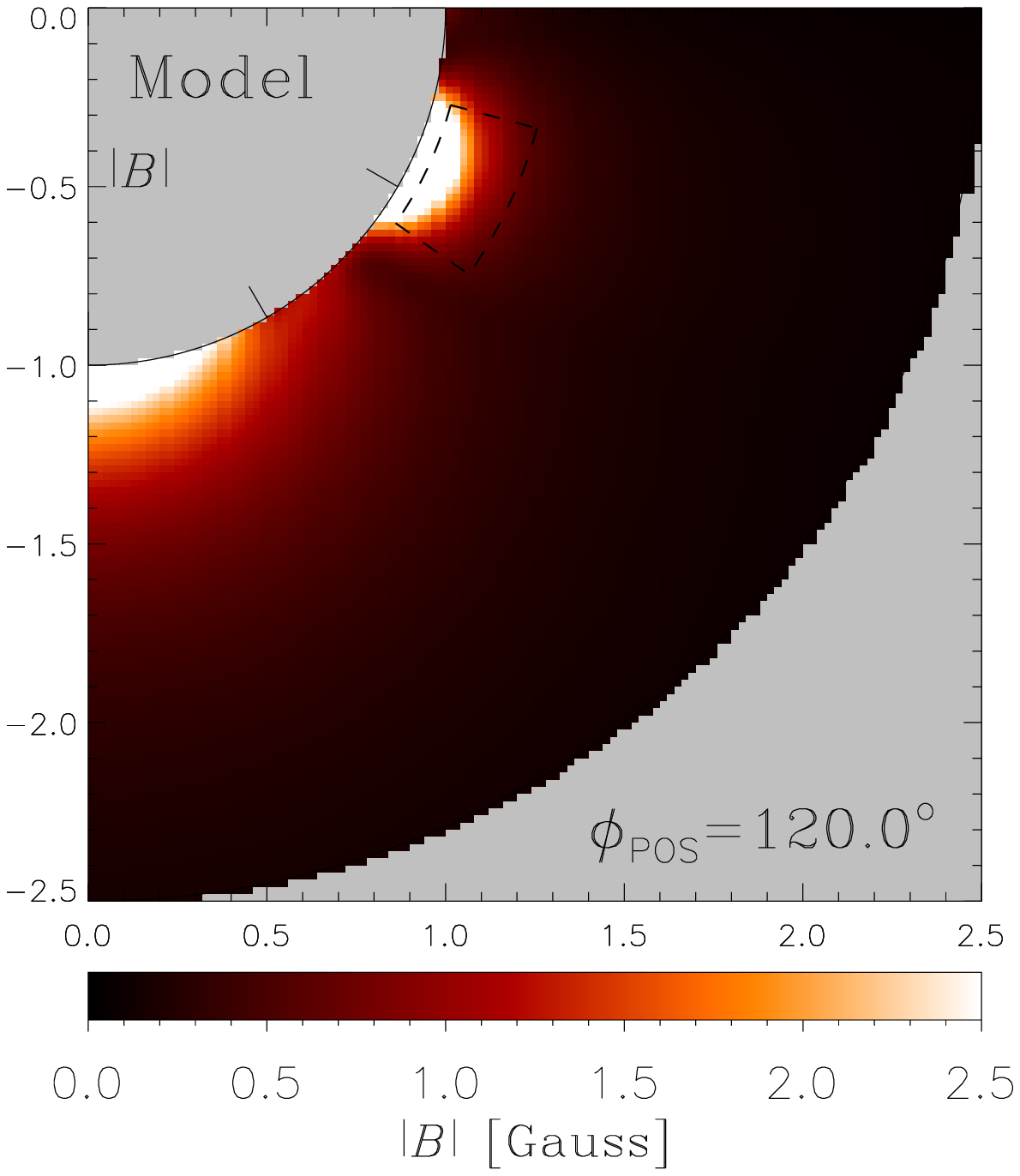}
\includegraphics*[bb=87 308 426 703,width=0.24\linewidth]{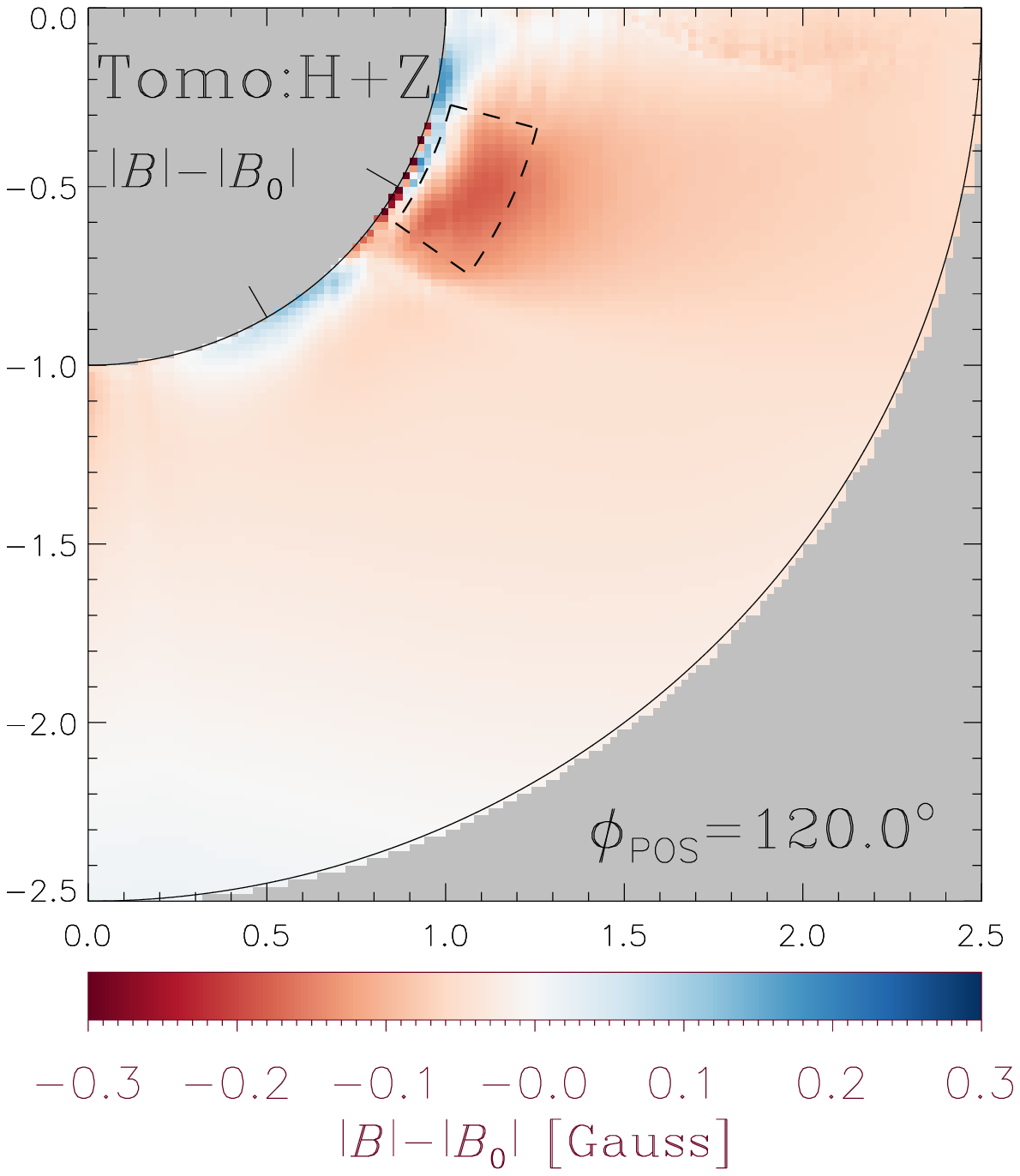}
\includegraphics*[bb=87 308 426 703,width=0.24\linewidth]{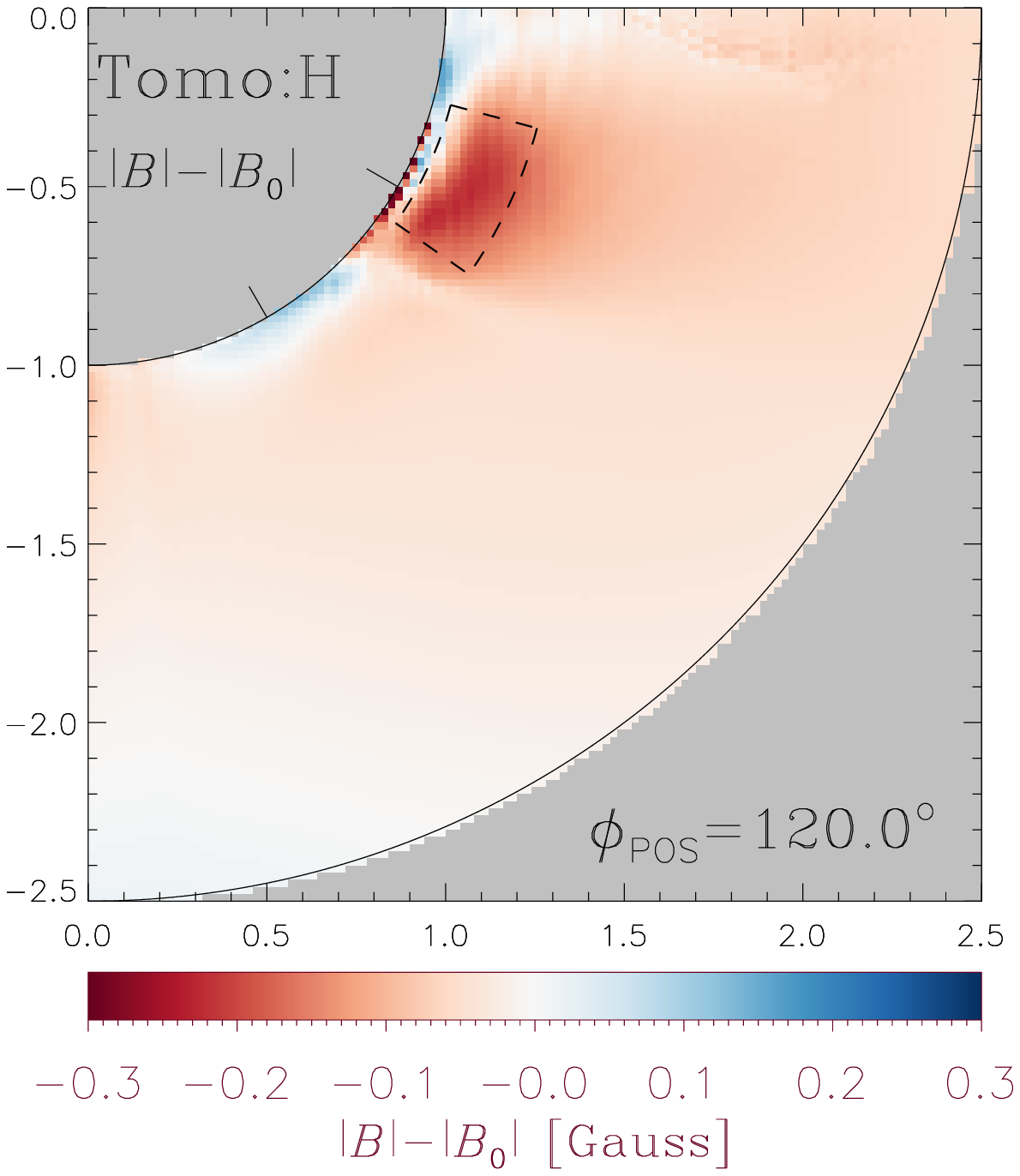}
\includegraphics*[bb=87 308 426 703,width=0.24\linewidth]{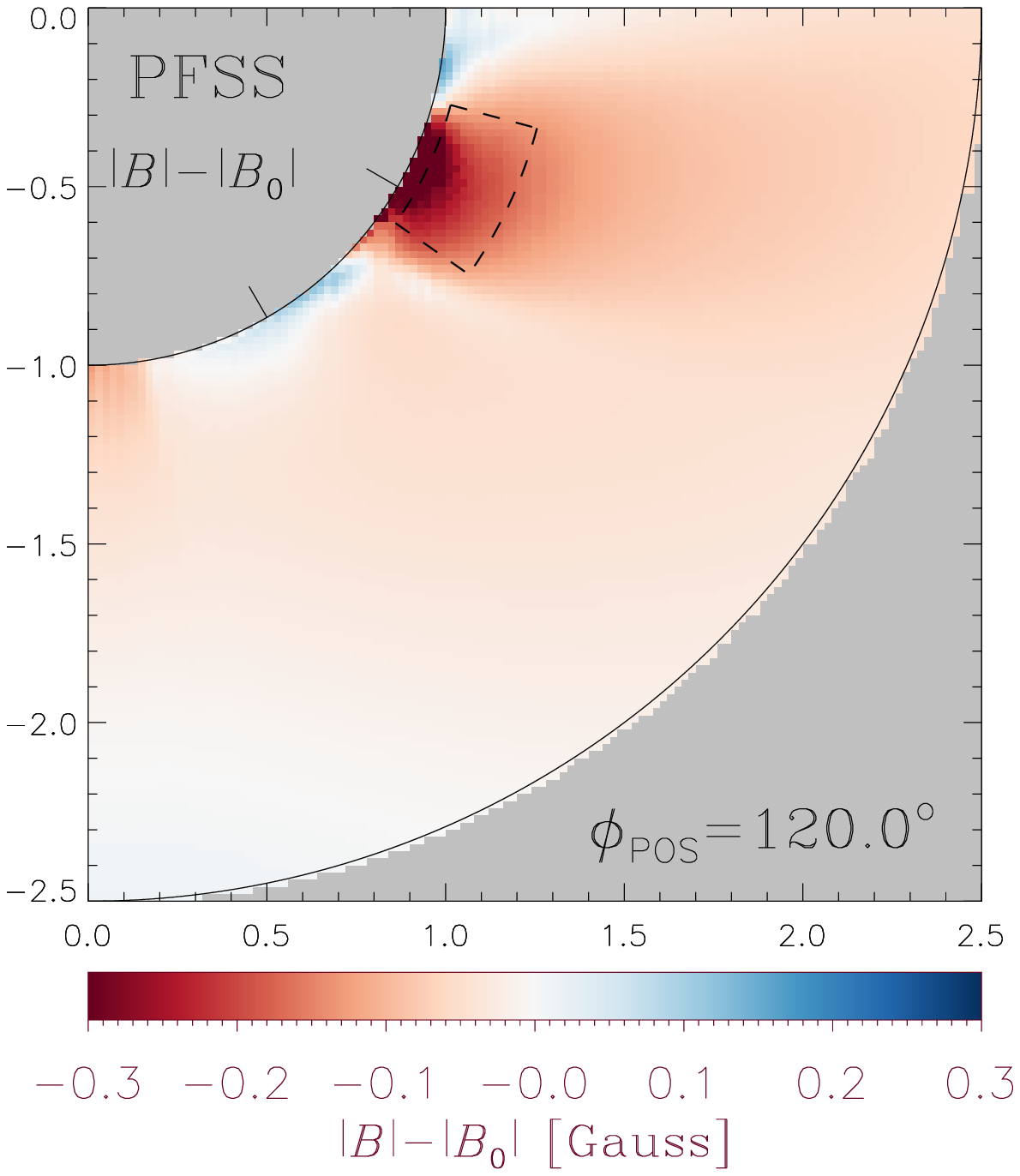}\\
\includegraphics*[bb=87 308 426 703,width=0.24\linewidth]{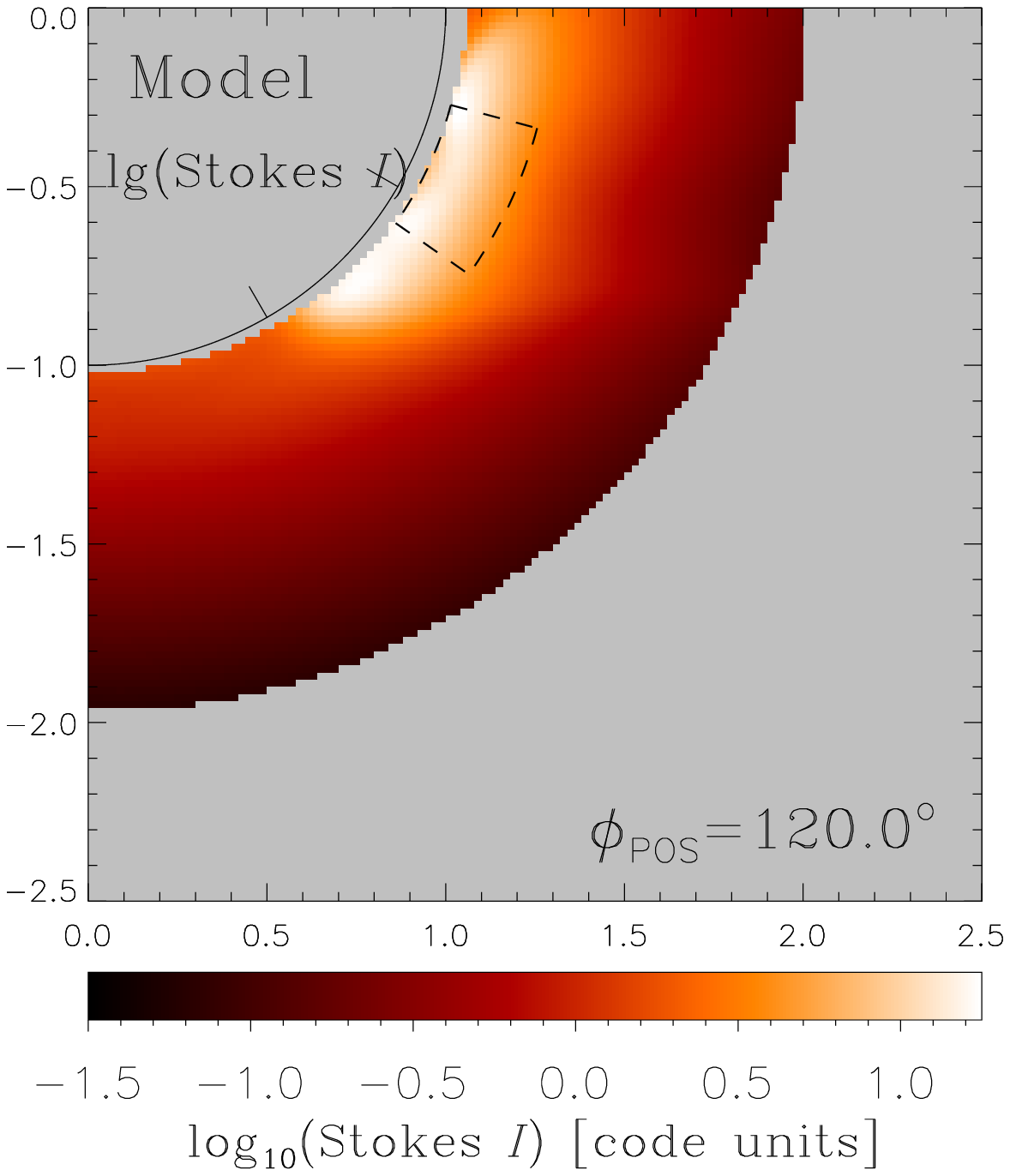}
\includegraphics*[bb=87 308 426 703,width=0.24\linewidth]{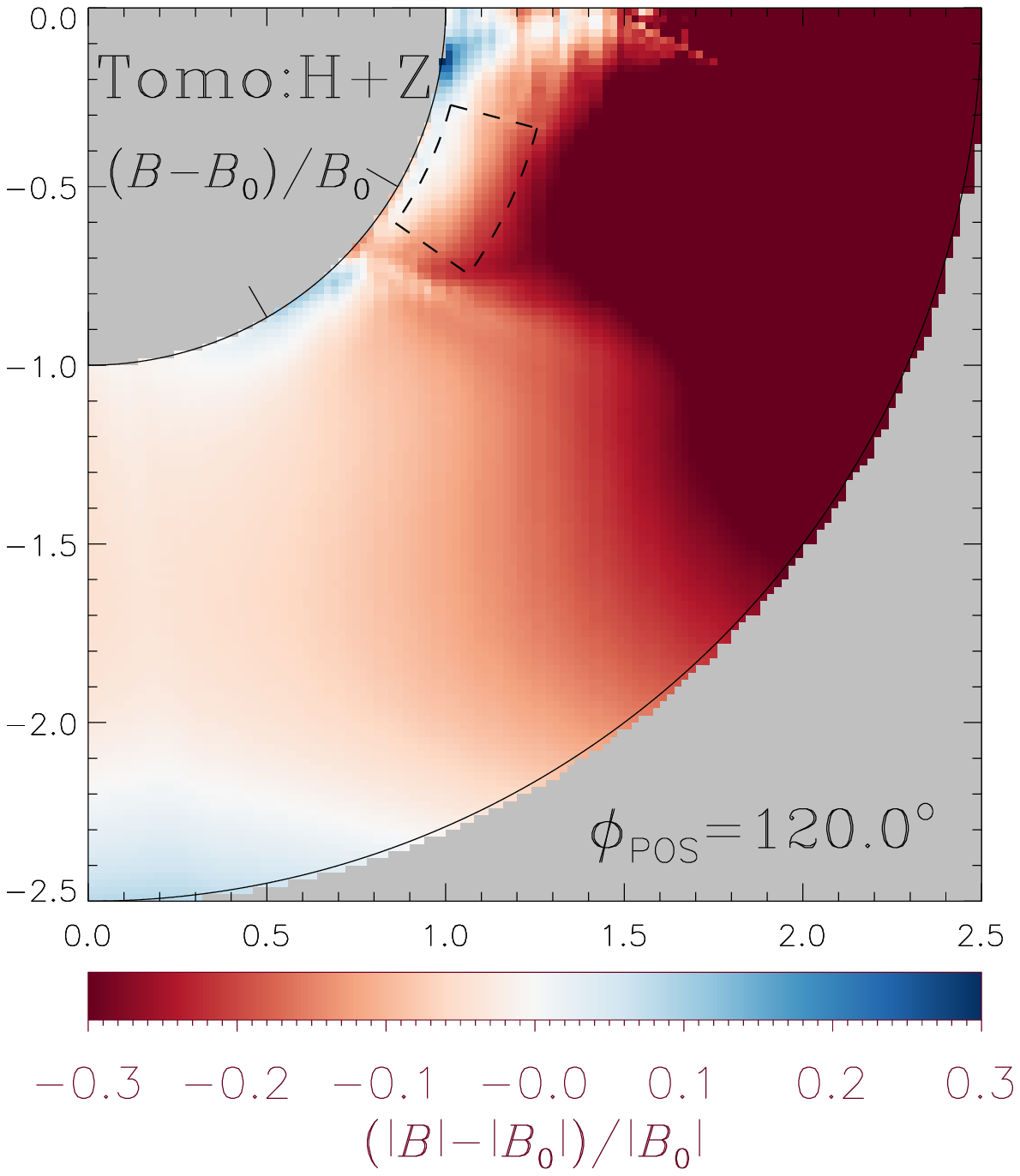}
\includegraphics*[bb=87 308 426 703,width=0.24\linewidth]{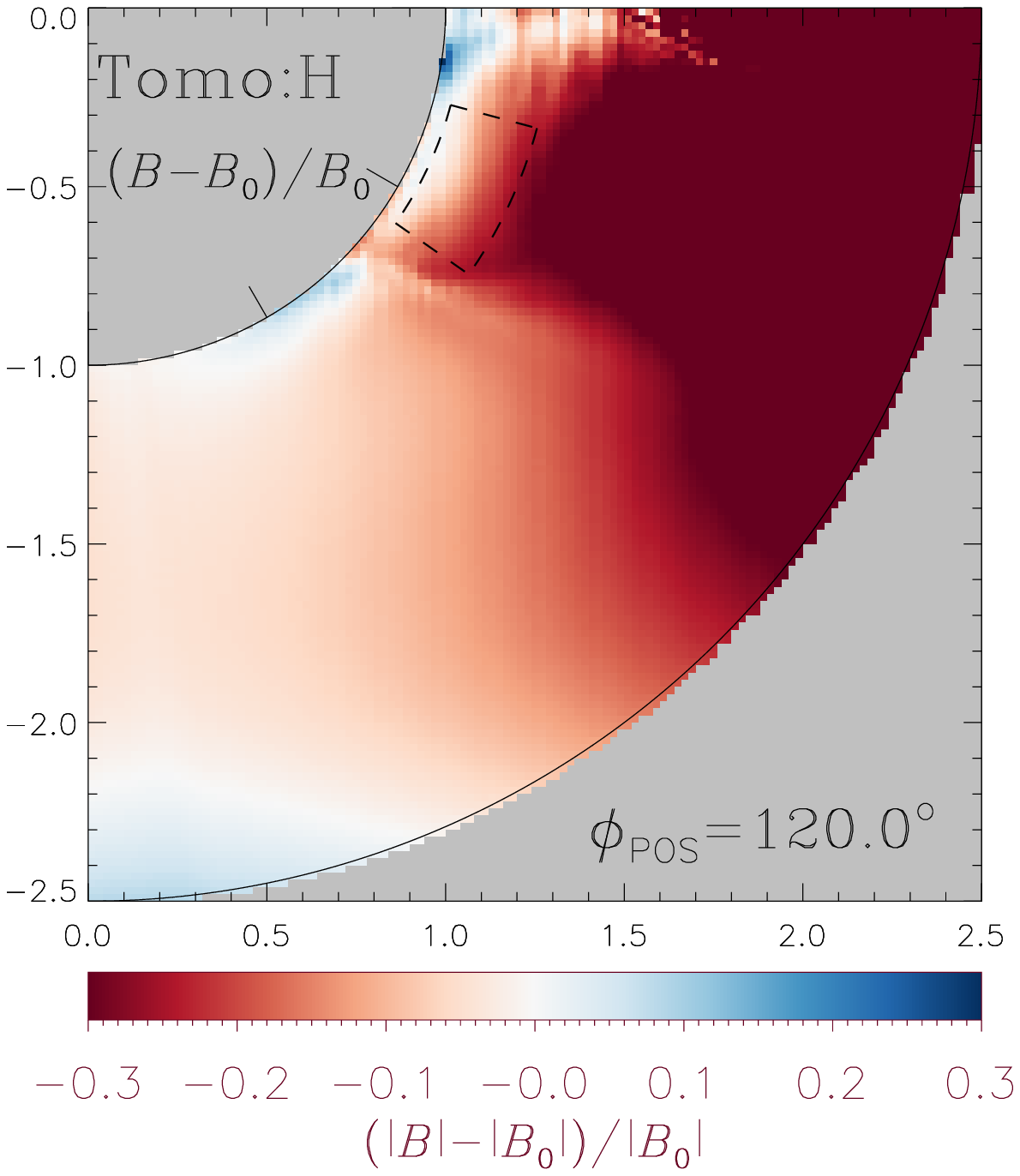}
\includegraphics*[bb=87 308 426 703,width=0.24\linewidth]{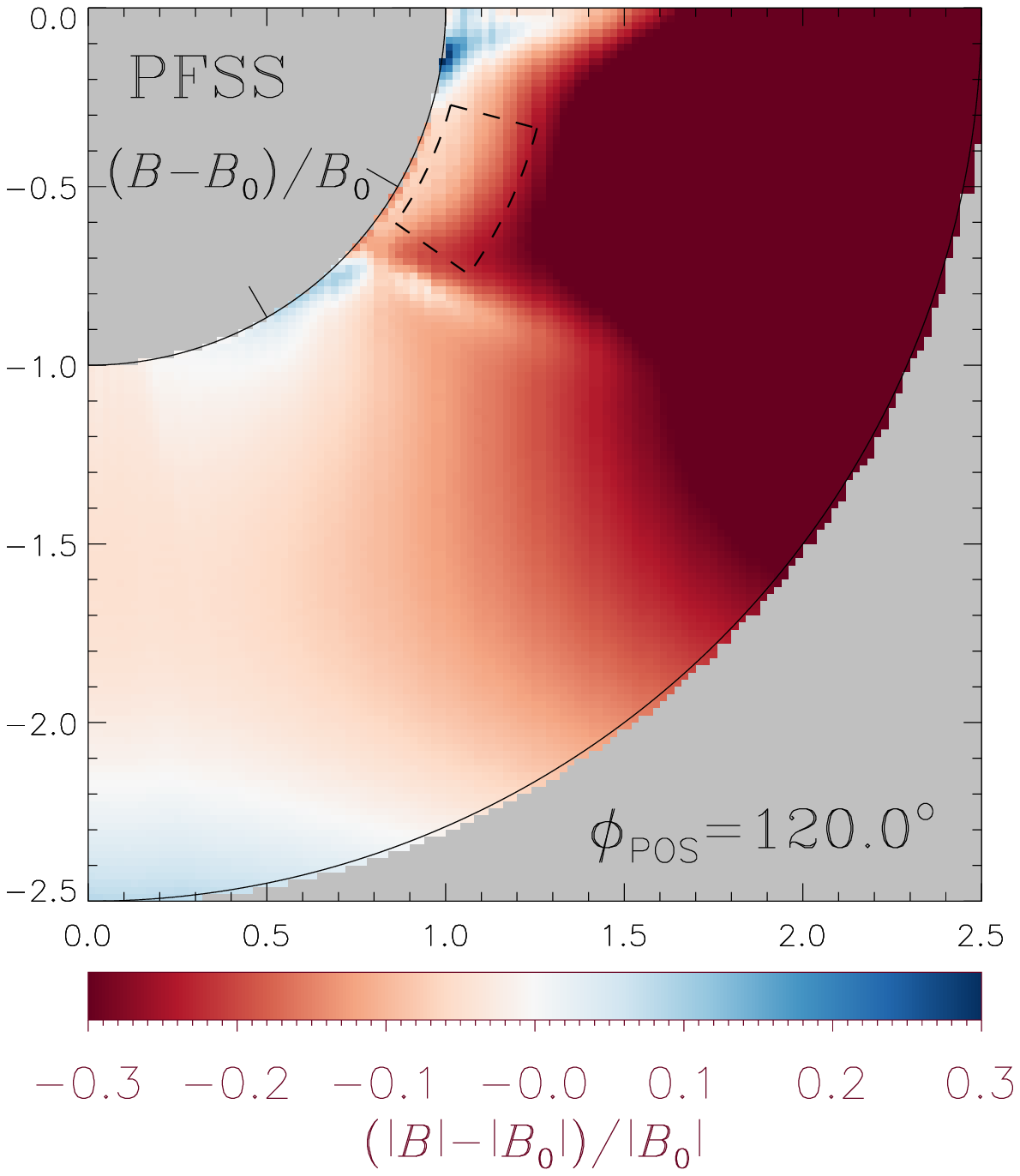}\\
\includegraphics*[bb=87 308 426 703,width=0.24\linewidth]{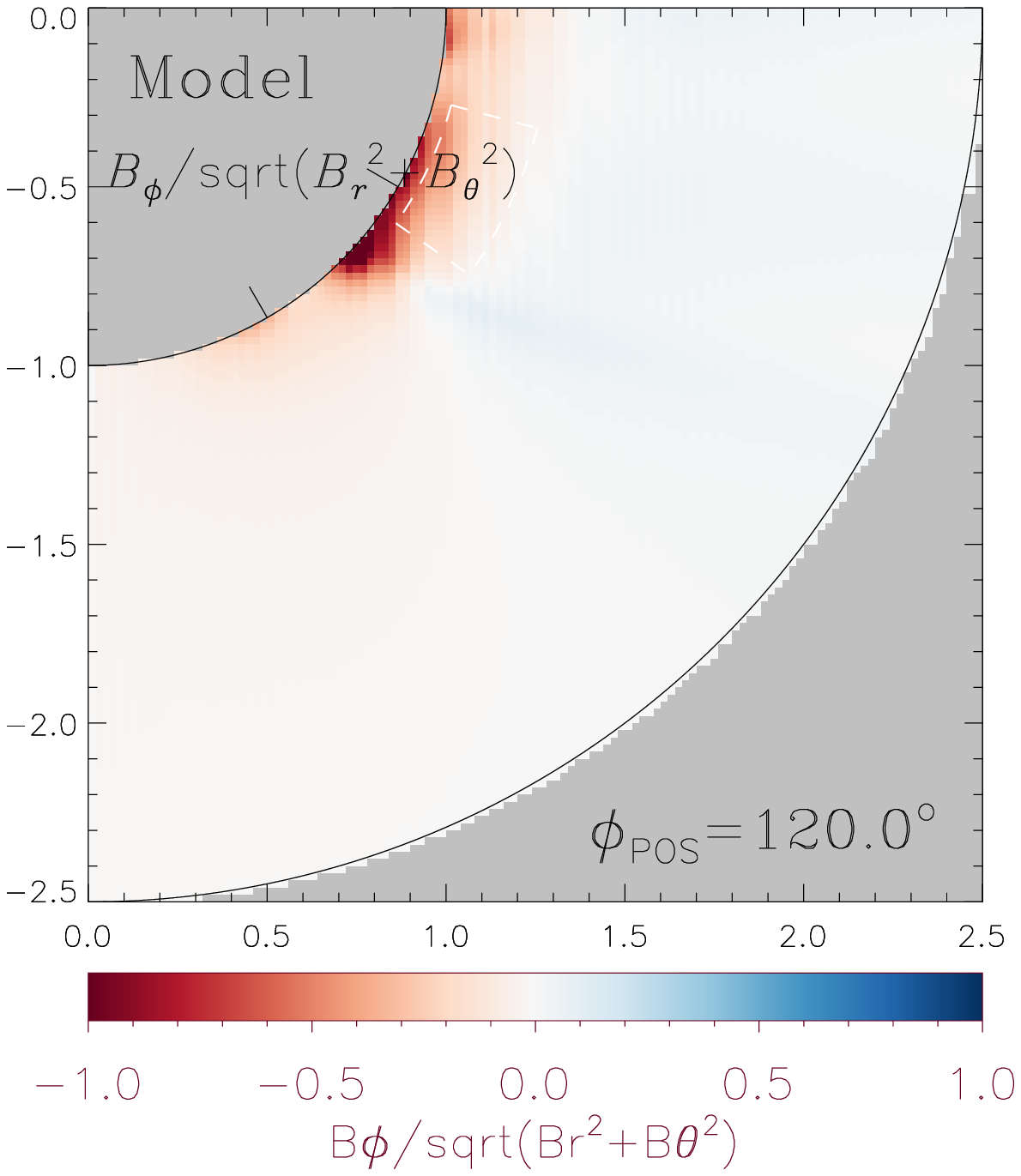}
\includegraphics*[bb=87 308 426 703,width=0.24\linewidth]{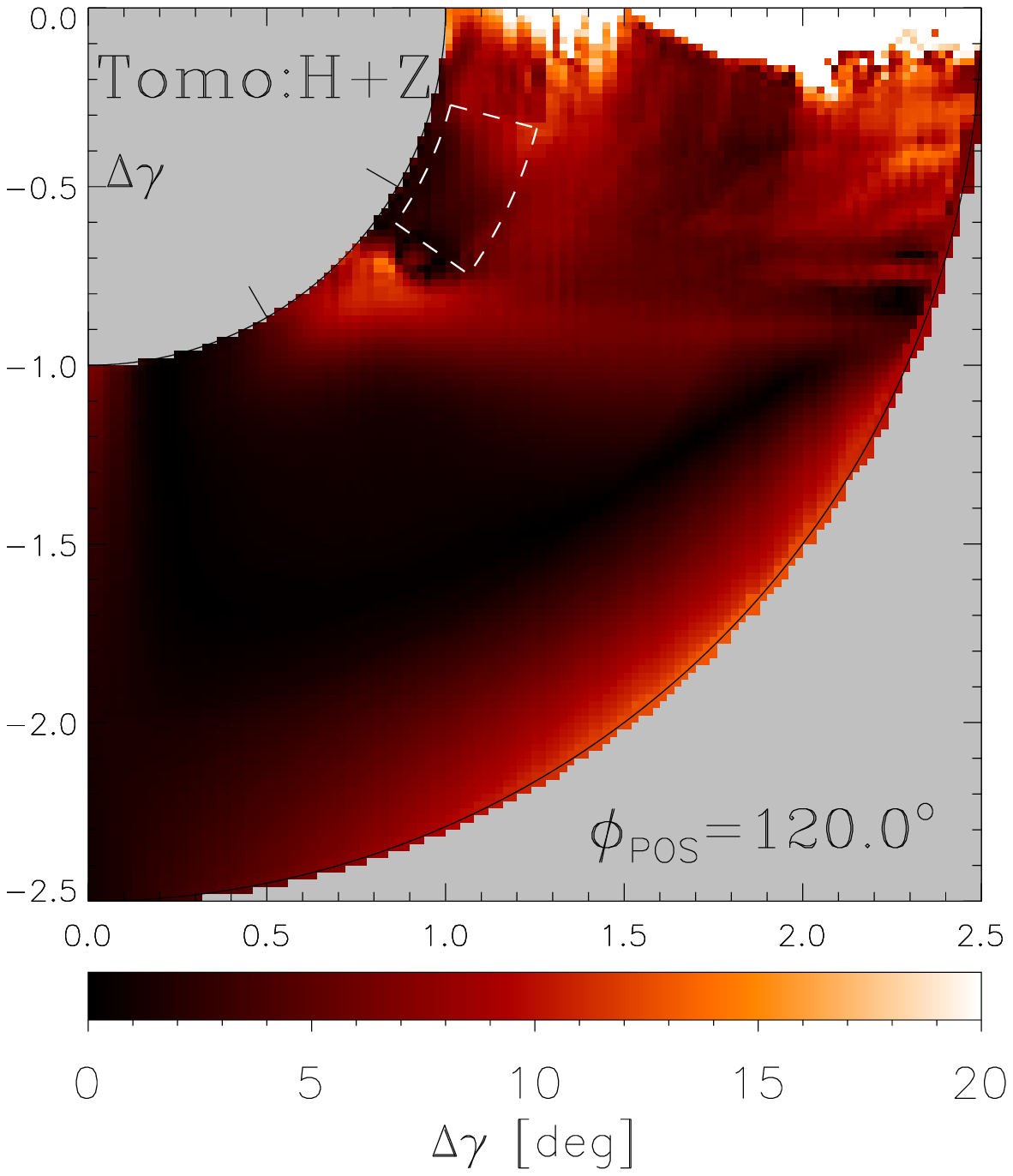}
\includegraphics*[bb=87 308 426 703,width=0.24\linewidth]{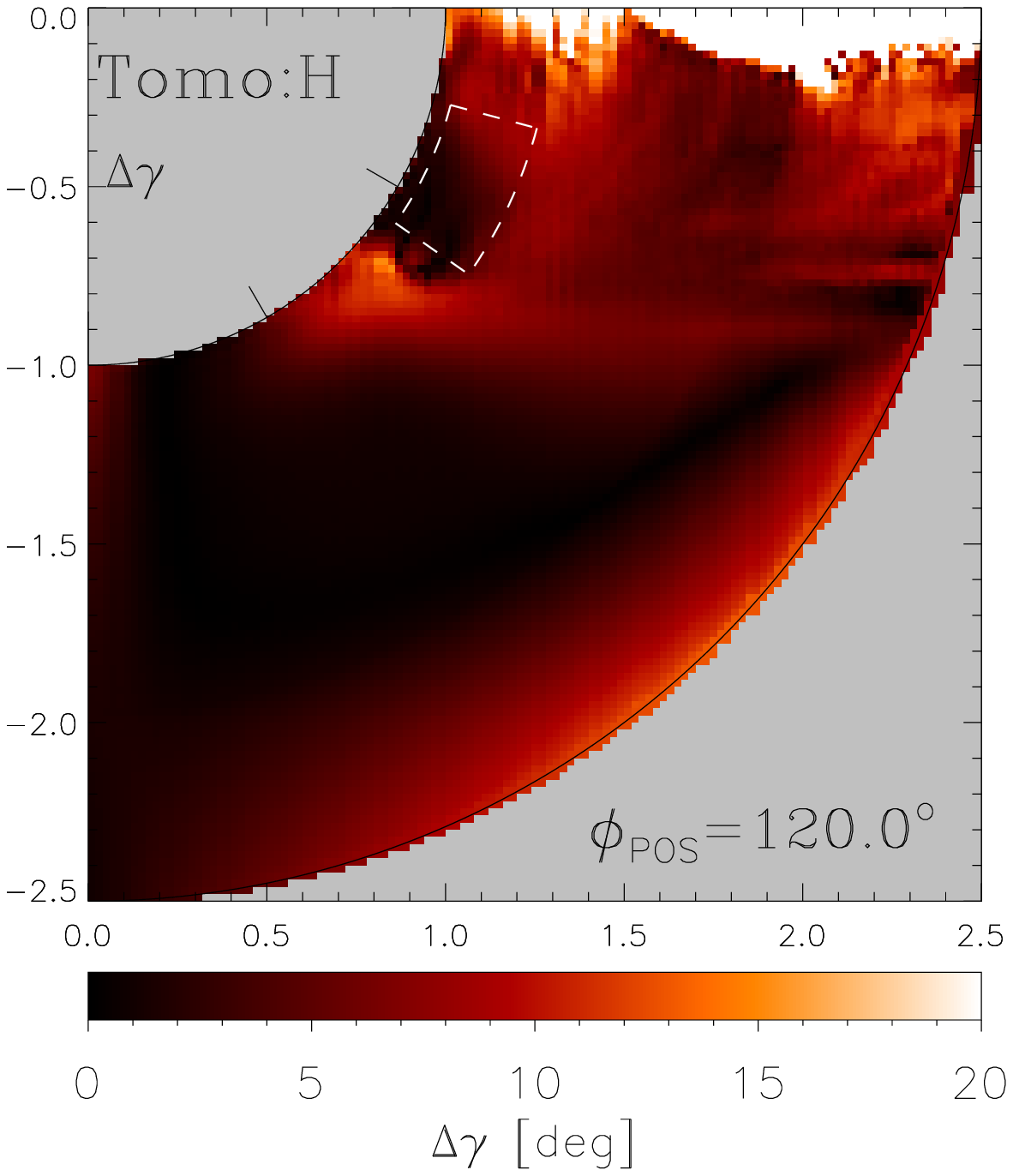}
\includegraphics*[bb=87 308 426 703,width=0.24\linewidth]{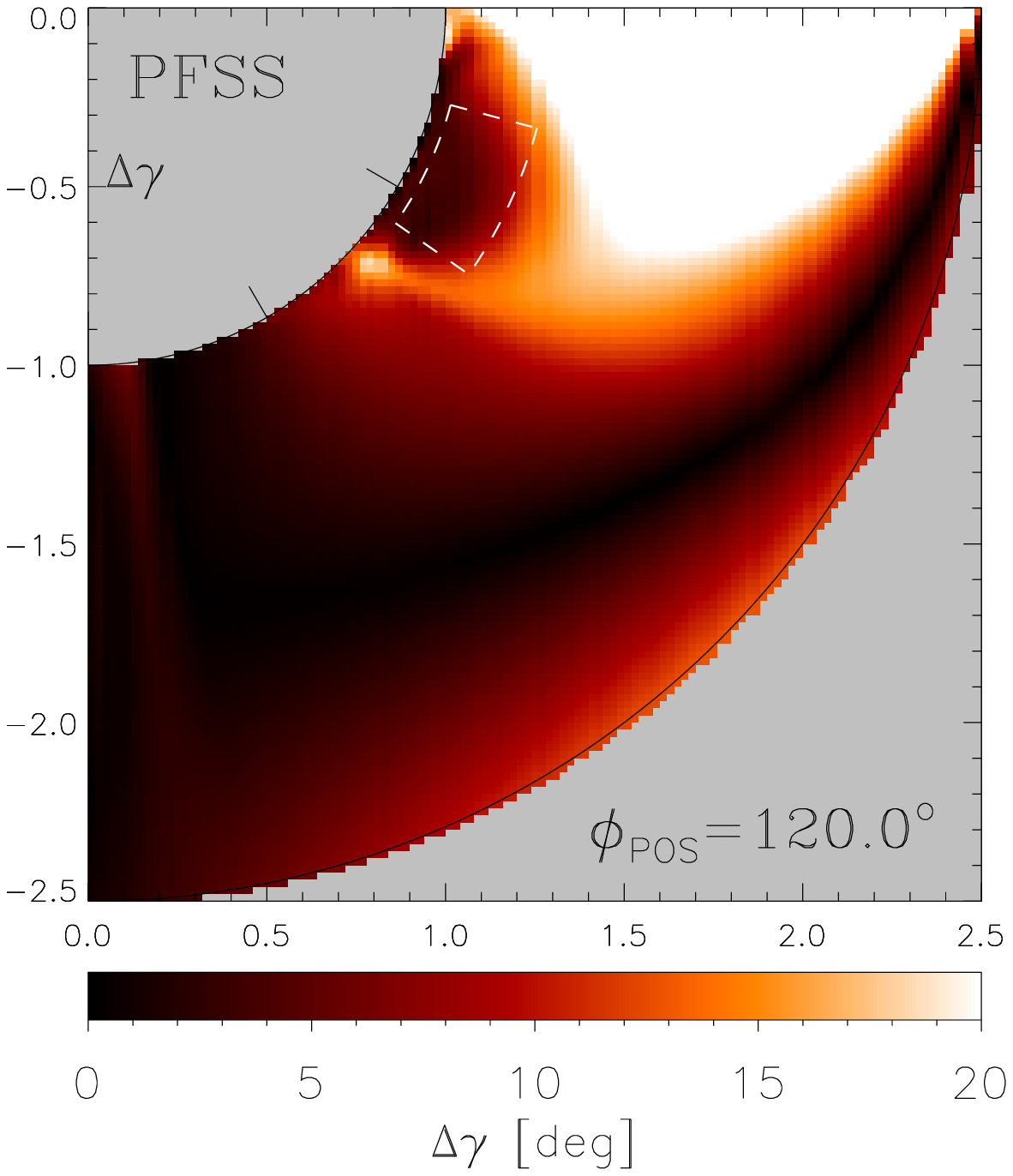}
\caption{Meridional cross-sections of the magnetic field properties for the GT MHD model and errors for the 
tomography reconstructions with full Stokes (second column) and only LP (third column) data, and for starting PFSS model (fourth column). First row shows maps of the magnetic field strength; second and third rows show maps of the absolute and relative field strength errors, respectively; fourth row shows maps of the difference angle between the GT model and the reconstructions. The cross-sections are for Carrington longitude of $120^\circ$.}
\label{Fig_Tomo_Err_Map_All}
\end{figure}

\begin{figure}[h!]
\includegraphics*[bb=87 308 426 703,width=0.24\linewidth]{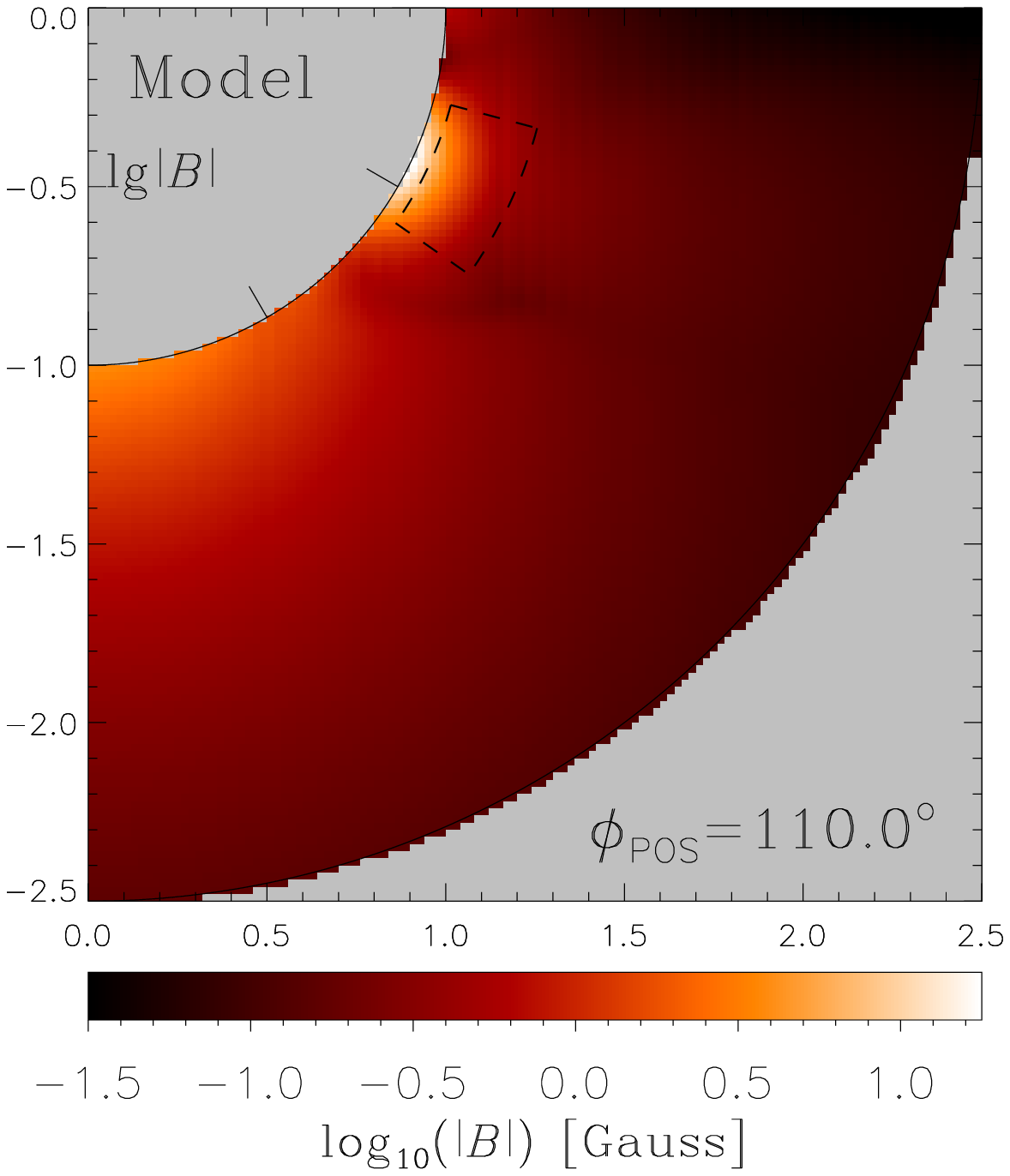}
\includegraphics*[bb=87 308 426 703,width=0.24\linewidth]{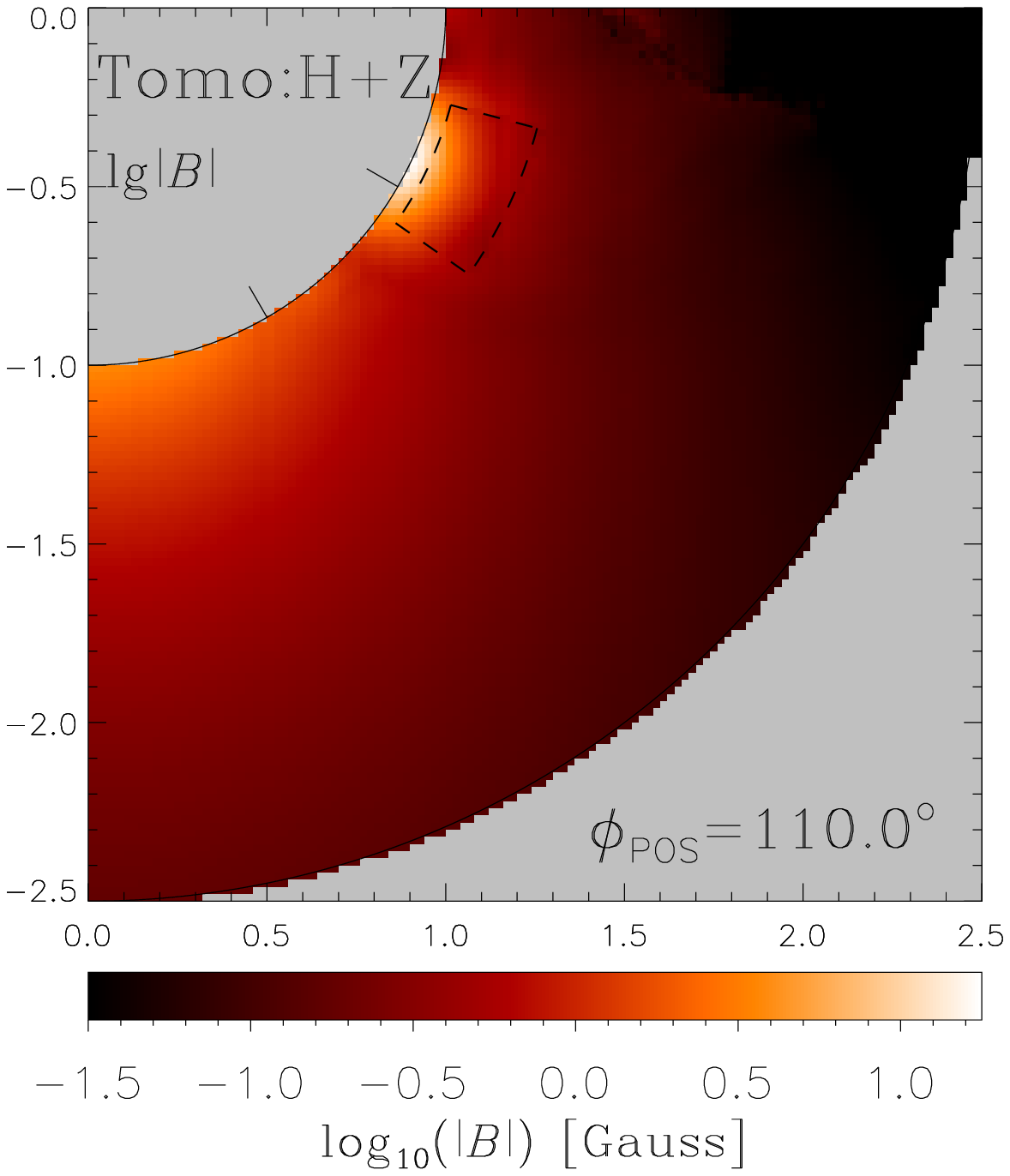}
\includegraphics*[bb=87 308 426 703,width=0.24\linewidth]{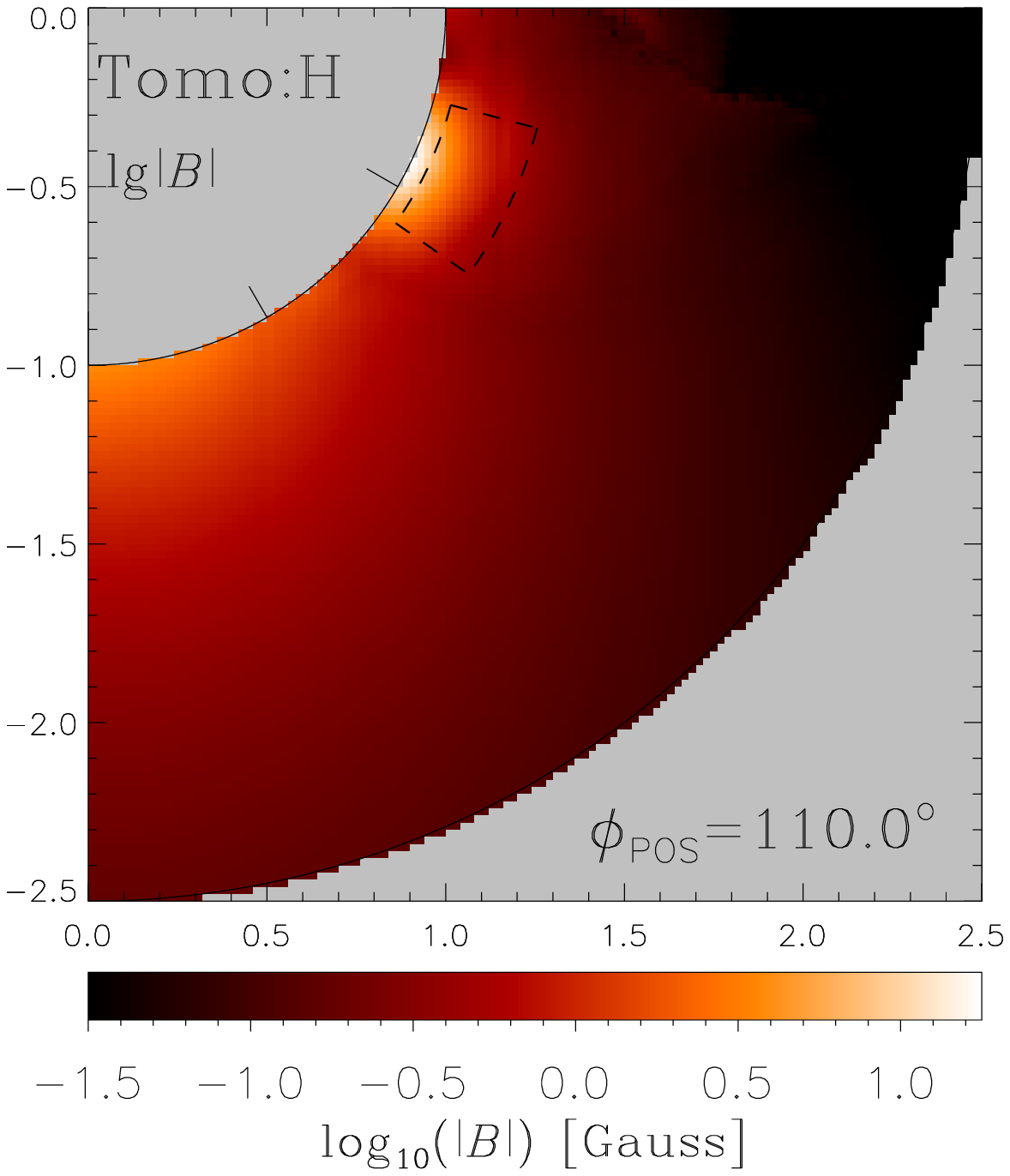}
\includegraphics*[bb=87 308 426 703,width=0.24\linewidth]{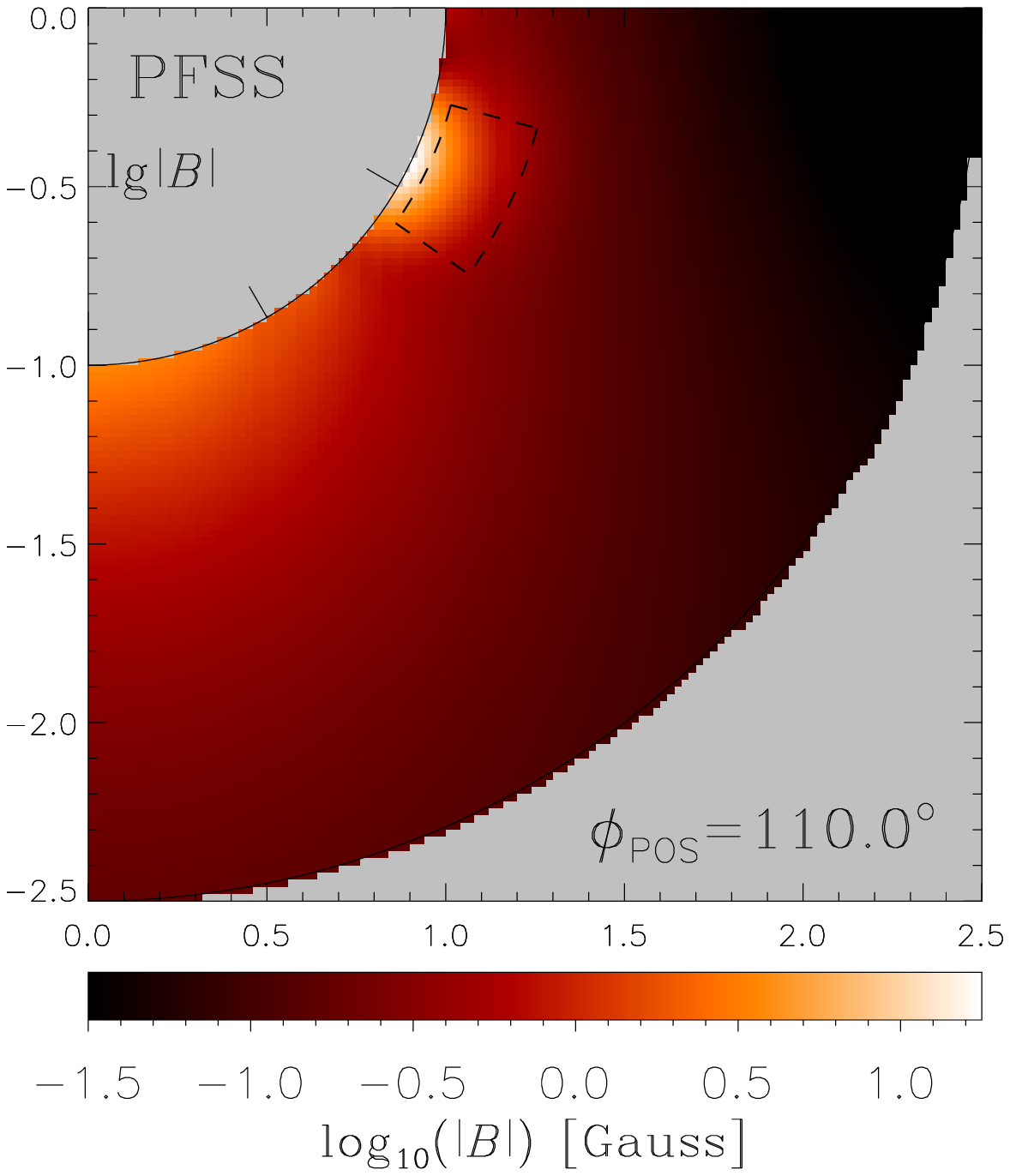}\\
\includegraphics*[bb=87 308 426 703,width=0.24\linewidth]{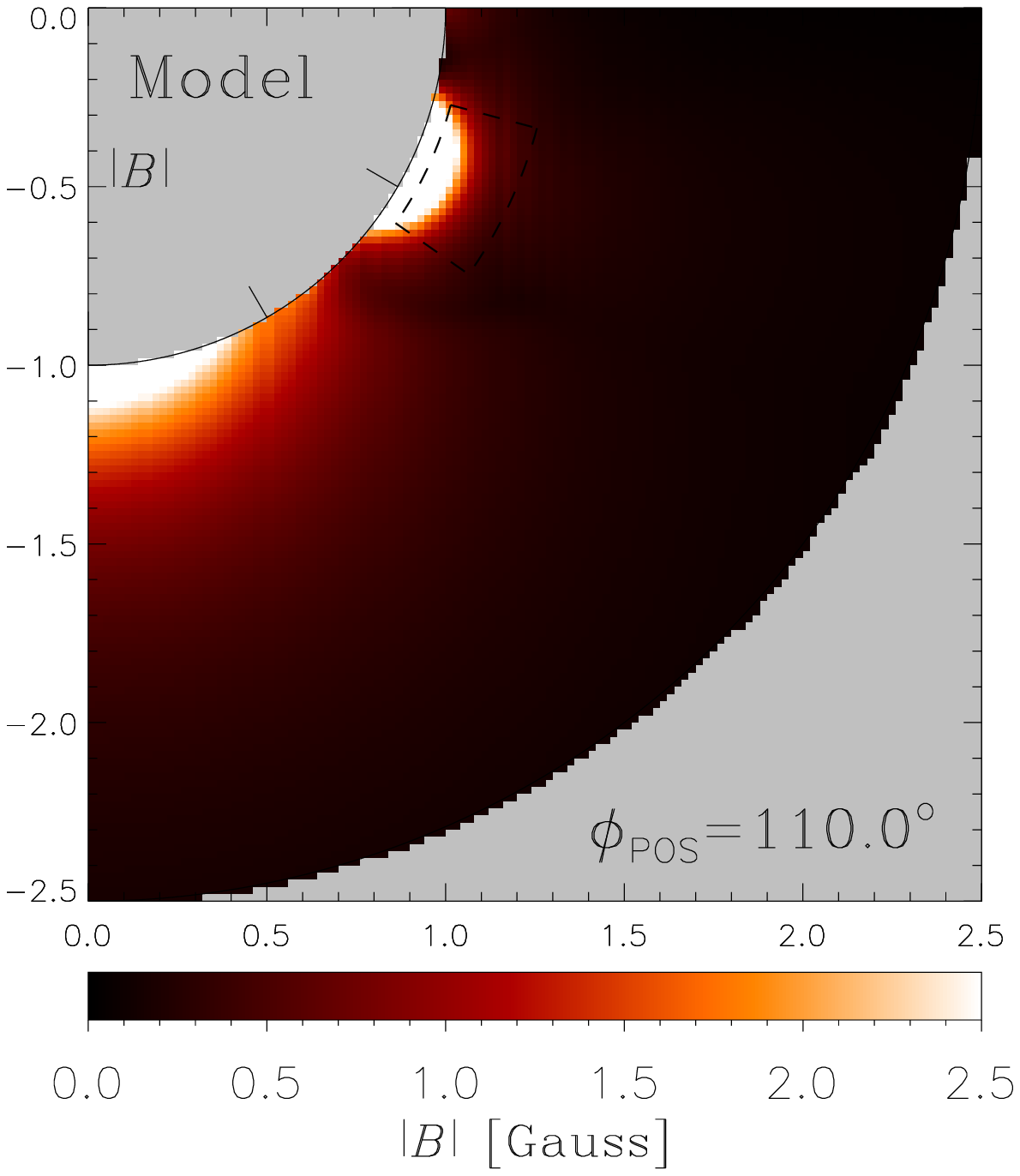}
\includegraphics*[bb=87 308 426 703,width=0.24\linewidth]{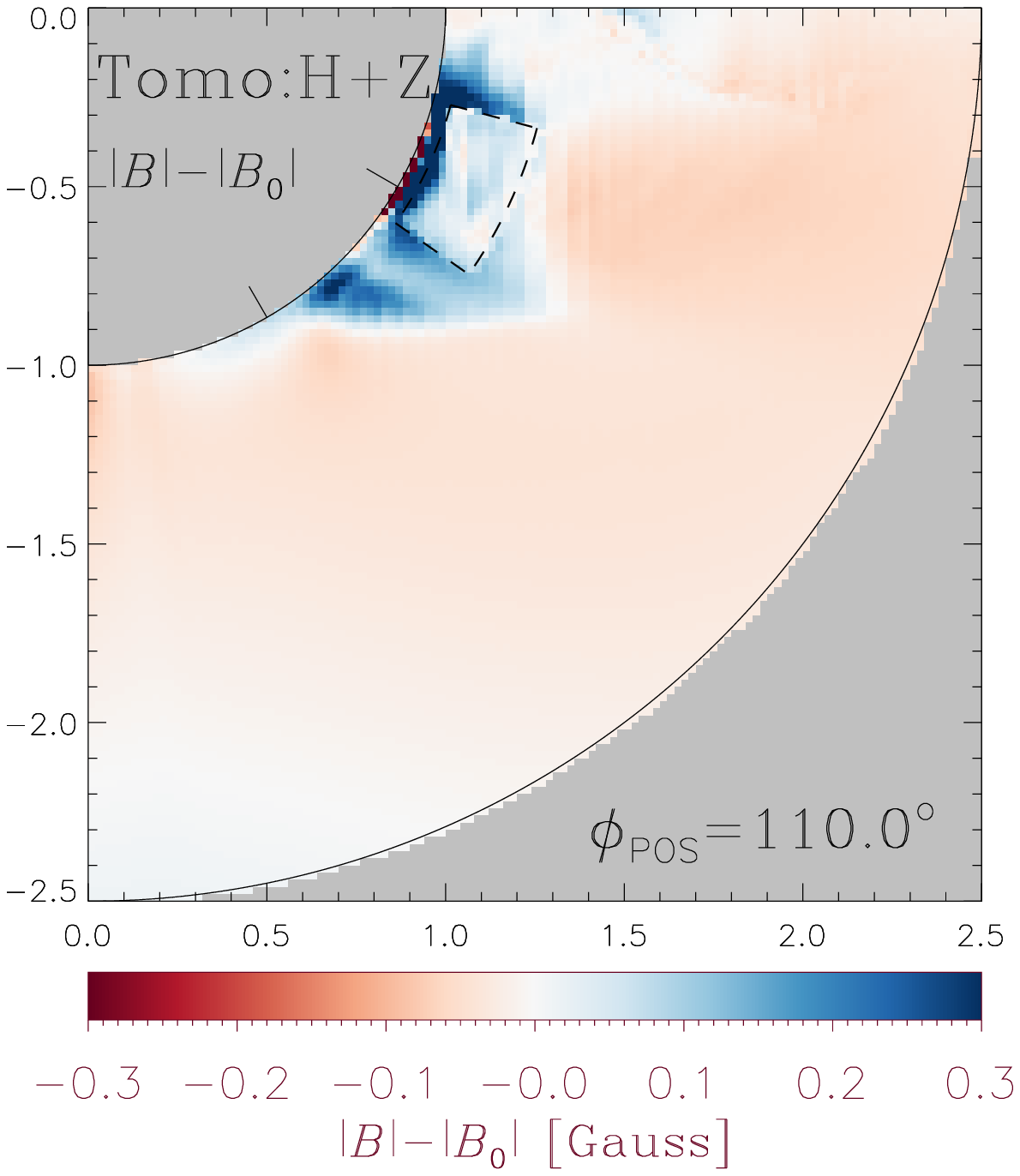}
\includegraphics*[bb=87 308 426 703,width=0.24\linewidth]{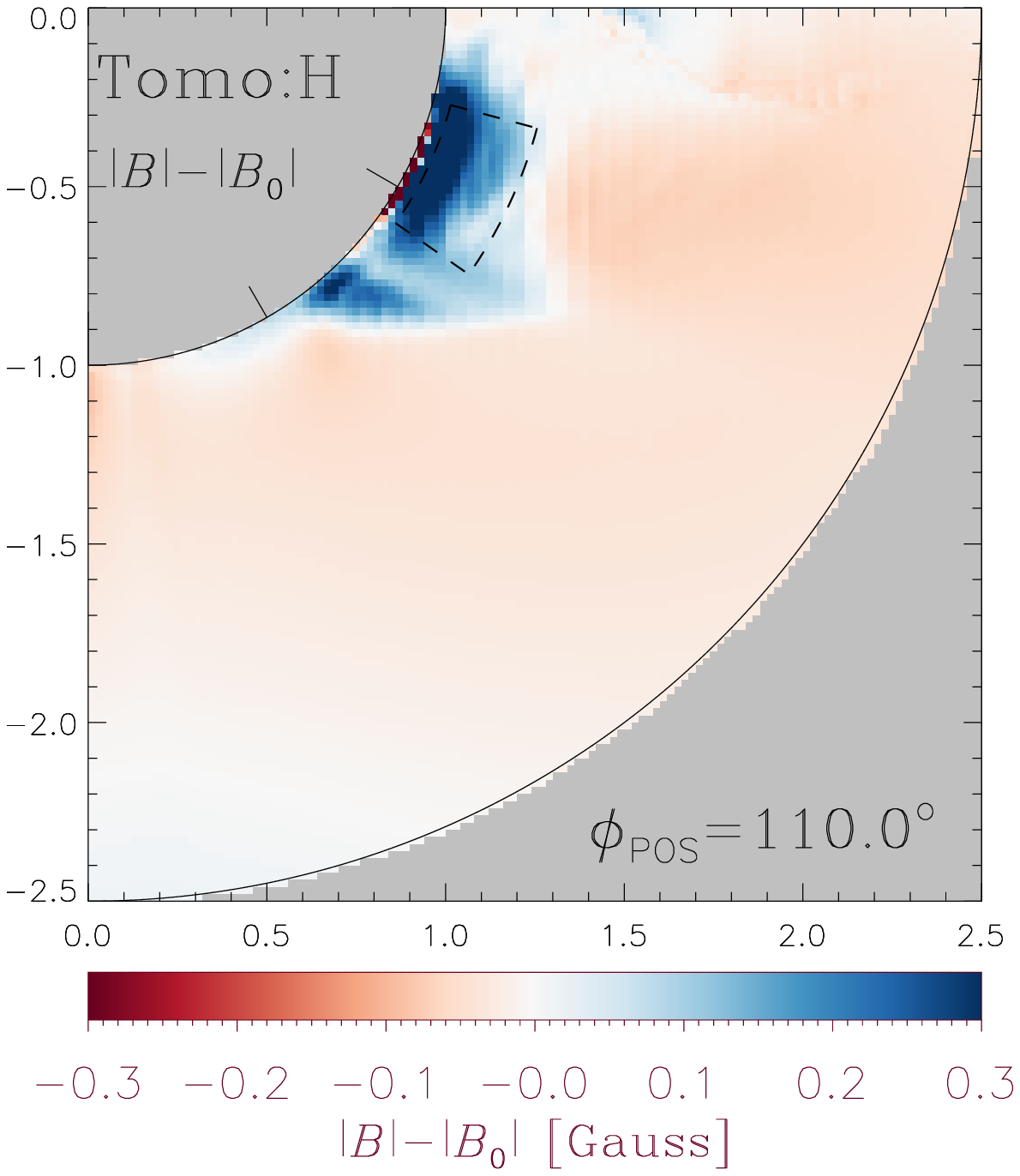}
\includegraphics*[bb=87 308 426 703,width=0.24\linewidth]{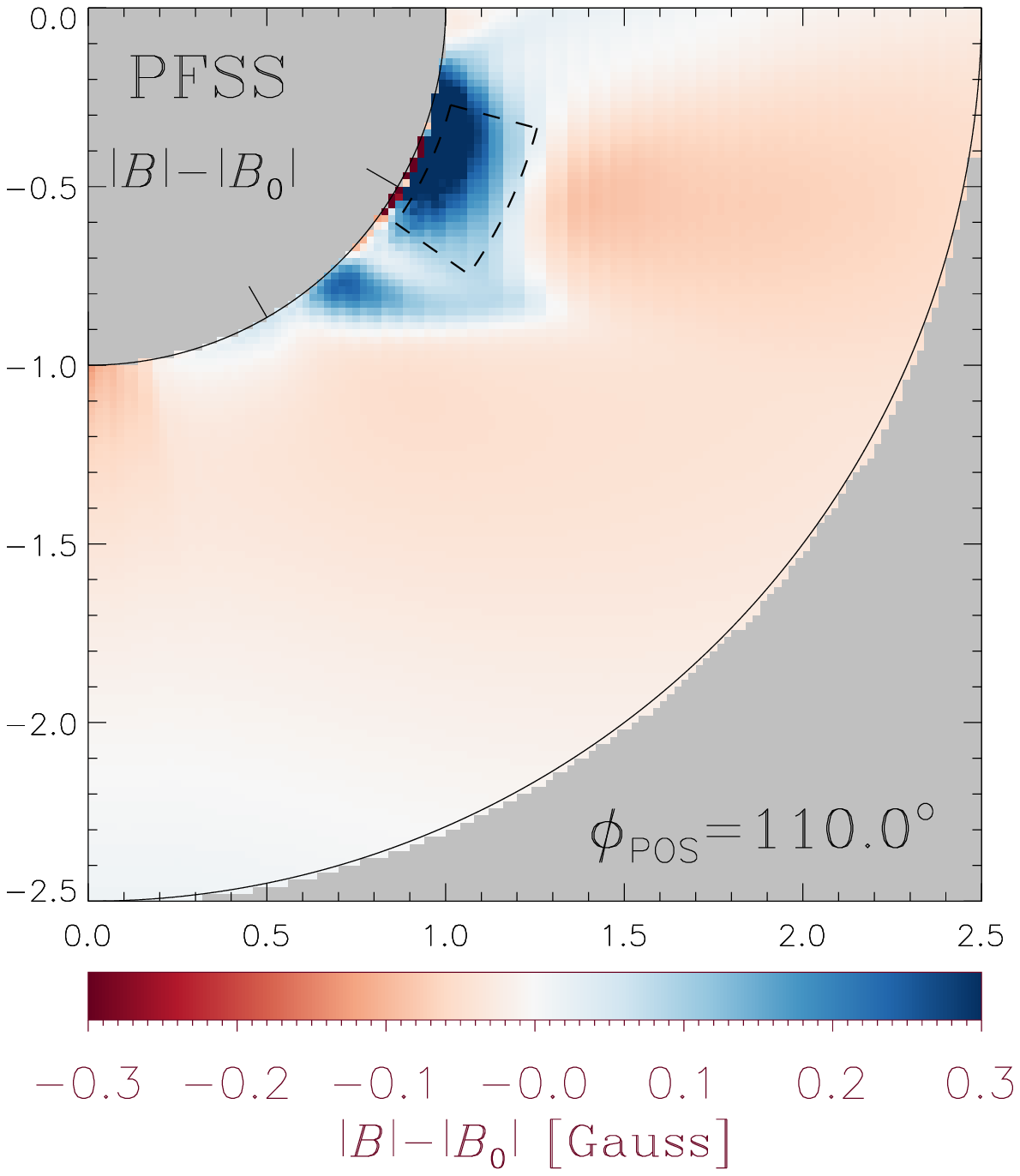}\\
\includegraphics*[bb=87 308 426 703,width=0.24\linewidth]{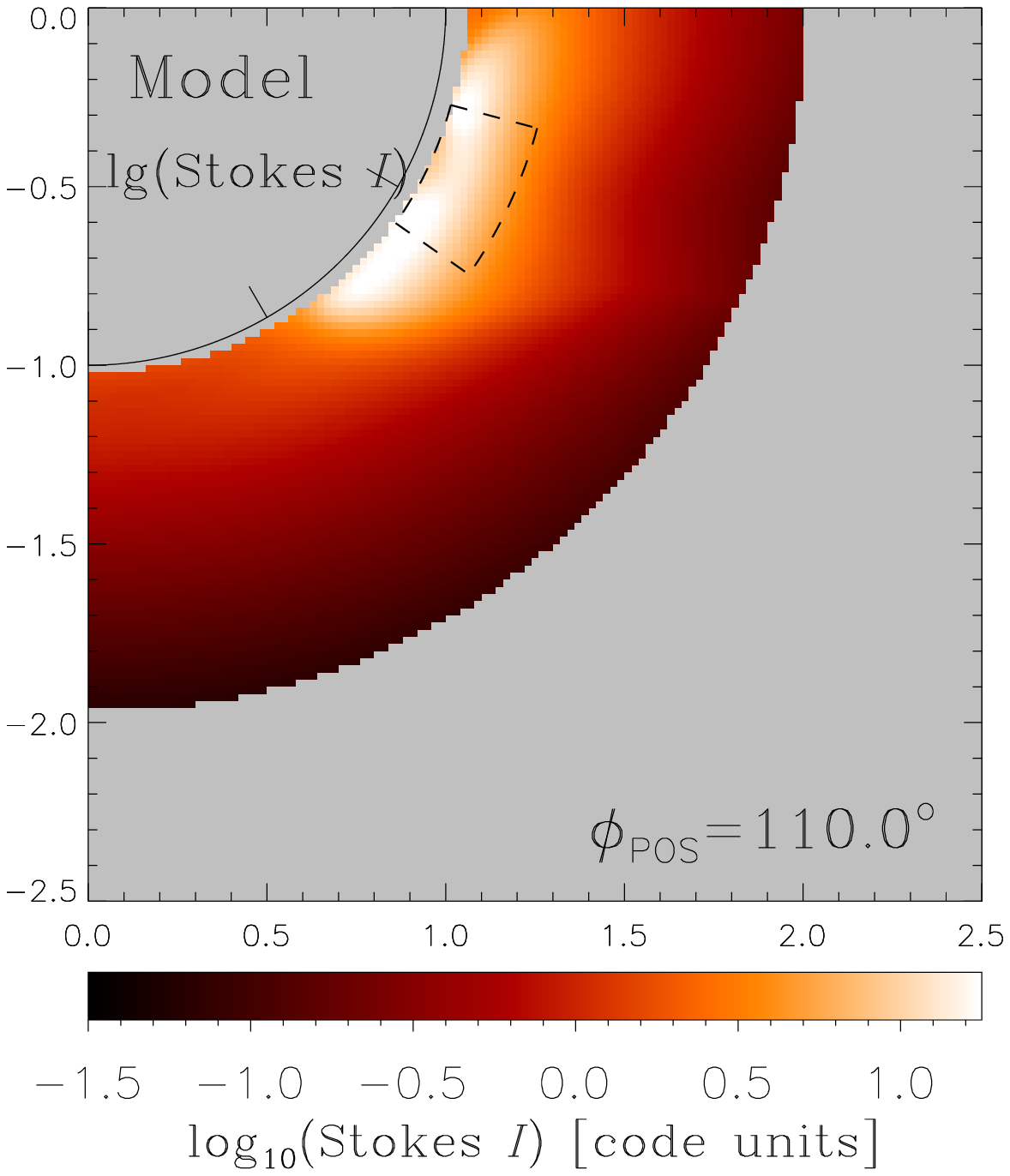}
\includegraphics*[bb=87 308 426 703,width=0.24\linewidth]{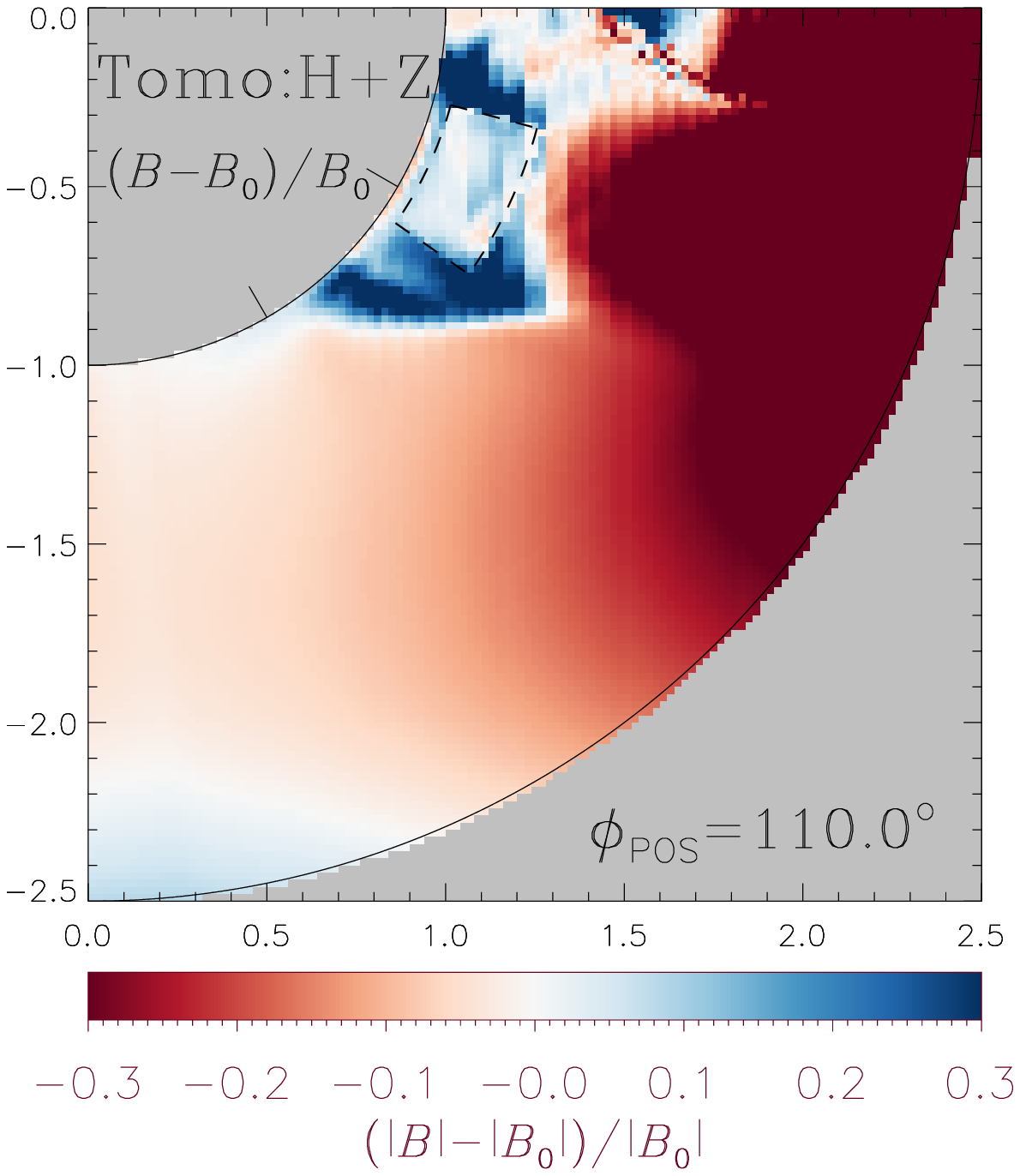}
\includegraphics*[bb=87 308 426 703,width=0.24\linewidth]{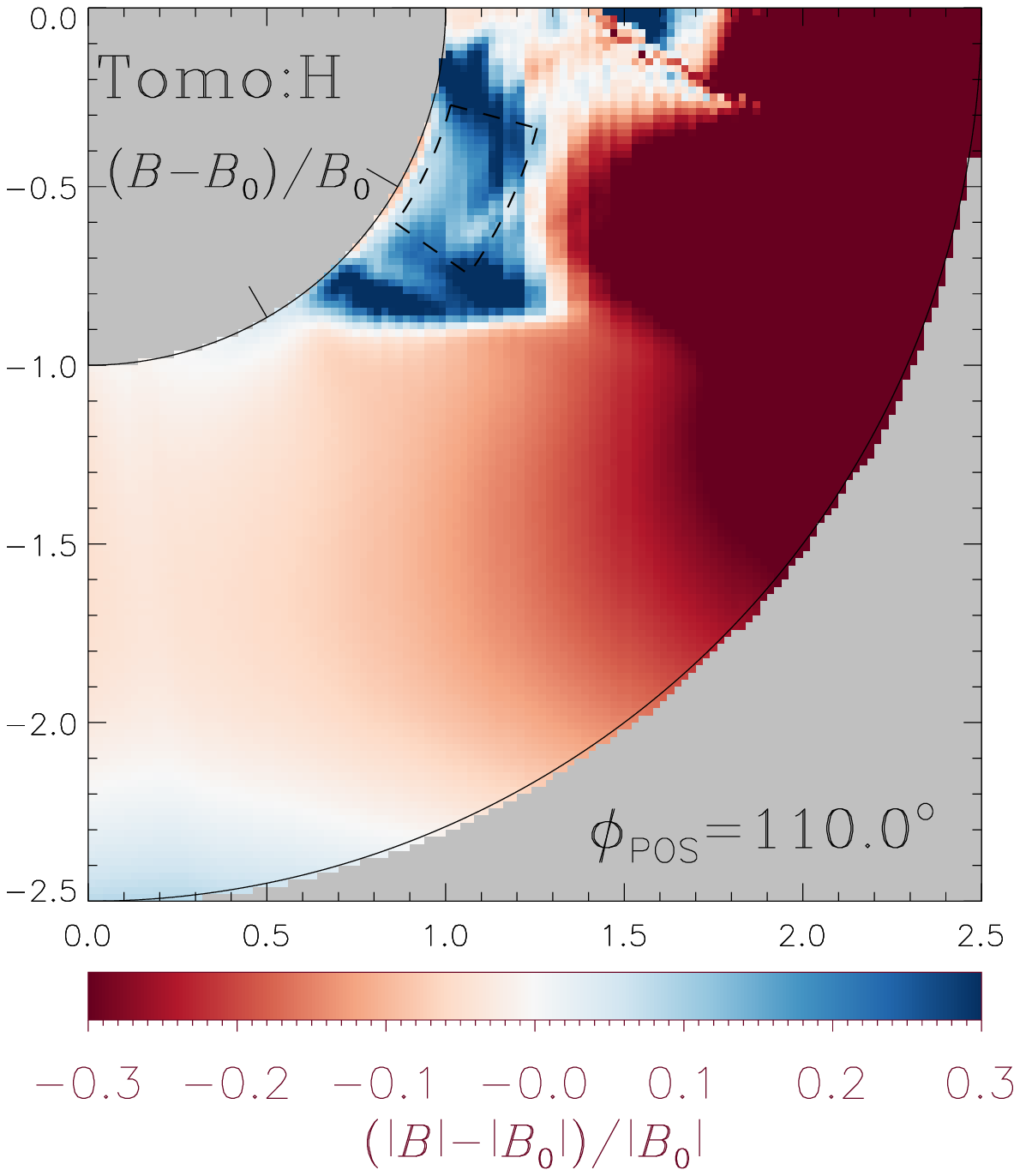}
\includegraphics*[bb=87 308 426 703,width=0.24\linewidth]{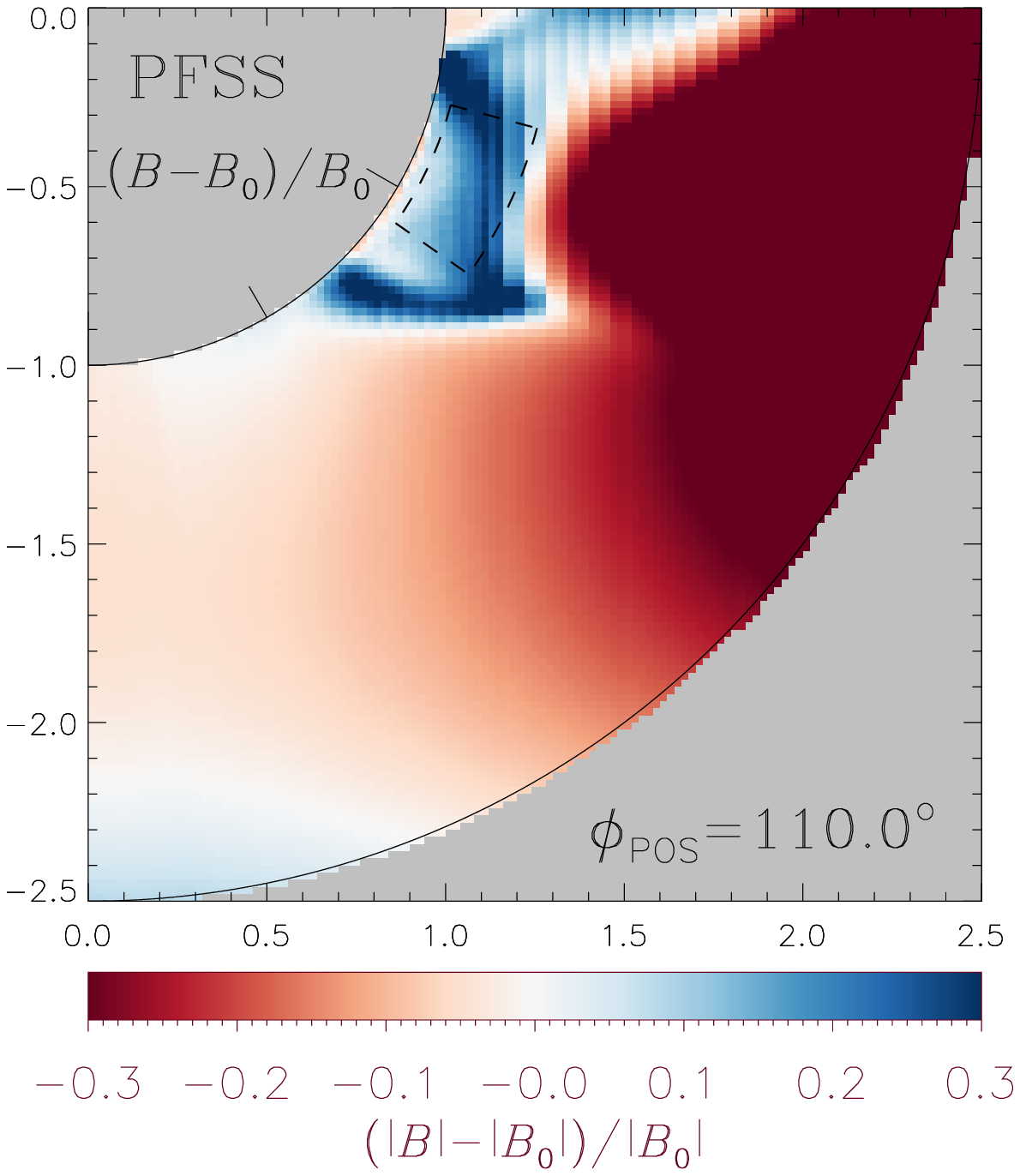}\\
\includegraphics*[bb=87 308 426 703,width=0.24\linewidth]{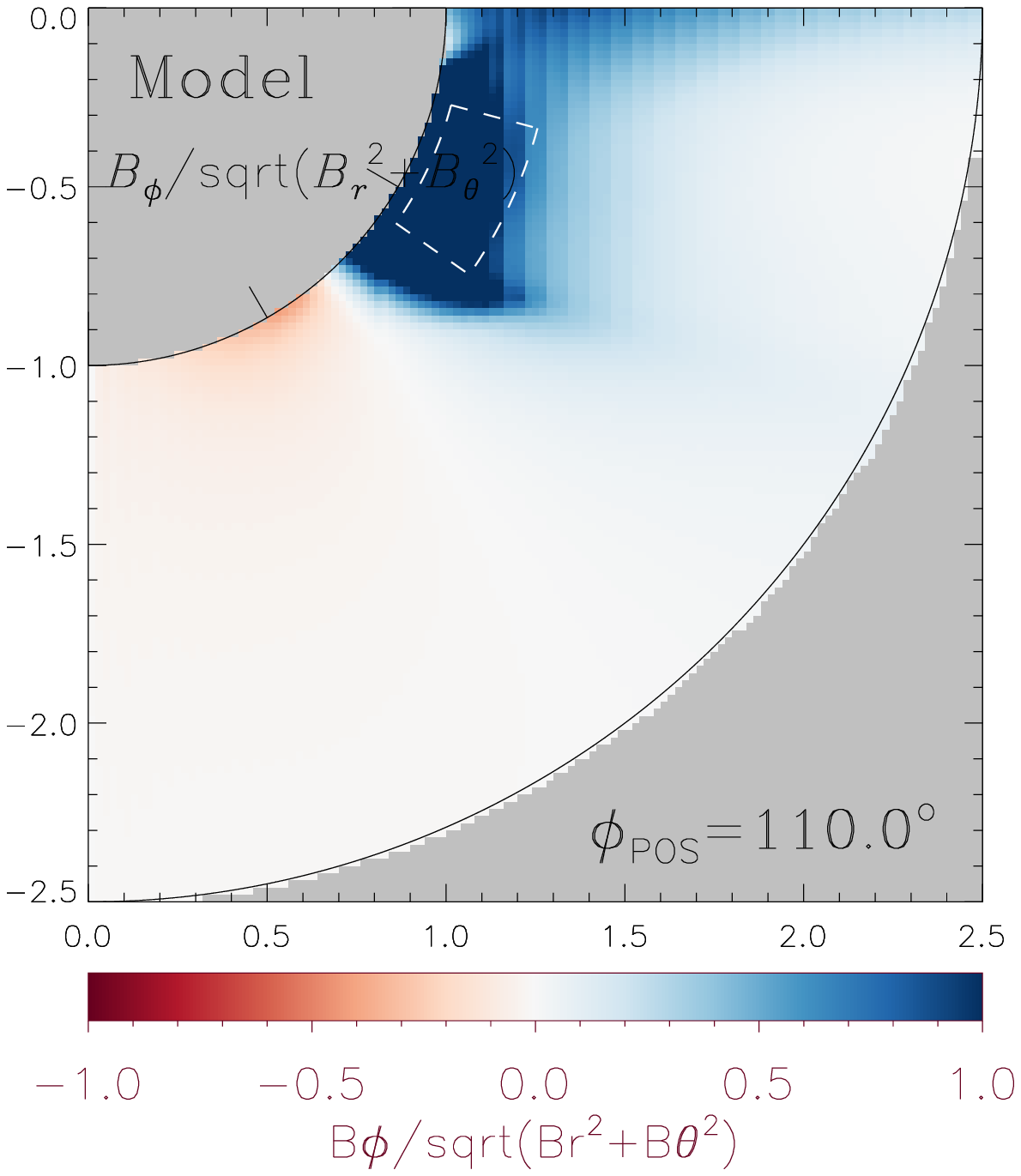}
\includegraphics*[bb=87 308 426 703,width=0.24\linewidth]{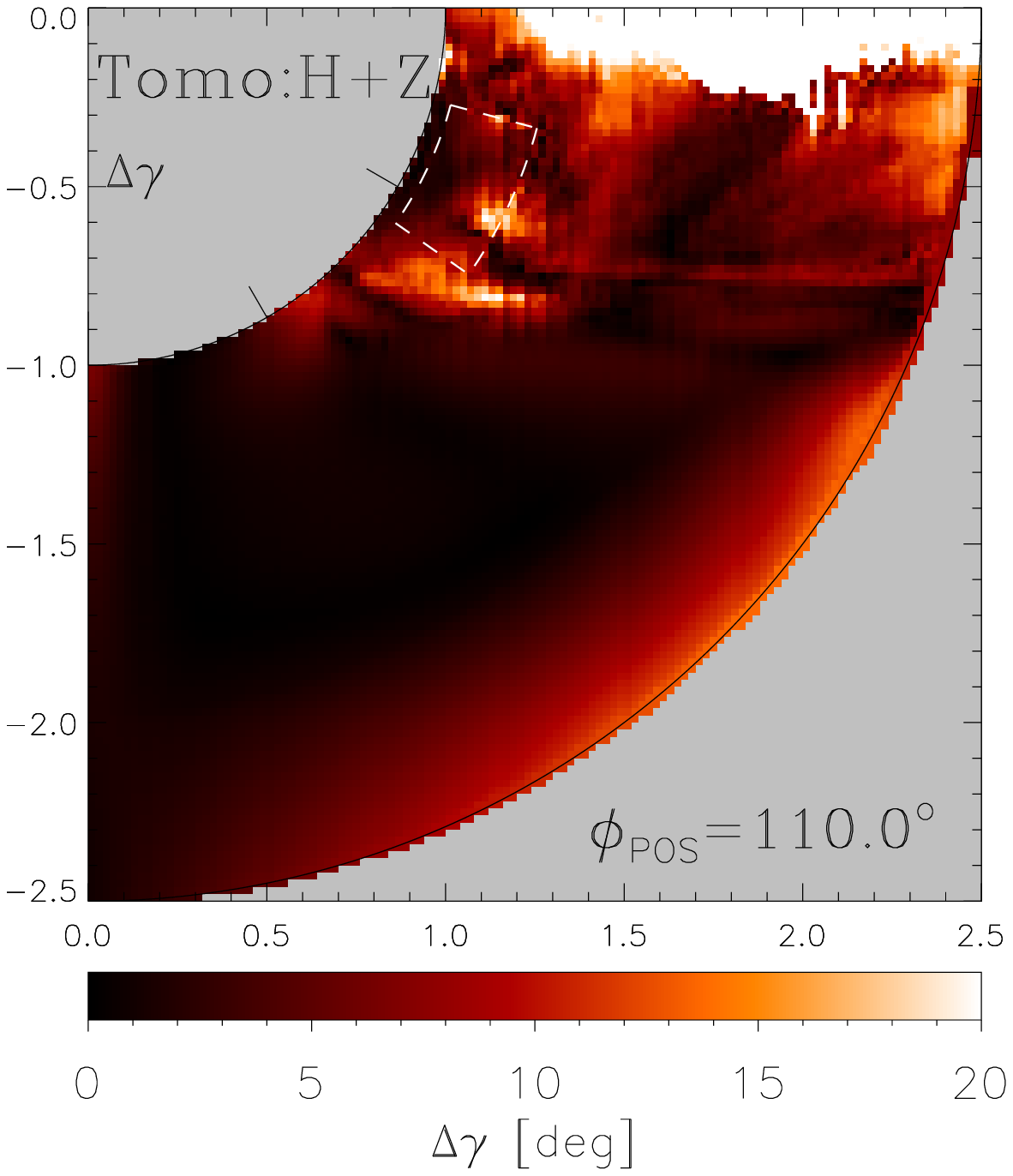}
\includegraphics*[bb=87 308 426 703,width=0.24\linewidth]{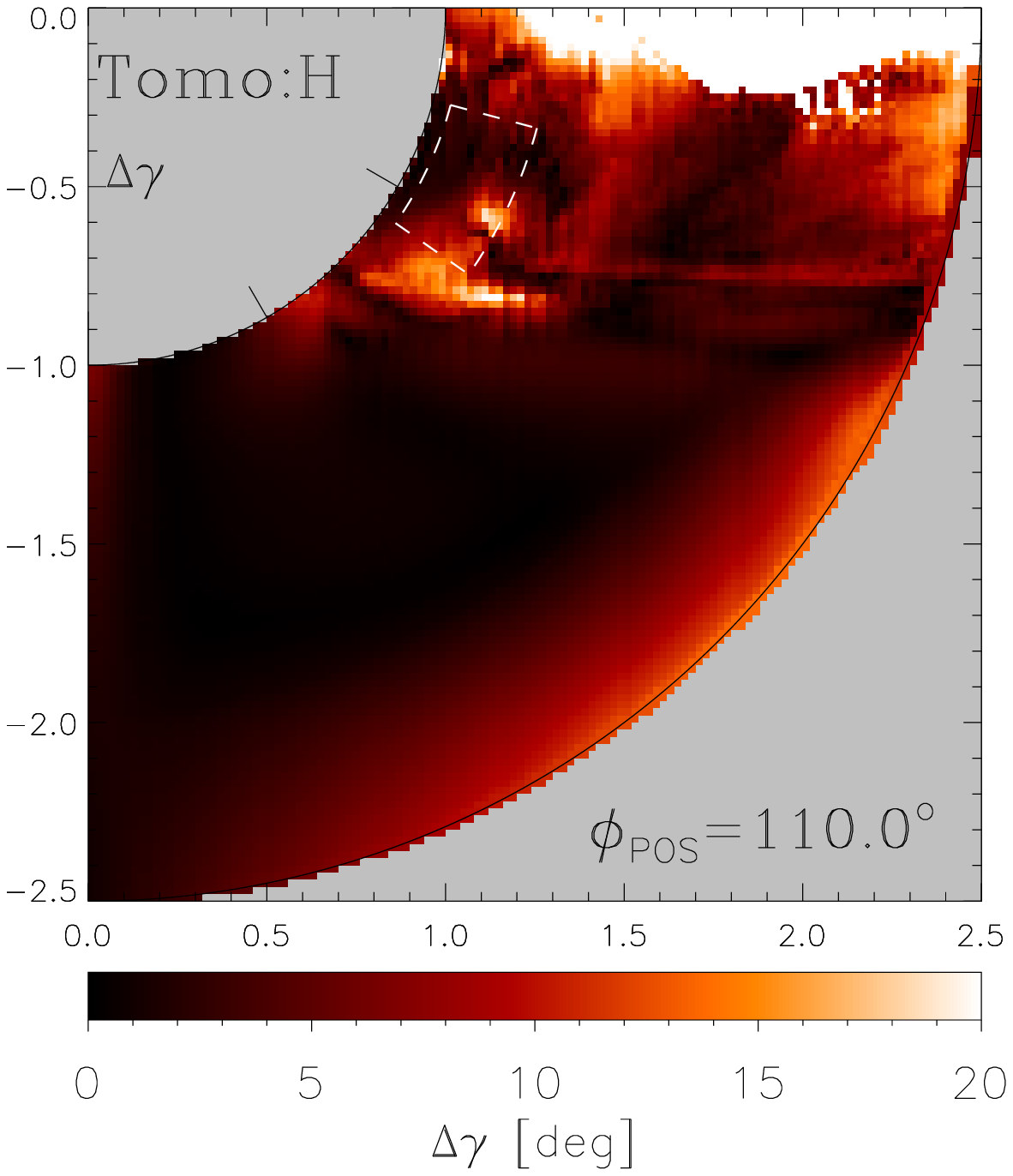}
\includegraphics*[bb=87 308 426 703,width=0.24\linewidth]{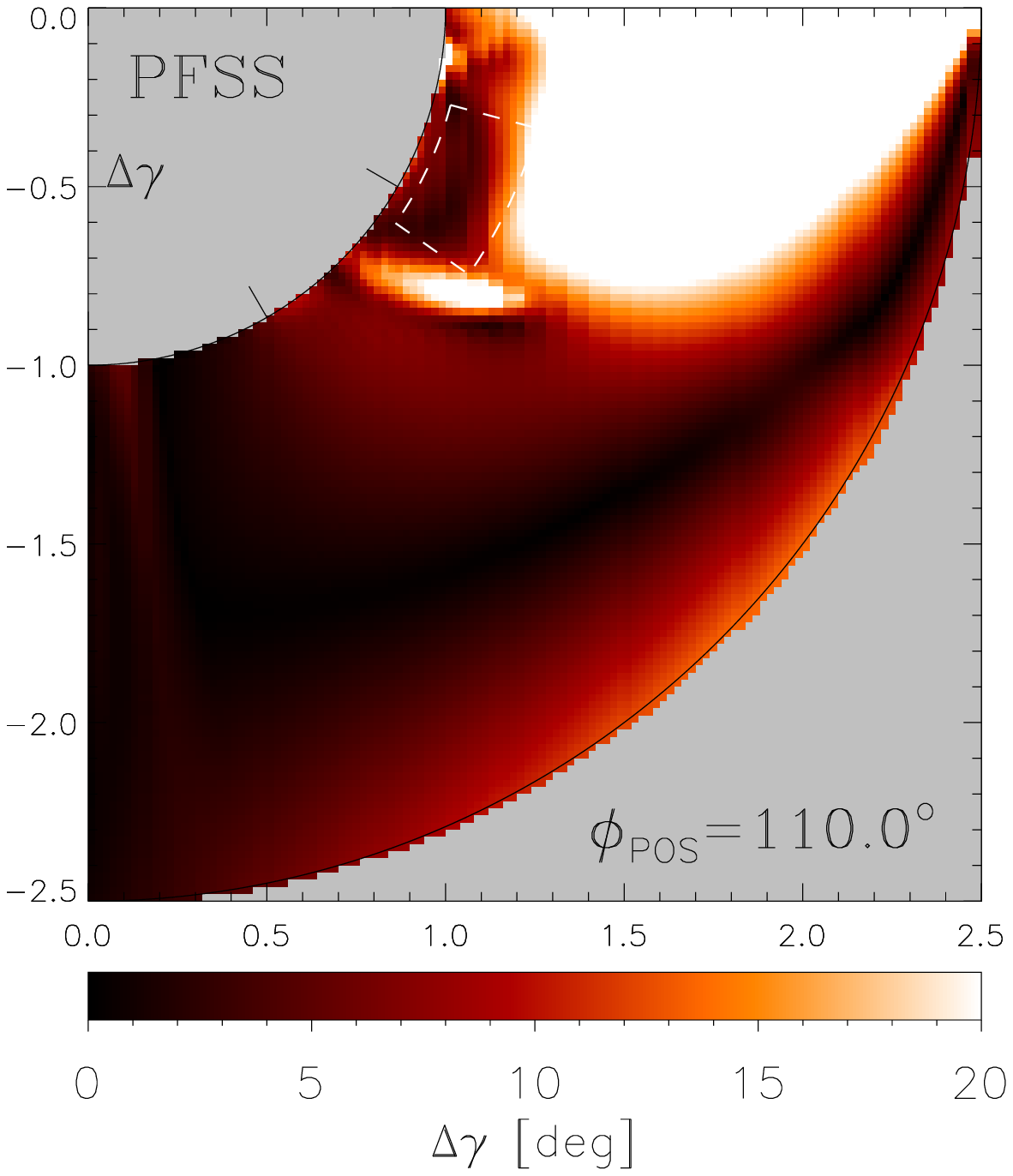}
\caption{Meridional cross-sections of the magnetic field properties for the GT MHD model and errors for the 
tomography reconstructions with full Stokes (second column) and only LP (third column) data, and for starting PFSS model (fourth column). First row shows maps of the magnetic field strength; second and third rows show maps of the absolute and relative field strength errors, respectively; fourth row shows maps of the difference angle between the GT model and the reconstructions. The cross-sections are for Carrington longitude of $110^\circ$.}
\label{Fig_Tomo_Err_Map_All2}
\end{figure}

\begin{figure}[h!]
\includegraphics*[bb=87 308 426 703,width=0.24\linewidth]{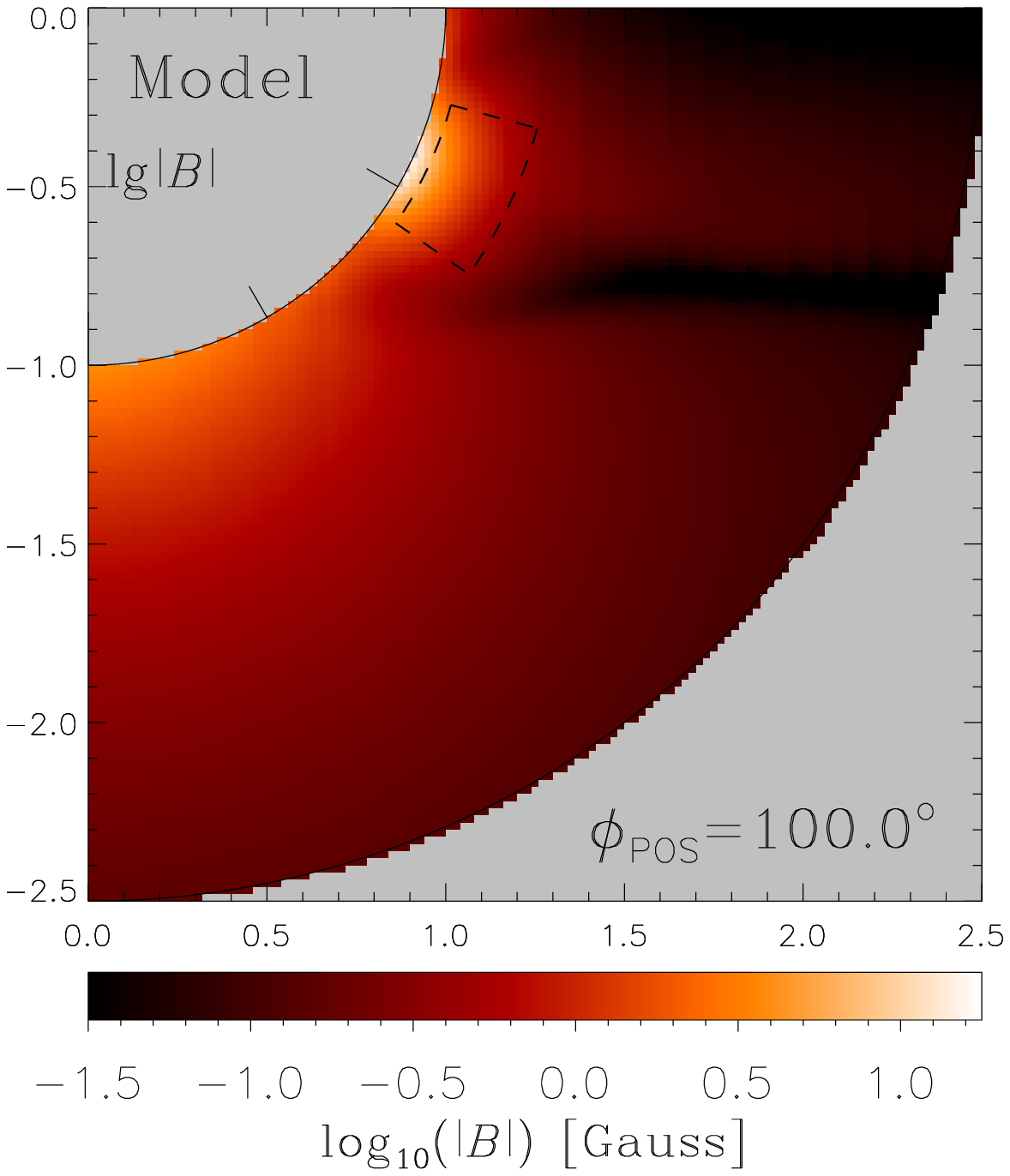}
\includegraphics*[bb=87 308 426 703,width=0.24\linewidth]{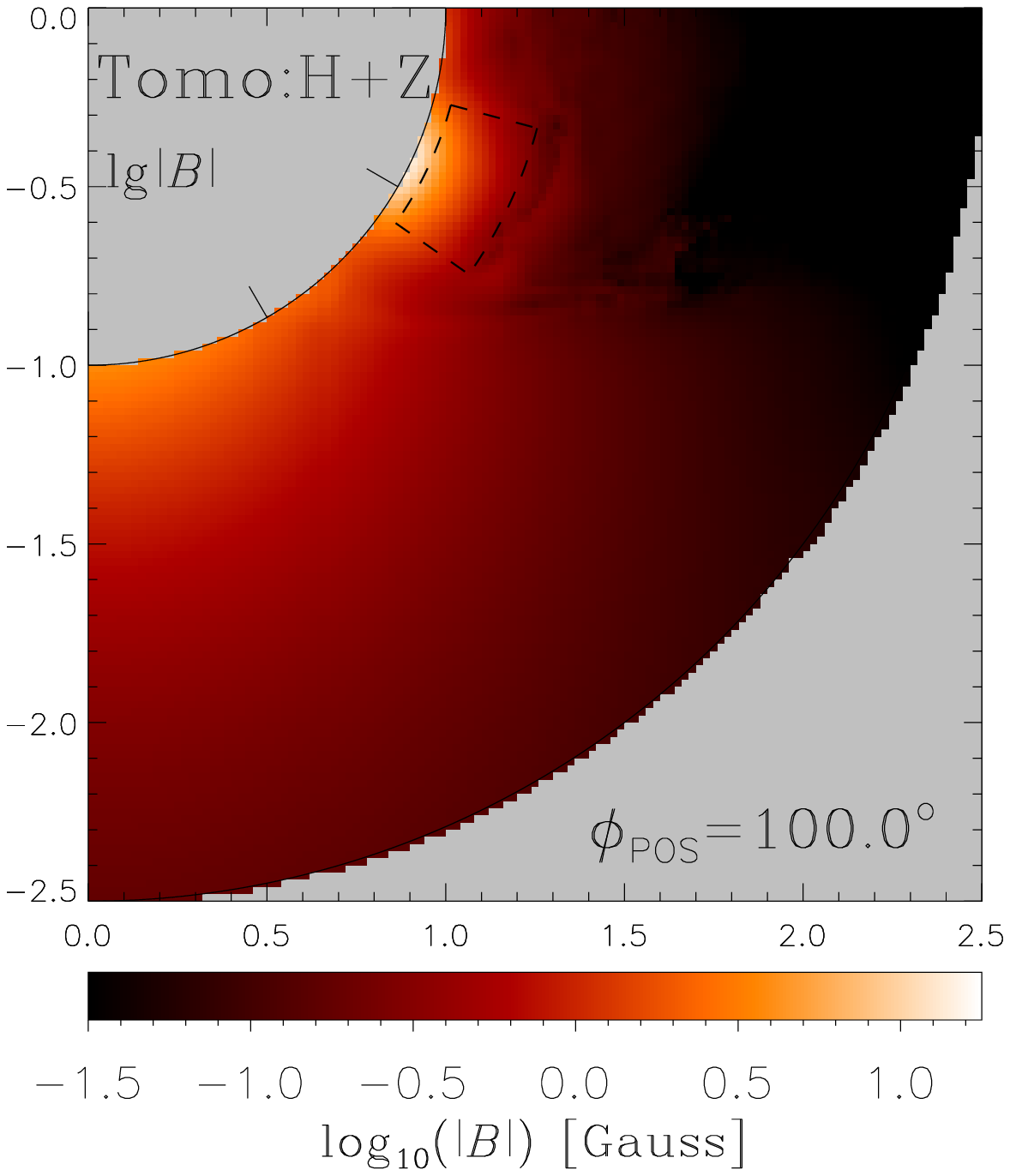}
\includegraphics*[bb=87 308 426 703,width=0.24\linewidth]{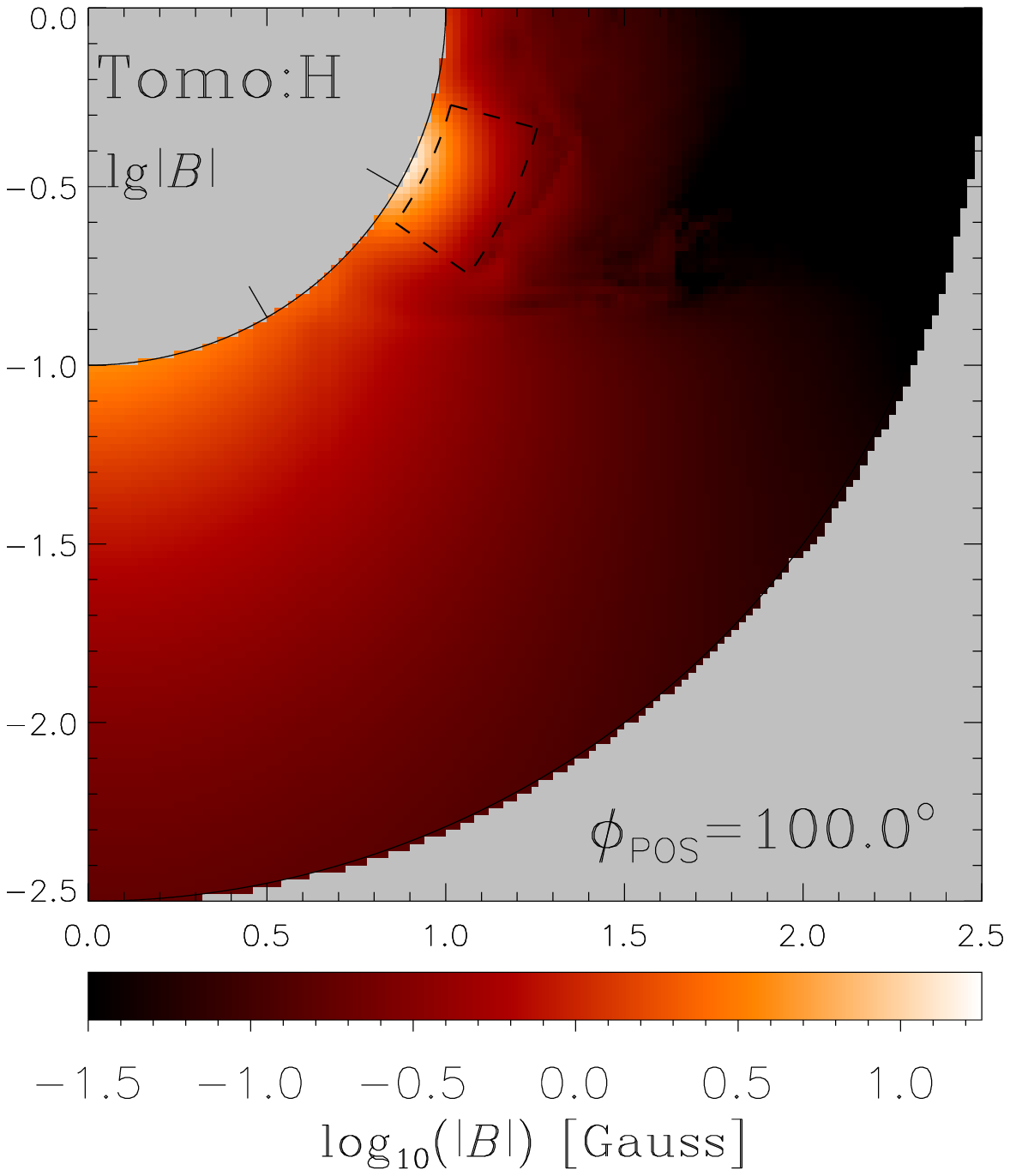}
\includegraphics*[bb=87 308 426 703,width=0.24\linewidth]{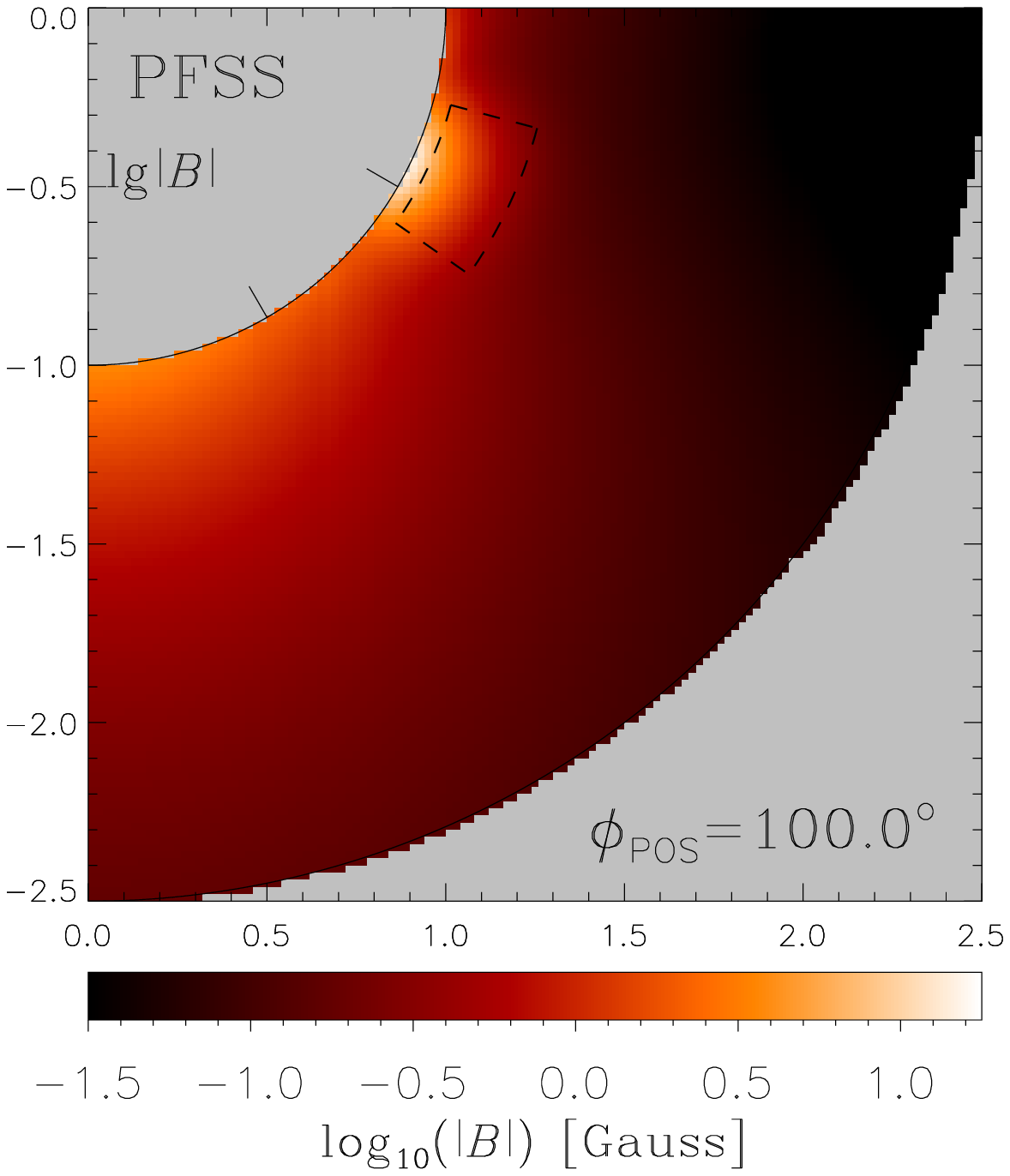}\\
\includegraphics*[bb=87 308 426 703,width=0.24\linewidth]{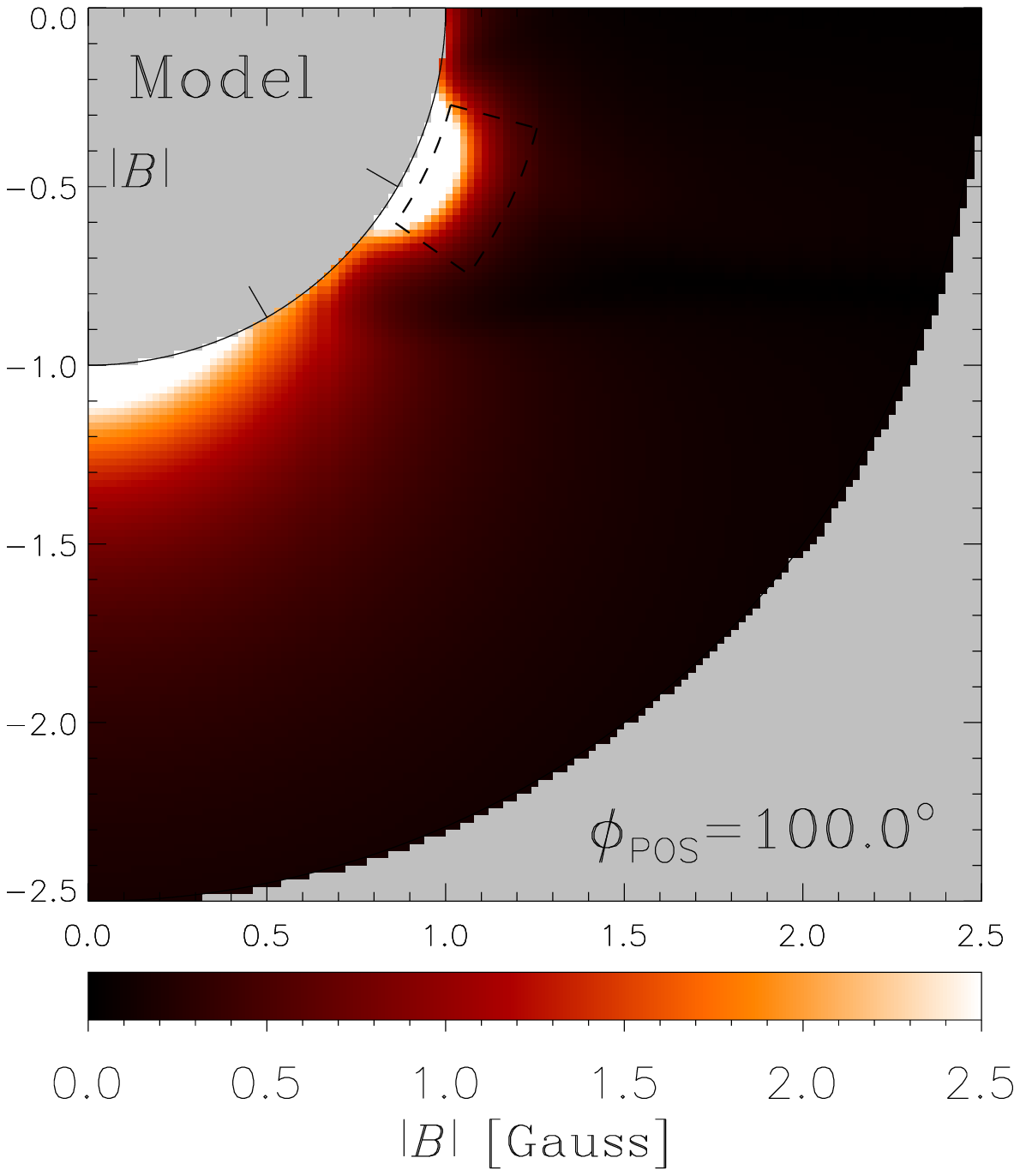}
\includegraphics*[bb=87 308 426 703,width=0.24\linewidth]{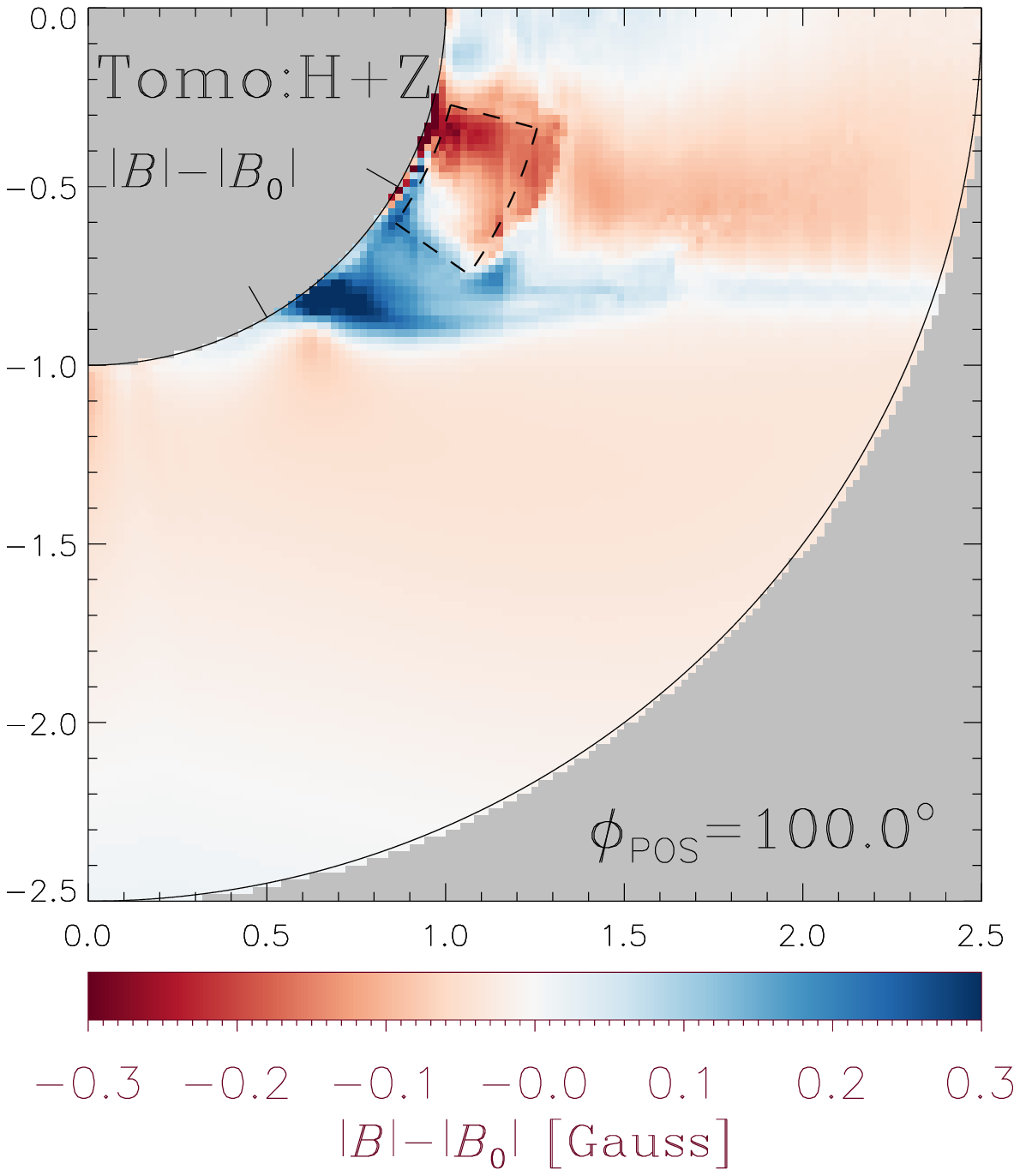}
\includegraphics*[bb=87 308 426 703,width=0.24\linewidth]{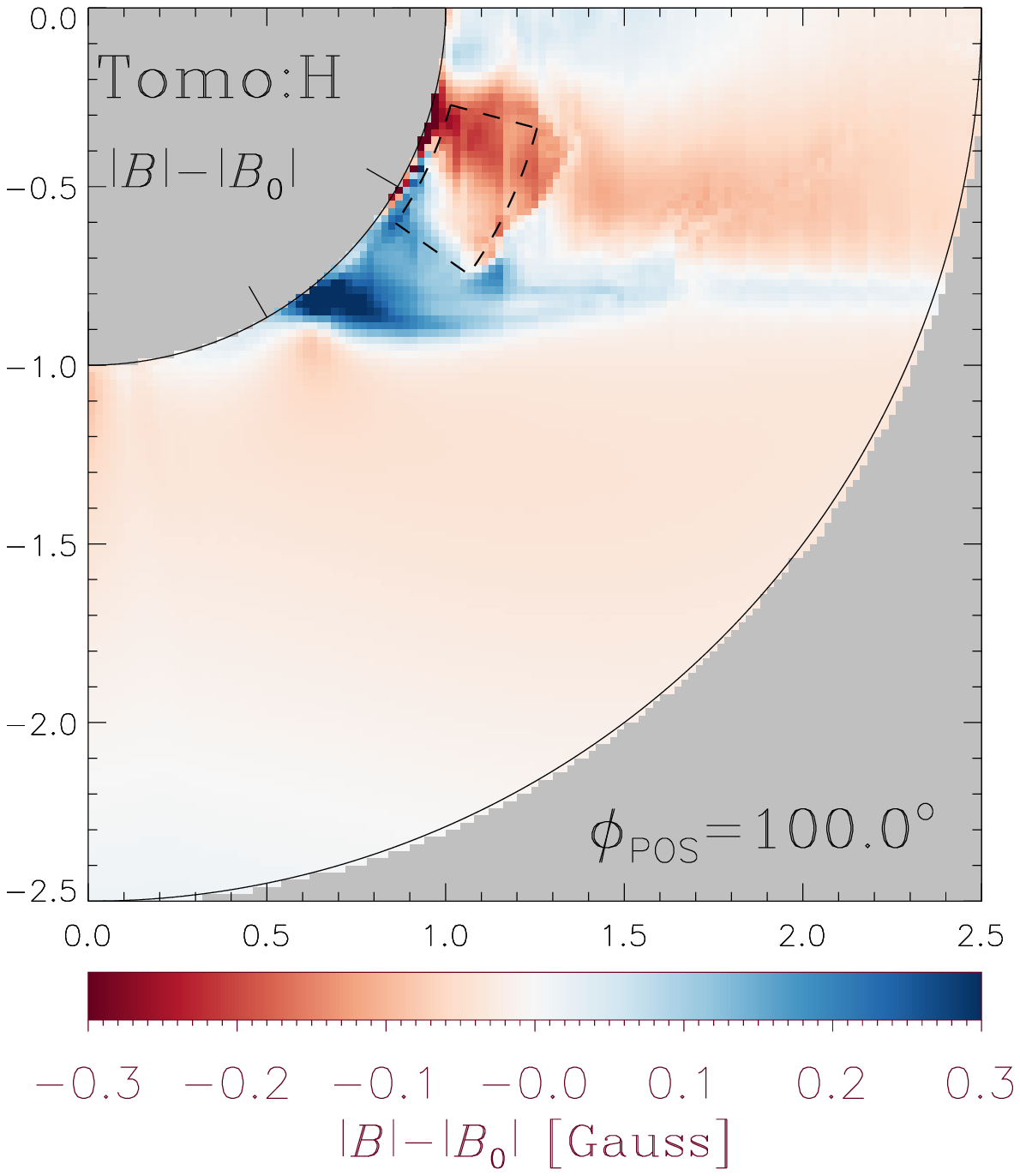}
\includegraphics*[bb=87 308 426 703,width=0.24\linewidth]{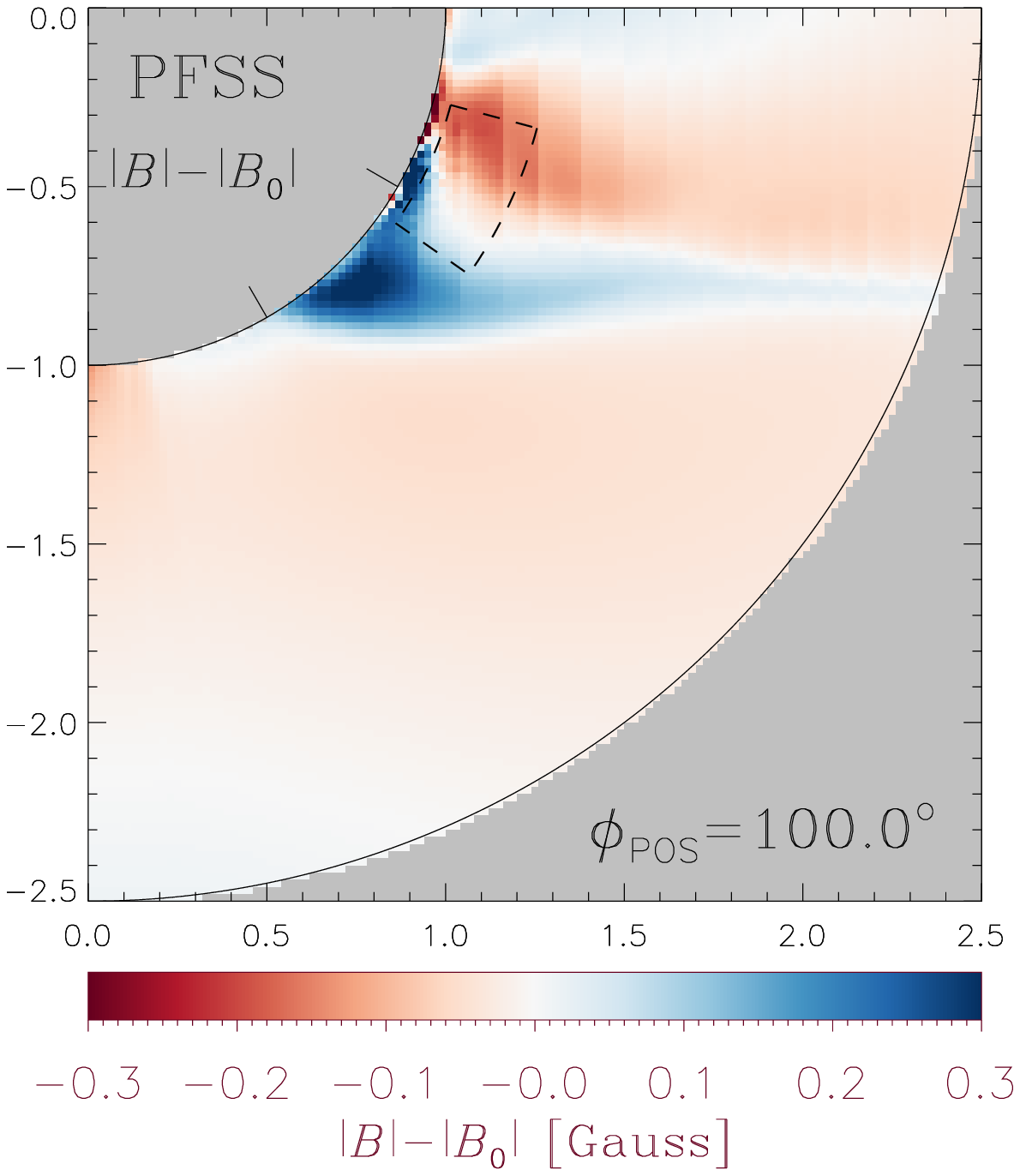}\\
\includegraphics*[bb=87 308 426 703,width=0.24\linewidth]{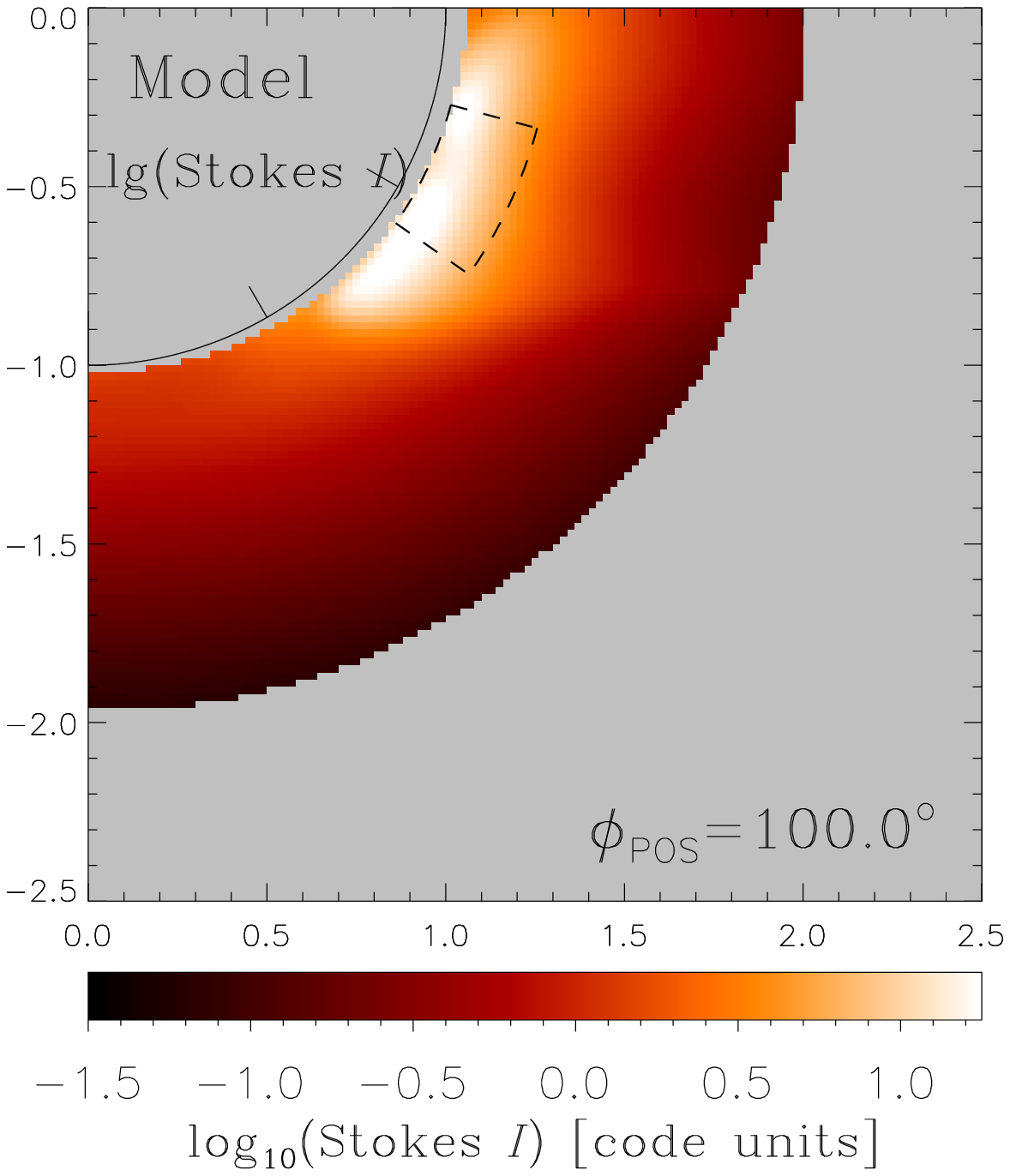}
\includegraphics*[bb=87 308 426 703,width=0.24\linewidth]{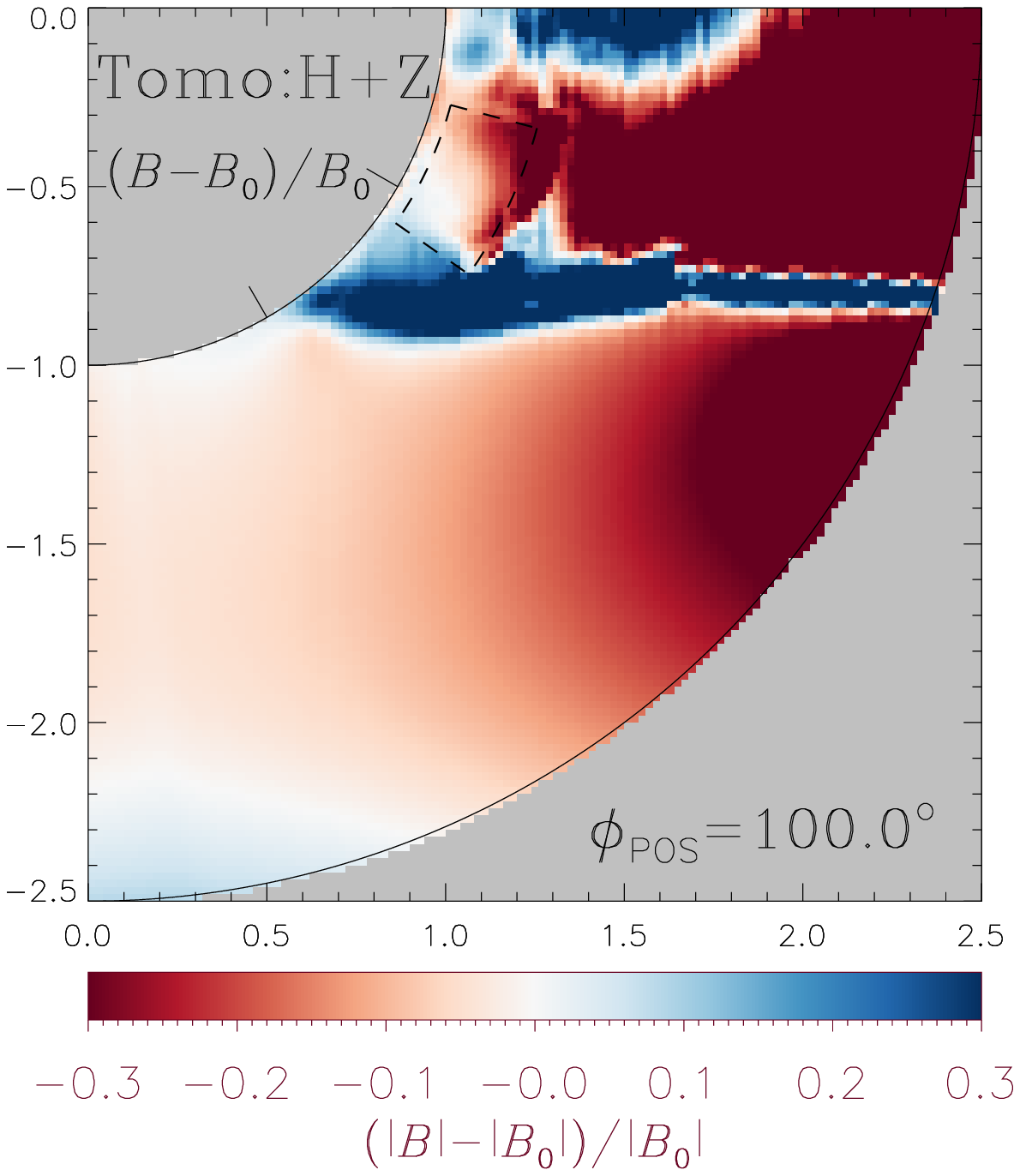}
\includegraphics*[bb=87 308 426 703,width=0.24\linewidth]{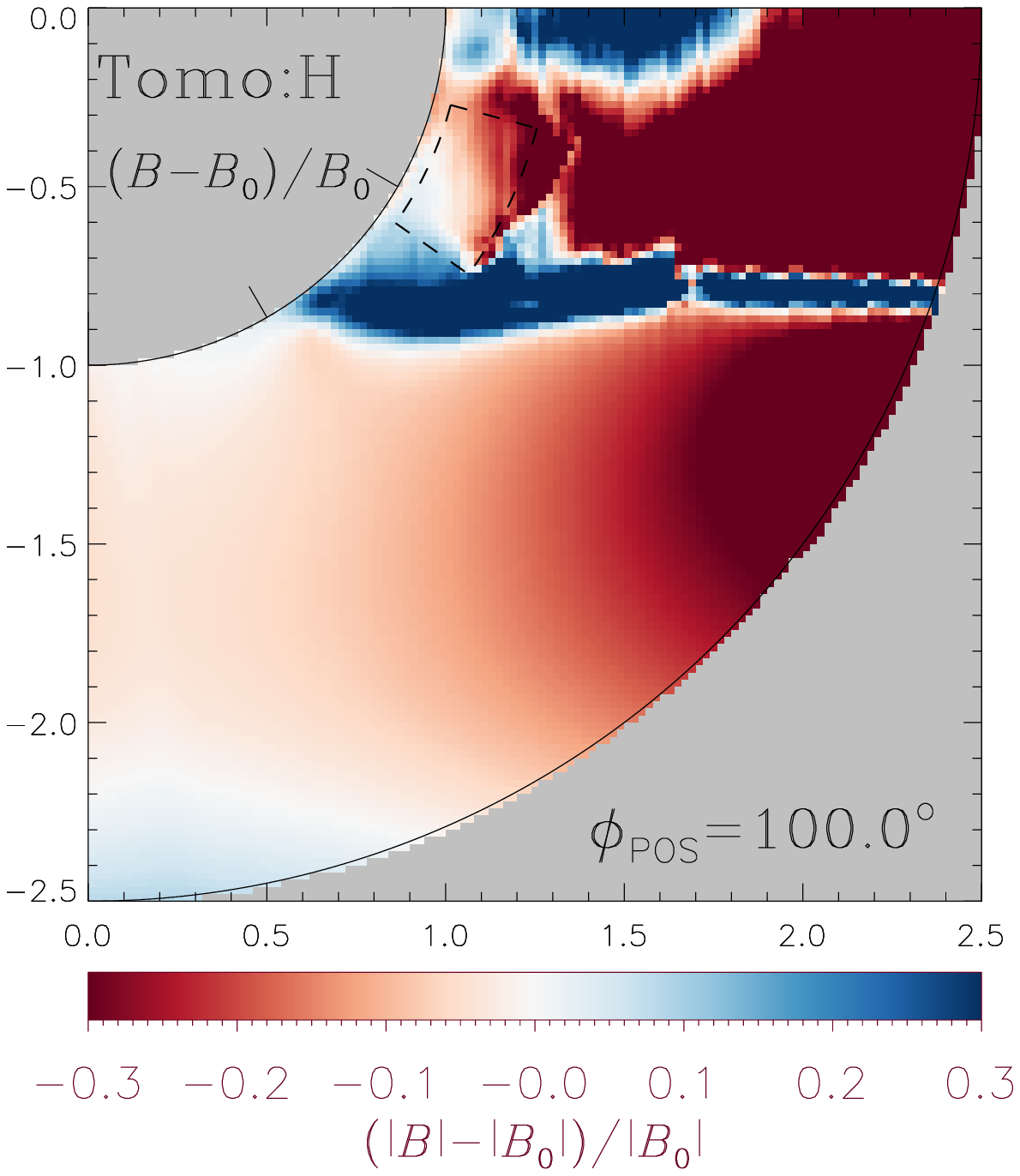}
\includegraphics*[bb=87 308 426 703,width=0.24\linewidth]{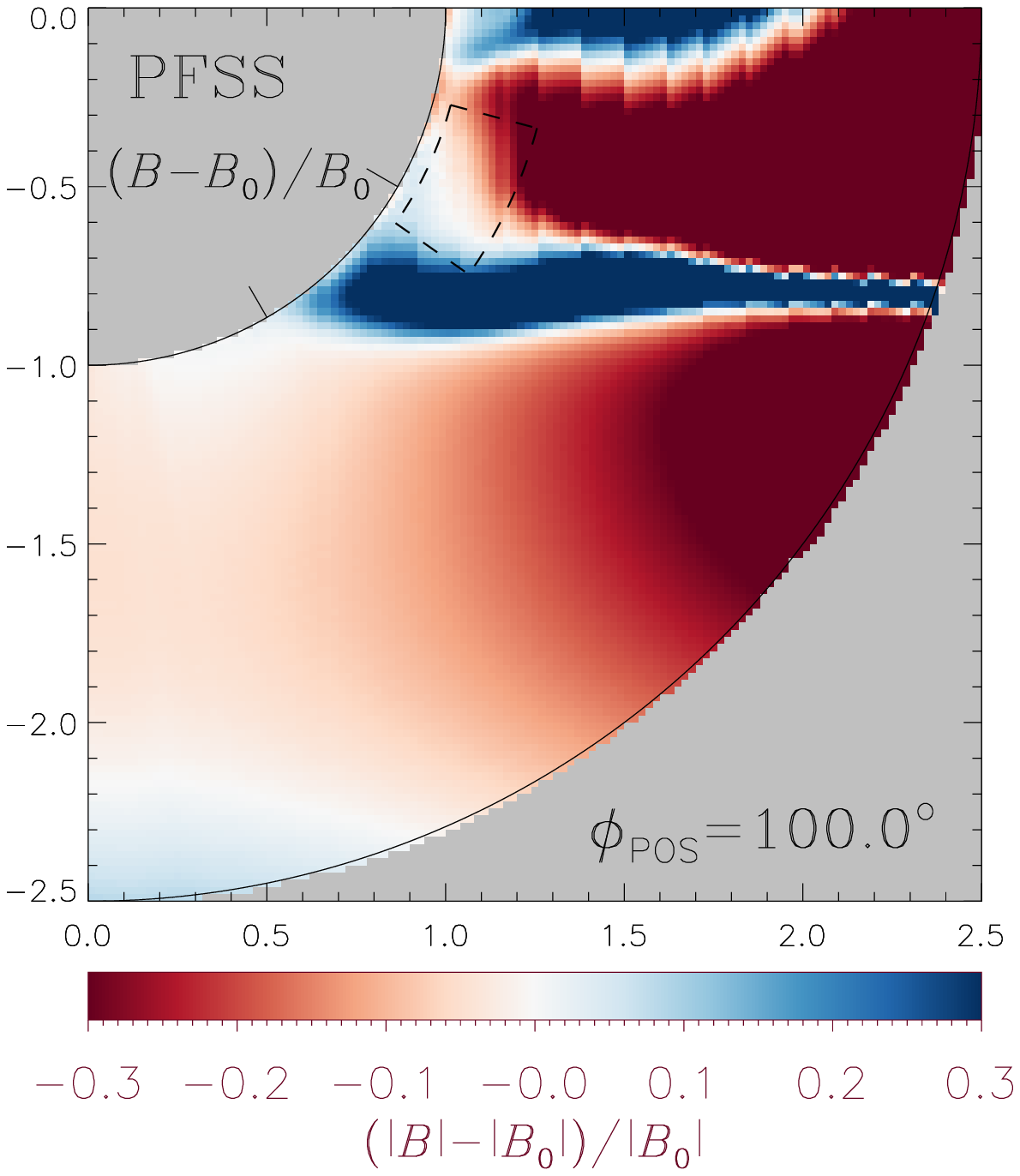}\\
\includegraphics*[bb=87 308 426 703,width=0.24\linewidth]{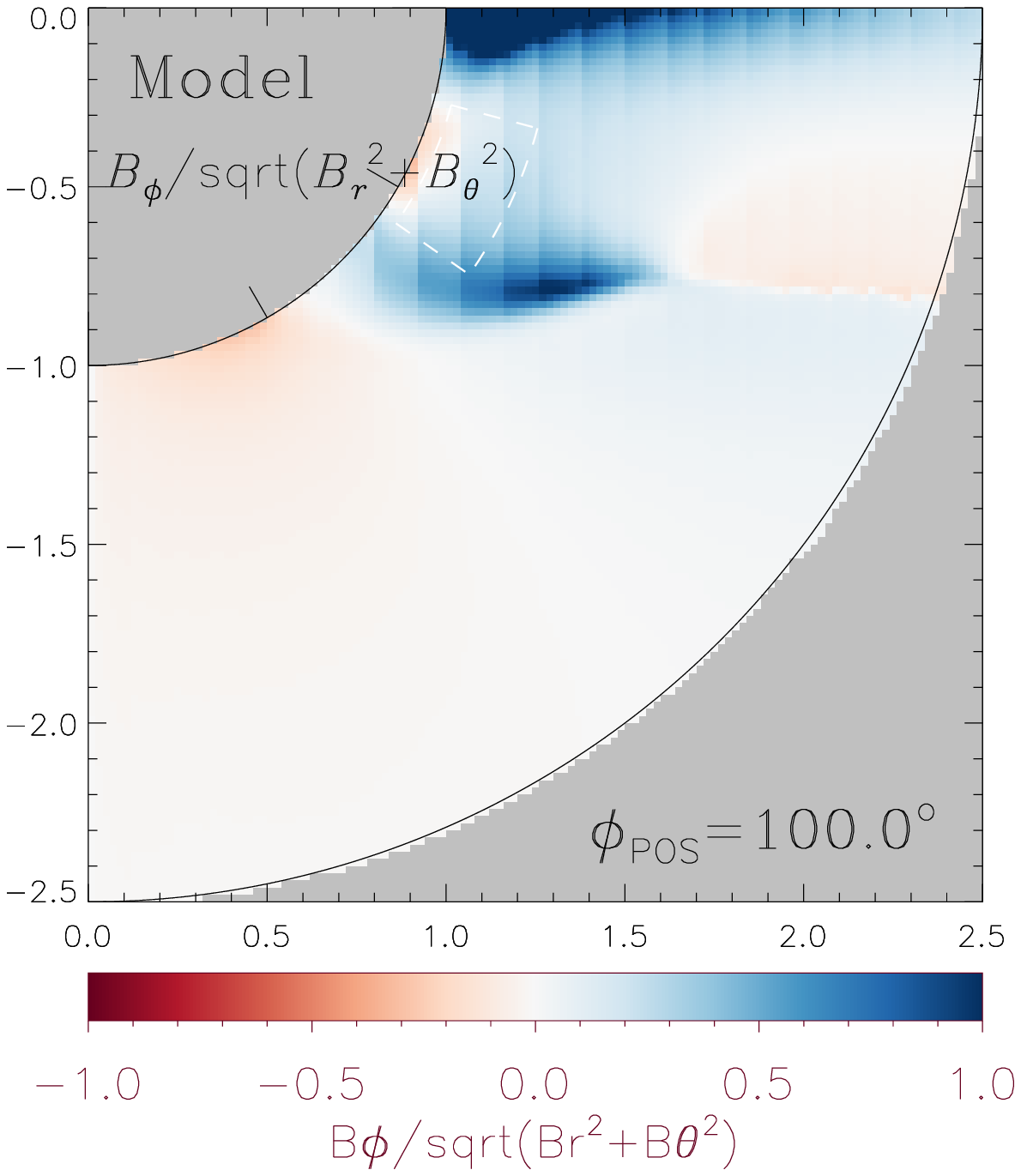}
\includegraphics*[bb=87 308 426 703,width=0.24\linewidth]{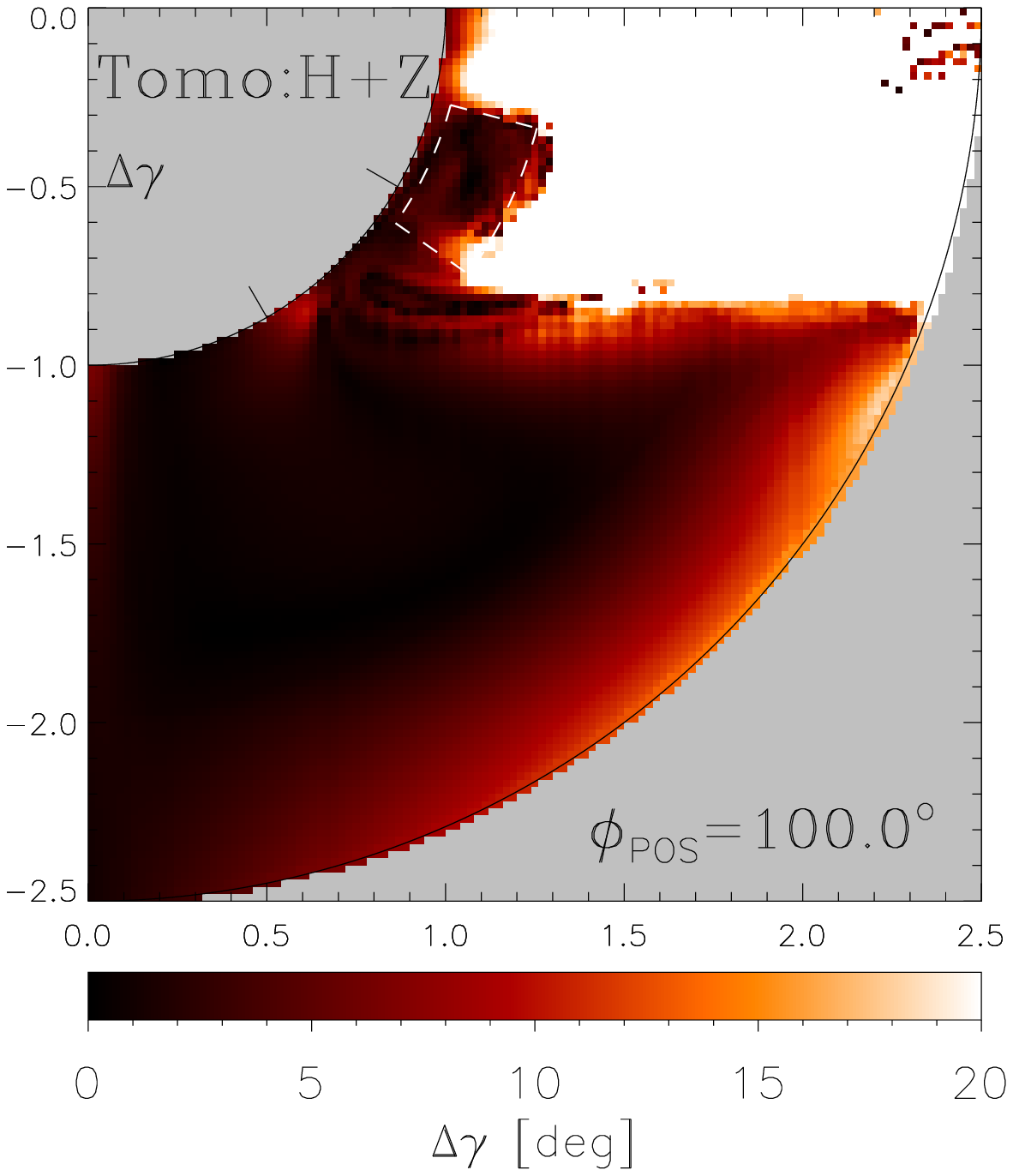}
\includegraphics*[bb=87 308 426 703,width=0.24\linewidth]{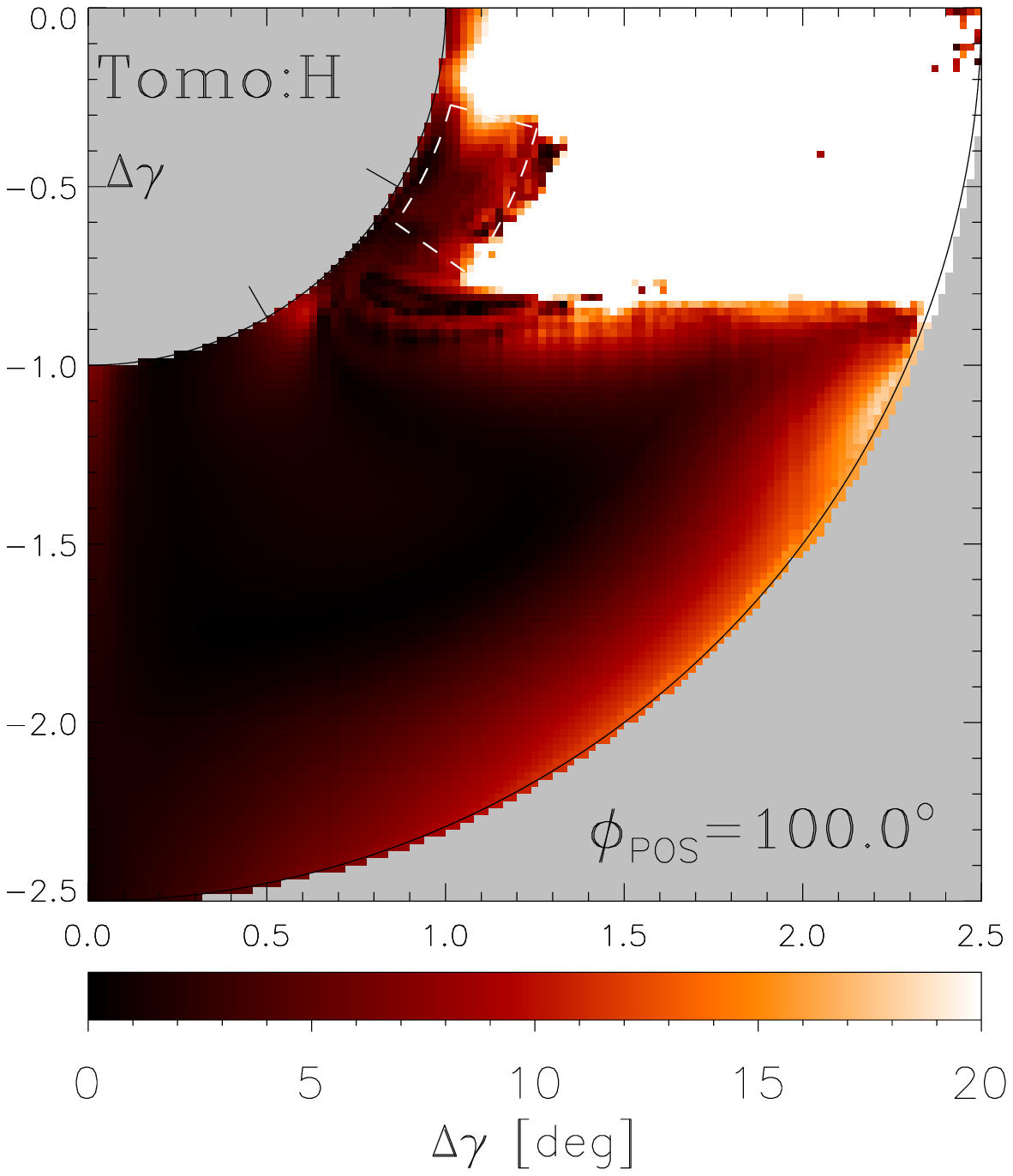}
\includegraphics*[bb=87 308 426 703,width=0.24\linewidth]{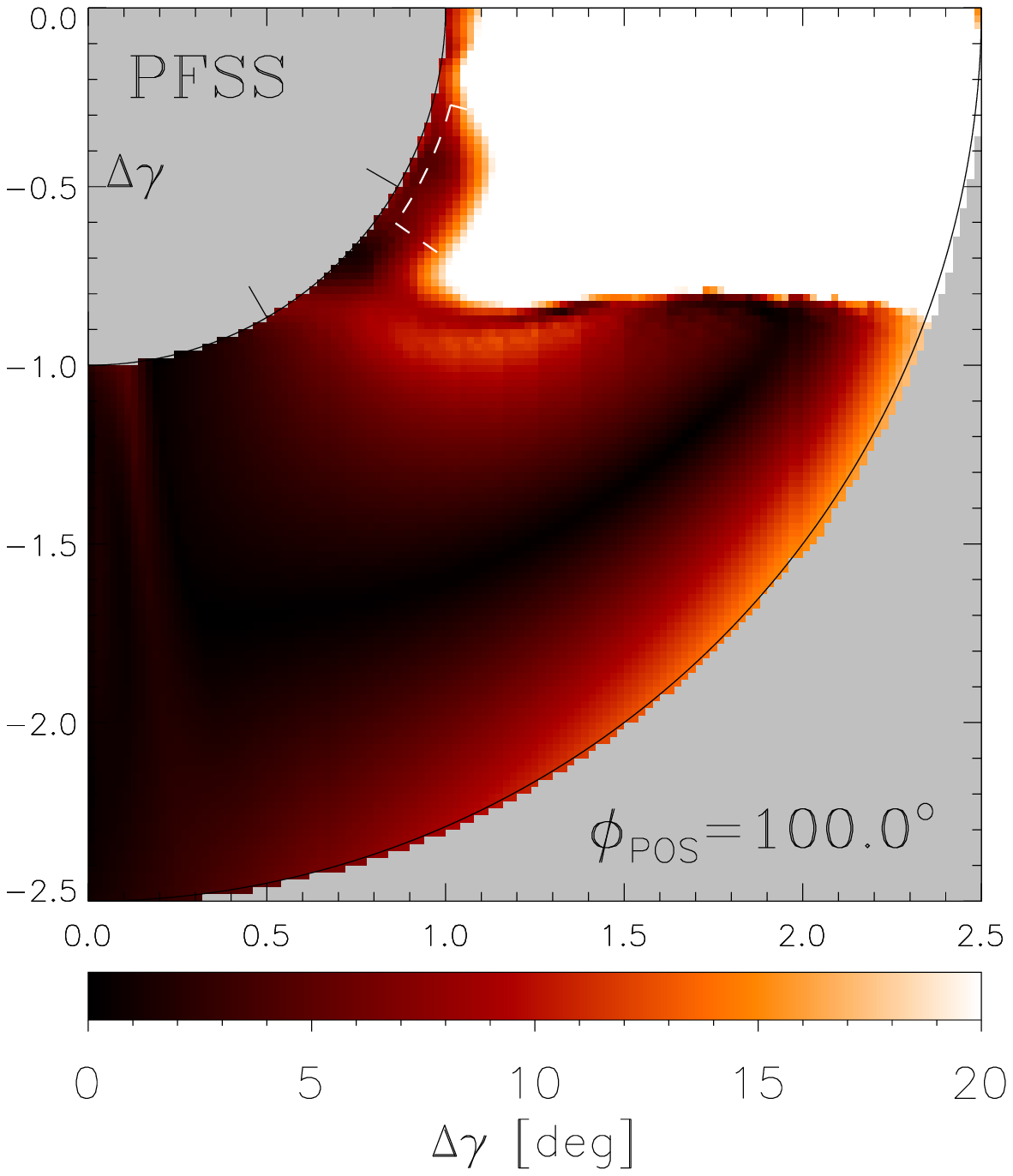}
\caption{Meridional cross-sections of the magnetic field properties for the GT MHD model and errors for the 
tomography reconstructions with full Stokes (second column) and only LP (third column) data, and for starting PFSS model (fourth column). First row shows maps of the magnetic field strength; second and third rows show maps of the absolute and relative field strength errors, respectively; fourth row shows maps of the difference angle between the GT model and the reconstructions. The cross-sections are for Carrington longitude of $100^\circ$.}
\label{Fig_Tomo_Err_Map_All3}
\end{figure}

This result is shown in more details on 
Figures \ref{Fig_Tomo_Err_Map_All}---\ref{Fig_Tomo_Err_Map_All3} which show the meridional cross-sections of the magnetic field properties for the GT MHD model and errors for the 
tomography reconstructions with full Stokes (second column) and only LP (third column) data, and for starting PFSS model (fourth column). 
The Stokes-V FOV is marked by a dashed line around the sector where the Stokes-V observations were taken as projected on to the corresponded plane of the sky (POS). Note, that the observed data are LOS integrated while the Figure show the meridional cross-sections.

\section{Conclusion}

In this assessment we investigated the accuracy of the solar coronal magnetic field reconstructions using the vector tomography method based on full Stokes and only LP measurements of the Fe XIII 10747 \AA \ emission line. 
All measurements were simulated by LOS integration over the Ground Truth (GT) MHD model from Predictive Science Inc (PSI). 
The LP observations (Stokes-Q,U) are "taken" by the UCoMP instrument for a half of solar rotation period (about two weeks) and has FOV up to $2\ R_\odot$. 
The CP observations (Stokes-V) are "taken" by the DKIST Cryo-NIRSP instrument for only two days. Each observational day consists of observation of the area over an active region with the FOV equal to about two maximal Cryo-NIRSP FOV scans (0.25x0.19 $R_\odot$). 

We demonstrated that even this small amount of Stokes-V observations can significantly improve the 3D coronal magnetic field reconstruction over a limited volume in the corona which can be associated with an active region.




\clearpage

\bibliographystyle{apj}

\tracingmacros=2
\bibliography{biblio}

\begin{thebibliography}{}
\expandafter\ifx\csname natexlab\endcsname\relax\def\natexlab#1{#1}\fi

\bibitem[{{Casini} \& {Judge}(1999)}]{Casini_1999}
{Casini}, R., \& {Judge}, P.~G. 1999, \apj, 522, 524

\bibitem[{{House}(1977)}]{House_1977}
{House}, L.~L. 1977, \apj, 214, 632

\bibitem[{{Judge}(1998)}]{Judge_1998}
{Judge}, P.~G. 1998, \apj, 500, 1009

\bibitem[{{Judge} \& {Casini}(2001)}]{Judge_Casini_2001ASPC}
{Judge}, P.~G., \& {Casini}, R. 2001, in Astronomical Society of the Pacific
  Conference Series, Vol. 236, Advanced Solar Polarimetry -- Theory,
  Observation, and Instrumentation, ed. M.~{Sigwarth}, 503

\bibitem[{{Keil} {et~al.}(2004){Keil}, {Rimmele}, {Oschmann}, {Hubbard},
  {Warner}, {Price}, {Dalrymple}, \& {Atst Team}}]{Keil_2004IAUS}
{Keil}, S.~L., {Rimmele}, T.~R., {Oschmann}, J., {et~al.} 2004, in IAU
  Symposium, Vol. 223, Multi-Wavelength Investigations of Solar Activity, ed.
  A.~V. {Stepanov}, E.~E. {Benevolenskaya}, \& A.~G. {Kosovichev}, 581--588

\bibitem[{{Kramar} {et~al.}(2014){Kramar}, {Airapetian}, {Miki{\'c}}, \&
  {Davila}}]{Kramar_2014}
{Kramar}, M., {Airapetian}, V., {Miki{\'c}}, Z., \& {Davila}, J. 2014,
  \solphys, 289, 2927

\bibitem[{{Kramar} \& {Inhester}(2007)}]{Kramar_2007}
{Kramar}, M., \& {Inhester}, B. 2007, \memsai, 78, 120

\bibitem[{{Kramar} {et~al.}(2013){Kramar}, {Inhester}, {Lin}, \&
  {Davila}}]{Kramar_2013}
{Kramar}, M., {Inhester}, B., {Lin}, H., \& {Davila}, J. 2013, \apj, 775, 25

\bibitem[{{Kramar} {et~al.}(2006){Kramar}, {Inhester}, \&
  {Solanki}}]{Kramar_2006}
{Kramar}, M., {Inhester}, B., \& {Solanki}, S.~K. 2006, \aap, 456, 665

\bibitem[{{Kramar} {et~al.}(2016){Kramar}, {Lin}, \&
  {Tomczyk}}]{Kramar_2016ApJL}
{Kramar}, M., {Lin}, H., \& {Tomczyk}, S. 2016, \apjl, 819, L36

\bibitem[{{Lin}(2016)}]{Lin_2016FrASS}
{Lin}, H. 2016, Frontiers in Astronomy and Space Sciences, 3, 9

\bibitem[{{Lin} \& {Casini}(2000)}]{Lin_Casini_2000}
{Lin}, H., \& {Casini}, R. 2000, \apj, 542, 528

\bibitem[{{Lin} {et~al.}(2004{\natexlab{a}}){Lin}, {Kuhn}, \&
  {Coulter}}]{Lin_Kuhn_Coulter_2004}
{Lin}, H., {Kuhn}, J.~R., \& {Coulter}, R. 2004{\natexlab{a}}, \apjl, 613, L177

\bibitem[{{Lin} {et~al.}(2004{\natexlab{b}}){Lin}, {Kuhn}, \&
  {Coulter}}]{Lin_2004}
---. 2004{\natexlab{b}}, \apjl, 613, L177

\bibitem[{{Lin} {et~al.}(2000){Lin}, {Penn}, \&
  {Tomczyk}}]{Lin_Penn_Tomczyk_2000}
{Lin}, H., {Penn}, M.~J., \& {Tomczyk}, S. 2000, \apjl, 541, L83

\bibitem[{{Lionello} {et~al.}(2009){Lionello}, {Linker}, \&
  {Miki{\'c}}}]{Lionello_2009}
{Lionello}, R., {Linker}, J.~A., \& {Miki{\'c}}, Z. 2009, \apj, 690, 902

\bibitem[{{Miki{\'c}} {et~al.}(2018){Miki{\'c}}, {}, {Downs}, {Linker},
  {Caplan}, {Mackay}, {Upton}, {Riley}, {Lionello}, {T{\"o}r{\"o}k}, {Titov},
  {Wijaya}, {Druckm{\"u}ller}, {Pasachoff}, \& {Carlos}}]{Mikic_2018NatAs}
{Miki{\'c}}, {}, Z., {Downs}, C., {et~al.} 2018, Nature Astronomy, 2, 913

\bibitem[{{Miki{\'c}} {et~al.}(2007){Miki{\'c}}, {Linker}, {Lionello}, {Riley},
  \& {Titov}}]{Mikic_2007}
{Miki{\'c}}, Z., {Linker}, J.~A., {Lionello}, R., {Riley}, P., \& {Titov}, V.
  2007, in Astronomical Society of the Pacific Conference Series, Vol. 370,
  Solar and Stellar Physics Through Eclipses, ed. O.~{Demircan}, S.~O. {Selam},
  \& B.~{Albayrak}, 299

\bibitem[{{Querfeld}(1982)}]{Querfeld_1982}
{Querfeld}, C.~W. 1982, \apj, 255, 764

\bibitem[{{Rimmele} {et~al.}(2020){Rimmele}, {Warner}, {Keil}, {Goode},
  {Kn{\"o}lker}, {Kuhn}, {Rosner}, {McMullin}, {Casini}, {Lin}, {W{\"o}ger},
  {von der L{\"u}he}, {Tritschler}, {Davey}, {de Wijn}, {Elmore}, {Fehlmann},
  {Harrington}, {Jaeggli}, {Rast}, {Schad}, {Schmidt}, {Mathioudakis},
  {Mickey}, {Anan}, {Beck}, {Marshall}, {Jeffers}, {Oschmann}, {Beard},
  {Berst}, {Cowan}, {Craig}, {Cross}, {Cummings}, {Donnelly}, {de Vanssay},
  {Eigenbrot}, {Ferayorni}, {Foster}, {Galapon}, {Gedrites}, {Gonzales},
  {Goodrich}, {Gregory}, {Guzman}, {Guzzo}, {Hegwer}, {Hubbard}, {Hubbard},
  {Johansson}, {Johnson}, {Liang}, {Liang}, {McQuillen}, {Mayer}, {Newman},
  {Onodera}, {Phelps}, {Puentes}, {Richards}, {Rimmele}, {Sekulic}, {Shimko},
  {Simison}, {Smith}, {Starman}, {Sueoka}, {Summers}, {Szabo}, {Szabo},
  {Wampler}, {Williams}, \& {White}}]{Rimmele_2020SoPh_DKIST}
{Rimmele}, T.~R., {Warner}, M., {Keil}, S.~L., {et~al.} 2020, \solphys, 295,
  172

\bibitem[{{Ru{\v s}in} {et~al.}(2010){Ru{\v s}in}, {Druckm{\"u}ller}, {Aniol},
  {Minarovjech}, {Saniga}, {Miki{\'c}}, {Linker}, {Lionello}, {Riley}, \&
  {Titov}}]{Rusin_2010}
{Ru{\v s}in}, V., {Druckm{\"u}ller}, M., {Aniol}, P., {et~al.} 2010, \aap, 513,
  A45

\bibitem[{{Schad} \& {Dima}(2021)}]{Schad_Dima_2021ascl_pycelp}
{Schad}, T.~A., \& {Dima}, G.~I. 2021, {pycelp: Python package for Coronal
  Emission Line Polarization}, Astrophysics Source Code Library, record
  ascl:2112.001, ascl:2112.001

\bibitem[{{Tomczyk} {et~al.}(2008){Tomczyk}, {Card}, {Darnell}, {Elmore},
  {Lull}, {Nelson}, {Streander}, {Burkepile}, {Casini}, \&
  {Judge}}]{Tomczyk_2008SoPh}
{Tomczyk}, S., {Card}, G.~L., {Darnell}, T., {et~al.} 2008, \solphys, 247, 411

\bibitem[{{T{\'o}th} {et~al.}(2011){T{\'o}th}, {van der Holst}, \&
  {Huang}}]{Toth_2011ApJ_FDIPS}
{T{\'o}th}, G., {van der Holst}, B., \& {Huang}, Z. 2011, \apj, 732, 102

\bibitem[{{Toth} {et~al.}(2016){Toth}, {van der Holst}, \&
  {Huang}}]{FDIPS_2016ascl_soft06011T}
{Toth}, G., {van der Holst}, B., \& {Huang}, Z. 2016, {FDIPS: Finite Difference
  Iterative Potential-field Solver}, Astrophysics Source Code Library, record
  ascl:1606.011, ascl:1606.011

\bibitem[{{Wiegelmann} {et~al.}(2017){Wiegelmann}, {Petrie}, \&
  {Riley}}]{Wiegelmann_2017SSRv}
{Wiegelmann}, T., {Petrie}, G. J.~D., \& {Riley}, P. 2017, \ssr, 210, 249

\bibitem[{{Woeger} {et~al.}(2021){Woeger}, {Rimmele}, {Casini}, {von der
  Luehe}, {Lin}, {Kuhn}, \& {Dkist Team}}]{Woeger_2021AAS_DKIST}
{Woeger}, F., {Rimmele}, T., {Casini}, R., {et~al.} 2021, in American
  Astronomical Society Meeting Abstracts, Vol.~53, American Astronomical
  Society Meeting Abstracts, 106.02

\bibitem[{{Yeates} {et~al.}(2018){Yeates}, {Amari}, {Contopoulos}, {Feng},
  {Mackay}, {Miki{\'c}}, {Wiegelmann}, {Hutton}, {Lowder}, {Morgan}, {Petrie},
  {Rachmeler}, {Upton}, {Canou}, {Chopin}, {Downs}, {Druckm{\"u}ller},
  {Linker}, {Seaton}, \& {T{\"o}r{\"o}k}}]{Yeates_2018SSRv}
{Yeates}, A.~R., {Amari}, T., {Contopoulos}, I., {et~al.} 2018, \ssr, 214, 99

\end{thebibliography}

\end{document}